\documentclass[aps,10pt,nofootinbib]{revtex4}
\begin{document}
\title{Antilinearity Rather than Hermiticity as a Guiding Principle for Quantum  Theory}
\author{Philip D. Mannheim}
\affiliation{Department of Physics, University of Connecticut, Storrs, CT 06269, USA.
email: philip.mannheim@uconn.edu}
\date{May 4, 2017}
\begin{abstract}

Currently there is much interest in Hamiltonians that are not Hermitian but instead possess an antilinear $PT$ symmetry. Here we seek to put such $PT$ symmetric theories into as general a context as possible. After providing a brief overview of the $PT$ symmetry program, we show that having an antilinear symmetry is the most general condition that one can impose on a quantum theory for which one can have an inner product that is time independent, have a Hamiltonian that is self-adjoint, and have energy eigenvalues that are all real. For each of these properties Hermiticity is only a sufficient condition but not a necessary one, with Hermiticity thus being the special case in which the Hamiltonian has both antilinearity and Hermiticity. As well as being the necessary condition for  the reality of energy eigenvalues, antilinearity in addition allows for the physically interesting cases of manifestly non-Hermitian but nonetheless self-adjoint Hamiltonians that have energy eigenvalues that appear in complex conjugate pairs, or that are Jordan block and cannot be diagonalized at all. We show that one can extend these ideas to quantum field theory, with the dual requirements of the existence of time independent inner products and invariance under complex Lorentz transformations forcing the antilinear symmetry to uniquely be $CPT$. We thus extend the $CPT$ theorem to non-Hermitian Hamiltonians.  For theories that are separately charge conjugation invariant, $PT$ symmetry then follows, with the case for the physical relevance of the $PT$-symmetry program thus being advanced. While $CPT$ symmetry can be defined at the classical level for every classical path in a path integral quantization procedure, in contrast, in such a path integral there is no reference at all to the Hermiticity of the Hamiltonian or the quantum Hilbert space on which it acts, as they are strictly quantum-mechanical concepts that can only be defined after the path integral quantization has been performed and the quantum Hilbert space has been constructed. $CPT$ symmetry thus goes beyond Hermiticity and has primacy over it, with our work  raising the question of how Hermiticity ever comes into quantum theory at all. To this end we show that whether or not a $CPT$-invariant theory has a Hamiltonian that is Hermitian is a property of the solutions to the theory and not of the Hamiltonian itself. Hermiticity thus never needs to be postulated at all.

\end{abstract}
\maketitle

\newpage

\section{Introduction to Antilinear Symmetry}
\label{intro}

\subsection{Overview of the Antilinear Symmetry Program}

Triggered by the fact that the eigenvalues of the non-Hermitian Hamiltonian $H=p^2+ix^3$ are all real \cite{Bender1998,Bender1999}, there has been much interest in the literature (see e.g. the reviews of  \cite{Bender2007,Special2012,Theme2013}) in Hamiltonians that are not Hermitian but have an antilinear $PT$ symmetry, where $P$ denotes parity and $T$ denotes time reversal. (Under $PT$: $p\rightarrow -p$, $x\rightarrow -x$, $i \rightarrow -i$, so that $p^2+ix^3\rightarrow p^2+ix^3$.) Even though the postulate of Hermiticity of a Hamiltonian has been an integral component of quantum mechanics ever since its inception, one can replace it by the more general requirement of antilinear symmetry (antilinearity) without needing to generalize or modify the basic structure of quantum mechanics in any way. Specifically, to construct a sensible Hilbert space description of quantum mechanics one needs to be able to define an inner product that is time independent, and one needs the Hamiltonian to be self-adjoint. There is no need for the inner product to be composed of a ket and its Hermitian conjugate or for the Hamiltonian to be Hermitian. The inner product can be composed of any choice of bra and ket states as long as it is time independent, and for the $PT$ case for instance the appropriate bra for time independence is the $PT$ conjugate of the ket rather than its Hermitian conjugate. And in regard to self-adjointness, it is not necessary that the Hamiltonian be Hermitian, it is only necessary that the Hamiltonian be well-enough behaved in some domain (known as a Stokes wedge) in the complex coordinate plane so that in an integration by parts one can throw away surface terms. And as we show here, the necessary condition for this to be the case is that the Hamiltonian possess an antilinear symmetry. In regard to eigenvalues, we note that while the eigenvalues of a Hermitian Hamiltonian are all real, Hermiticity of a Hamiltonian is only a sufficient condition for such reality but not a necessary one. And again, the necessary condition is  that the Hamiltonian possess an antilinear symmetry, and we note that this condition is in a sense surprising since it involves an operator that acts antilinearly in the space of states rather than linearly, and is thus not ordinarily considered in linear algebra studies.

While antilinear symmetry of a Hamiltonian is the necessary condition for the time independence of inner products, for self-adjointness, and for the reality of eigenvalues, antilinearity goes further as it encompasses physically interesting cases that cannot be achieved with Hermitian Hamiltonians,  while of course also encompassing Hermitian ones since a Hamiltonian can both have an antilinear symmetry and be Hermitian. In general, antilinear symmetry requires that Hamiltonians have energy eigenvalues that all real or have some or all eigenvalues appear in complex conjugate pairs ($E=E_R\pm iE_I$). In addition, antilinear symmetry admits of Jordan-block Hamiltonians that cannot be diagonalized at all. 

The complex conjugate pair case corresponds to the optical cavity gain ($E=E_R+i E_I$) plus loss ($E=E_R-iE_I$) systems that have been explored experimentally in the $PT$ literature \cite{Guo2009} and reviewed in \cite{Special2012,Theme2013}. In the presence of complex conjugate pairs of energy eigenvalues  one still has a time independent inner product, with the only allowed transitions being between the decaying and growing states. In consequence, when a state $|A\rangle$ (the state whose energy has a negative imaginary part) decays into some other state $|B\rangle$ (the one whose energy has a positive imaginary part), as the population of state $|A\rangle$ decreases that of $|B\rangle$ increases in proportion. Thus despite the presence of the growing state $\langle B|$,  the $\langle B|A\rangle$ transition matrix element never grows in time \cite{Mannheim2013}. In contrast, in the standard approach to decays, one has just the decaying mode alone.

As regards Hamiltonians that are not diagonalizable, this is not just of abstract interest since systems have been constructed that expressly correspond to the Jordan-block case for specific values of the parameters in a Hamiltonian \cite{Special2012,Theme2013}, these values being referred to as exceptional points in the $PT$ literature. The Jordan-block case has also been found to occur in the  fourth-order derivative Pais-Uhlenbeck two-oscillator model when the two oscillator frequencies are equal, with the relevant Hamiltonian being shown \cite{Bender2008a} to not be Hermitian but to instead be $PT$ symmetric  (actually $CPT$ symmetric since charge conjugation plays no role here) and non-diagonalizable \cite{Bender2008b}. The fourth-order derivative conformal gravity theory (viz. gravity based on the action $I_{\rm W}=-\alpha_g\int d^4x (-g)^{1/2}C_{\lambda\mu\nu\kappa} C^{\lambda\mu\nu\kappa}$ where $C^{\lambda\mu\nu\kappa}$ is the Weyl conformal tensor) that has been offered \cite{Mannheim2011,Mannheim2012}  as a candidate alternate to the standard Einstein gravity theory also falls into this category, and is able to be ghost free and unitary at the quantum level because of it \cite{Bender2008a,Bender2008b}. 

The Jordan-block case is particularly interesting since for any Jordan-block Hamiltonian the eigenvalues all have to be equal. Jordan-block Hamiltonians that have a total of two eigenvalues and have an antilinear symmetry cannot have all eigenvalues be equal if one is the complex conjugate pair realization, and thus Jordan-block Hamiltonians must fall into the antilinear realization in which all eigenvalues are real.  Jordan-block Hamiltonians with antilinear symmetry thus  provide a direct demonstration of the fact that while Hermiticity implies the reality of eigenvalues, reality does not imply Hermiticity. It will be shown here that in both the Jordan-block case and in the complex conjugate pair realizations of antilinear symmetry the Hamiltonian is still self-adjoint. These two realizations thus provide a direct demonstration of the fact that while Hermiticity implies self-adjointness, self-adjointness does not imply Hermiticity.

With the exception of isolated studies such as the conformal gravity study,  most of the study of Hamiltonians with an antilinear symmetry has been made within the context of non-relativistic quantum mechanics, a domain where one can in principle use any appropriate antilinear symmetry. While a study of general non-relativistic systems is of value for developing understanding of the implications of antilinear symmetry, for any given non-relativistic quantum theory to be of physical relevance it has to be the non-relativistic limit of a relativistically invariant theory. (Even if the system of interest might be composed of slow moving components the observer is free to move with any velocity up to just below the speed of light, and the physics cannot depend on the velocity of the observer.) With a $CPT$ transformation having a direct connection to relativity since its linear part is a specific complex Lorentz transformation, when combined solely with the requirement of the time independence of inner products, through use of complex Lorentz invariance  the allowed antilinear symmetry is uniquely fixed to be $CPT$.  With the $CPT$ theorem previously only having been established for Hermitian Hamiltonians, the $CPT$ theorem is thus extended to the non-Hermitian case. $CPT$ is thus the uniquely favored antilinear symmetry for nature, and any physically relevant theory has to possess it. Since one is below the threshold for particle creation at non-relativistic energies, in non-relativistic quantum mechanics $CPT$ symmetry reduces to $PT$ symmetry, to thus put the $PT$ symmetry program on a quite secure theoretical foundation. Thus for non-relativistic quantum mechanics antilinearity is more basic than Hermiticity, while for relativistic quantum field theory $CPT$ symmetry is uniquely selected as the antilinear symmetry, with antilinearity again being more basic than Hermiticity. In this paper we shall explore antilinearity per se as an interesting concept in and of itself, and shall explore its connection to $CPT$ symmetry. In order to see how the requirement of antilinearity works in practice, for the benefit of the reader we provide a straightforward example.

\subsection{Antilinear Symmetry for Matrices}

A simple model in which one can illustrate the basic features of antilinear symmetry is  the matrix given in \cite{Bender2007}:
\begin{eqnarray}
M(s)=\left(\matrix{1+i&s\cr s&1-i\cr}\right),
\label{H1ab}
\end{eqnarray}
where the parameter $s$ is real and positive. The matrix $M(s)$ does not obey the Hermiticity condition $M_{ij}=M^*_{ji}$. However, if we set $P=\sigma_1$ and  $T=K$, where $K$ denotes complex conjugation we obtain $PTM(s)T^{-1}P^{-1}=M(s)$, with $M(s)$ thus being  $PT$ symmetric for any value of  the real parameter $s$. With the eigenvalues of $M(s)$ being given by $E_{\pm}=1 \pm (s^2-1)^{1/2}$, we see that both of these eigenvalues are real if $s$ is either greater or equal to one, and form a complex conjugate pair if $s$ is less than one. And while the energy eigenvalues would be real and degenerate (both eigenvalues being equal to one) at the crossover point where $s=1$, at this point the matrix becomes of non-diagonalizable Jordan-block form \cite{Mannheim2013}. Neither of the $s=1$ or $s<1$  possibilities is achievable with Hermitian Hamiltonians.

As regards the Jordan-Block case, we recall that in matrix theory Jordan showed that via a sequence of similarity transformations any matrix can be brought either to a diagonal form or to the Jordan canonical form in which all the eigenvalues are on the diagonal, in which the only non-zero off-diagonal elements fill one of the diagonals next to the leading diagonal, and in which all non-zero elements in the matrix are all equal to each other. To see this explicitly for our example, when $s=1$ we note that by means of  a similarity transformation we can bring $M(s=1)$ to the Jordan-block  form
\begin{eqnarray}
\left(\matrix{1&0\cr i&1\cr}\right)\left(\matrix{1+i&1\cr 1&1-i\cr}\right)\left(\matrix{1&0\cr -i&1\cr}\right)=\left(\matrix{1&1\cr 0&1\cr}\right),
\label{H2ab}
\end{eqnarray}
and on noting that 
\begin{eqnarray}
\left(\matrix{1&1 \cr 0&1}\right)\left(\matrix{p \cr q}\right)=\left(\matrix{p+q\cr q}\right)=\left(\matrix{p
\cr q}\right),\qquad \left(\matrix{1&1 \cr 0&1}\right)\left(\matrix{1 \cr 0}\right)=\left(\matrix{1\cr 0}\right)
\label{H3ab}
\end{eqnarray}
for eigenvalue equal to one, we see that the transformed $M(s=1)$ is found to only possess one eigenvector, viz.  the $\widetilde{(1,0)}$ one with $q=0$, where the tilde denotes transpose. Thus even though the secular equation $|M(s=1)-\lambda I|=0$ has two solutions (each with $\lambda=1$), there is only one eigenvector and $M(s=1)$ cannot be diagonalized. (Since the energy eigenvalues have to share the only eigenvector available in the Jordan-block case, they must be degenerate.) Such lack of diagonalizability cannot occur for Hermitian matrices, to show that antilinear symmetry is richer than Hermiticity, with the above $M(s=1)$ being a clearcut example of a non-Hermitian matrix whose eigenvalues are all real, and thus the simplest demonstration of the fact that while Hermiticity implies the reality of eigenvalues, reality does not imply Hermiticity.

To understand why a $PT$-symmetric Hamiltonian must be Jordan block at a transition point  such as $s=1$, we note that in the region where the energy eigenvalues are in complex conjugate pairs their eigenfunctions are given by $\exp(-i(E_R+iE_I)t)$ and $\exp(-i(E_R-iE_I)t)$. Then, as we adjust the parameters in the Hamiltonian so that we approach the transition point from the complex energy region (cf. letting $s$ approach one from below), not only do the two energy eigenvalues become equal, their eigenvectors become equal too, Thus at the transition point there is only one eigenvector, with the Hamiltonian then necessarily being Jordan block. While the Hamiltonian loses an eigenvector at the transition point the Hilbert space on which it acts must still contain two wave functions since it did so before the limit was taken. The combination that becomes the eigenvector in the limit is given by the $E_I\rightarrow 0$ limit 
\begin{eqnarray}
\frac{\exp(-i(E_R+iE_I)t)+\exp(-i(E_R-iE_I)t}{2}\rightarrow \exp(-iE_Rt).
\label{H4ab}
\end{eqnarray}
The second combination is given by the $E_I\rightarrow 0$ limit
\begin{eqnarray}
\frac{\exp(-i(E_R+iE_I)t)-\exp(-i(E_R-iE_I)t}{2iE_I}\rightarrow t\exp(-iE_Rt),
\label{H5ab}
\end{eqnarray}
to thus behave as the non-stationary $t\exp(-iE_Rt)$. The Hilbert space on which the Hamiltonian acts is still complete, it is just the set of stationary states that is not \cite{Bender2008b}. Because of this, wave packets have to be constructed out of the complete set of stationary and non-stationary states combined, with the associated inner products still being preserved in time \cite{Bender2008b}. For the matrix given on the right-hand side of (\ref{H2ab}) for instance, the  right- and left-Schr\"odinger equation wave functions are non-stationary, being given by  
\begin{eqnarray}
i\frac{\partial}{\partial_t}\left(\matrix{(1-it)\exp(-it)
\cr \exp(-it)}\right)&=&\left(\matrix{1&1\cr 0&1\cr}\right)\left(\matrix{(1-it)\exp(-it)
\cr \exp(-it)}\right),\nonumber\\
- i\frac{\partial}{\partial_t}\left(\exp(it), (1+it)\exp(it)\right)&=&\left(\exp(it), (1+it)\exp(it)\right)\left(\matrix{1&1\cr 0&1\cr}\right).
\label{H6ab}
\end{eqnarray}
Their overlap is given by 
\begin{eqnarray}
\left(\exp(it), (1+it)\exp(it)\right)\left(\matrix{(1-it)\exp(-it)
\cr \exp(-it)}\right)=1-it+1+it=2.
\label{H7ab}
\end{eqnarray}
Thus, despite the presence of terms  linear in $t$, their overlap is time independent. In this paper we will have occasion to return to Jordan-block Hamiltonians, and especially to discuss theories such as the illustrative Pais-Uhlenbeck two-oscillator model, whose Hamiltonian appears to be Hermitian but in fact is not.

For the complex conjugate eigenvalue case we can also construct a time-independent inner product. As we shall show in detail in Sec. \ref{antilinearity}, to do this we need to introduce an operator $V$ that effects $VHV^{-1}=H^{\dagger}$. Thus for $M(s<1)$, if we set $\sinh\beta=(1-s^2)^{1/2}/s=\nu/s$, the needed $V$ operator and the right-eigenvectors of $M(s<1)$ are given by \cite{Mannheim2013}
\begin{eqnarray}
V&=&\frac{1}{i\sinh\beta}\left(\sigma_0+ \sigma_2\cosh\beta\right),
\nonumber\\
u_+&=&\frac{e^{-it+\nu t}}{ (2\sinh\beta)^{1/2}}\left(\matrix{e^{\beta/2}\cr -ie^{-\beta/2}\cr}\right),
\qquad
u_-=\frac{e^{-it-\nu t} }{(2\sinh\beta)^{1/2}}\left(\matrix{ie^{-\beta/2}\cr e^{\beta/2}\cr}\right),
\label{H8ab}
\end{eqnarray}
The $V$-operator based inner products obey the expressly time-independent orthogonality and closure relations 
\begin{eqnarray}
&&u_{\pm}^{\dagger}Vu_{\pm}=0,\qquad u_{-}^{\dagger}Vu_{+}=+ 1,\qquad u_{+}^{\dagger}Vu_{-}=- 1,
\qquad u_{+}u^{\dagger}_{-}V-u_{-}u^{\dagger}_{+}V=I,
\label{H9ab}
\end{eqnarray}
with the associated propagator then being given by \cite{Mannheim2013}
\begin{eqnarray}
D(E)=\frac{u_{-}^{\dagger}Vu_{+}}{E-(E_0-i\Gamma)}+\frac{u_{+}^{\dagger}Vu_{-}}{E-(E_0+i\Gamma)}
\label{H10ab}
\end{eqnarray}
(This propagator is the analog of the $\langle \Omega_{+}|\phi(0,t)\phi(0,0)|\Omega_{-}\rangle +\langle \Omega_{-}|\phi(0,t)\phi(0,0)|\Omega_{+}\rangle$ Green's function discussed in Sec. \ref{euclidean} below.) 

Now we recall that in the conventional quantum-mechanical discussion of potential scattering, near a resonance one can parametrize the energy-dependent phase shift as $\tan \delta=\Gamma/(E_0-E)$,  so that $\delta=\pi/2$  at $E=E_0$. With this phase shift the scattering amplitude behaves as
\begin{eqnarray}
f(E)\sim e^{i\delta}\sin\delta= \frac{\Gamma}{E_0-i\Gamma-E},
\label{H11ab}
\end{eqnarray}
and the propagator has the standard Breit-Wigner form
\begin{eqnarray}
D_{\rm BW}(E)=\frac{1}{E-(E_0-i\Gamma)}
=\frac{E-E_0-i\Gamma}{(E-E_0)^2+\Gamma^2},
\label{H12ab}
\end{eqnarray}
with both $f(E)$ and $D_{\rm BW}(E)$ only possessing a decaying mode that behaves as $\exp(-i(E_0-i\Gamma)t/\hbar)$. This decaying mode is associated with a time delay of order $\hbar/\Gamma$ due to the scattered wave being held by the potential. 

In contrast, in the complex conjugate pair case, one has both growing and decaying modes, with the scattering amplitude having poles at both $E_0+i\Gamma$ and $E_0-i\Gamma$, corresponding to both time advance and time delay. In the presence of both types of poles the propagator $D(E)$ is as given in (\ref{H10ab}), and we note that because of the relative minus sign between the residues of the two pole terms as expressly required by (\ref{H9ab}), $D(E)$ takes the form
\begin{eqnarray}
D(E)=\frac{1}{E-(E_0-i\Gamma)}-\frac{1}{E-(E_0+i\Gamma)}
=\frac{-2i\Gamma}{(E-E_0)^2+\Gamma^2}.
\label{H13ab}
\end{eqnarray}
With the imaginary part of $D(E)$ automatically having the same sign as that of the imaginary part of a standard Breit-Wigner, and with it behaving the same way as a Breit-Wigner at the resonance peak where $E=E_0$, the interpretation of $D(E)$ as a probability is thus the standard one that is associated with decays. For our purposes here, we note that the utility in having a complex conjugate pair of energy eigenvalues is that even with states that decay or grow one still has an inner product that is time independent since, as shown in (\ref{H9ab}), the only non-trivial transitions are matrix elements that connect the decaying and growing modes. Thus with a time-independent inner product, the presence of a time advance does not lead to a propagator that violates probability conservation, and the complex conjugate pair realization of antilinear symmetry is fully viable. In passing we note that the interplay between the two complex conjugate poles exhibited in (\ref{H13ab}) has a pre-$PT$ symmetry theory antecedent in the Lee-Wick analysis of the complex conjugate pair realization of the Lee model \cite{Lee1969}, where one has the same $D(E)$ and no violation of probability conservation.

While we can make contact between the antilinear symmetry $D(E)$ propagator and the Breit-Wigner $D_{\rm BW}(E)$ propagator, there are still some key difference between the two cases. For the complex conjugate case there exist experimentally established processes that exhibit both gain and loss, while for the standard Breit-Wigner case one only has loss. Also, as we show in the Appendix, even in the complex conjugate pair case one can still construct a propagator that is causal, i.e. one that does not take support outside the light cone, with the presence of the time advance that accompanies the time delay not violating causality.

In analyzing the eigenspectrum of $M(s>1)$, even though $M(s>1)$ does not obey $M_{ij}=M^*_{ji}$, we should not characterize the $s>1$ situation as being a non-Hermitian case in which all energy eigenvalues are real. The reason for this is that on setting $\sin\alpha=(s^2-1)^{1/2}/s$, we can write
\begin{eqnarray}
&&S(s>1)\left(\matrix{1+i&s \cr s&1-i}\right)S^{-1}(s>1)=\left(\matrix{A&-iB \cr iB&A}\right)\left(\matrix{1+i&s \cr s&1-i}\right)\left(\matrix{A&iB \cr -iB&A}\right)=\left(\matrix{1&\tan\alpha \cr \tan\alpha&1}\right),
\label{H14ab}
\end{eqnarray}
where 
\begin{eqnarray}
A=\left(\frac{1+\sin\alpha}{2\sin\alpha}\right)^{1/2},\qquad
B=\left(\frac{1-\sin\alpha}{2\sin\alpha}\right)^{1/2},
\label{H15ab}
\end{eqnarray}
Thus under the $S(s>1)$ similarity transformation we can bring $M(s>1)$ to a form $S(s>1)M(s>1)S^{-1}(s>1)=M^{\prime}$ which does obey $M^{\prime}_{ij}=M^{\prime *}_{ji}$. With $M^{\prime}$ being Hermitian the matrix $M(s>1)$ is actually Hermitian in disguise. The similarity transformation needed to bring $M(s>1)$ to a Hermitian form is not unitary and is thus a transformation from a skew basis to an orthogonal one. The definition of Hermiticity as the condition $M^{\prime}_{ij}=M^{\prime *}_{ji}$ is not a basis-independent definition. To be specific, consider a Hamiltonian $H$ that obeys $H_{ij}=H_{ji}^*$ in some given basis. Now apply a similarity transformation $S$ to a new  basis to construct $H^{\prime}=SHS^{-1}$. In the new  basis we have 
\begin{eqnarray}
[H^{\prime}]^{\dagger}=[S^{-1}]^{\dagger}H^{\dagger}S^{\dagger}=[S^{-1}]^{\dagger}HS^{\dagger}=[S^{-1}]^{\dagger}S^{-1}H^{\prime}SS^{\dagger}.
\label{F16ab} 
\end{eqnarray}
As we see,  $[H^{\prime}]^{\dagger}$ is not in general equal to $H^{\prime}$, being so only if $S$ is unitary. Thus to say that a Hamiltonian is Hermitian is to say that one can find a  basis  in which $H_{ij}$ is equal to $H_{ji}^*$, with the basis-independent statement being that the eigenvalues of a Hermitian operator are all real and the eigenvectors are complete. And if a Hamiltonian with these properties is not in  a basis in which $H_{ij}=H_{ji}^*$, the Hamiltonian is Hermitian in disguise. In consequence, matrices such as $M(s>1)$ are Hermitian in disguise even though they do not appear to be so, and are in the quasi-Hermitian class of operators discussed in \cite{Scholtz1992}. With the Hamiltonian $H=p^2+ix^3$ possessing an energy eigenspectrum that is real and complete, $H=p^2+ix^3$ is also Hermitian in disguise. The utility of antilinear symmetry is that since it is the necessary condition for the reality of eigenvalues, if a Hamiltonian does not possess an antilinear symmetry one can conclude immediately that not all of its eigenvalues can be real, and one is able to make such a claim without needing to actually determine any single eigenvalue at all or seek a similarity transformation that could establish that the Hamiltonian is Hermitian in disguise. For any Hamiltonian that does descend from a relativistic theory, the required antilinear symmetry is uniquely prescribed to be $CPT$,  so one only has to check whether to not it might be $CPT$ invariant. (We will show below that this in fact the case for $H=p^2+ix^3$.) 

On recognizing the matrix $M(s)$ as being Hermitian in disguise when $s>1$, we see that whether or not a Hamiltonian is Hermitian or Hermitian in disguise is a property of the solutions to the theory, and is something that cannot be determined by inspection. While we have seen that a Hamiltonian can be Hermitian (in disguise) even if does not appear to be so, below we will find examples of Hamiltonians that are not Hermitian (and not even Hermitian in disguise) even though they do appear to be so.

Even though a non-linear condition such as $H=H^{\dagger}$ is not preserved under a similarity transformation, we should note that in contrast commutation relations are preserved under similarity transformations. While standard for linear operators, a relation such as $[H,A]=0$ where $A=LK$ ($A$ antilinear, $L$ linear) is also preserved when $A$ is antilinear, though, as noted in \cite{Mannheim2013},  under a similarity transformation it would be  the transformed $L$  that would obey $[H^{\prime},L^{\prime}K]=0$. Specifically, if we set $H^{\prime}=SHS^{-1}$, $L^{\prime}=SL[S^{-1}]^*$,  then 
\begin{eqnarray}
[L^{\prime}K,H^{\prime}]=SL[S^{-1}]^*KSHS^{-1}-SHS^{-1}SL[S^{-1}]^*K=S[LK,H]S^{-1}. 
\label{F17ab}
\end{eqnarray}
In consequence, the commutation relation $[CPT,H]=0$ is preserved under a similarity transform, a very powerful constraint, with the linear part of $CPT$  transforming as would be needed.

The same is true for the $PT$ operator. Specifically, if we want to maintain the discrete properties of $P$ and $T$, we set $P=\pi$, $T=\tau K =K \tau^*$  and require that $P^2=I$, $T^2=I$, $[P,T]=0$, to then  obtain $\pi^2=I$, $\tau\tau^*=I$, $\pi\tau=\tau\pi^*$. If we now make a similarity transform $SPS^{-1}=P^{\prime}$, $STS^{-1}=T^{\prime}$ and set $P^{\prime}=\pi^{\prime}$, $T^{\prime}=\tau^{\prime}K$, then  with $\pi^{\prime}=S\pi S^{-1}$, $\tau^{\prime}=S\tau (S^{-1})^*$ we obtain $P^{\prime 2}=I$, $T^{\prime 2}=I$, $[P^{\prime},T^{\prime}]=0$, and $\pi^{\prime 2}=I$, $\tau^{\prime} \tau^{{\prime}*}=I$, $\pi^{\prime}\tau^{\prime}=\tau^{\prime}\pi^{{\prime}*}$. If we transform a Hamiltonian $H$ obeying $H=PTHTP=\pi\tau H^*\tau^*\pi$, we find that 
\begin{eqnarray}
H^{\prime}=SHS^{-1}=P^{\prime}T^{\prime}H^{\prime}T^{\prime}P^{\prime}=\pi^{\prime}\tau^{\prime} H^{{\prime} *}\tau^{{\prime}*}\pi^{\prime}, 
\label{F18ab}
\end{eqnarray}
with $PT$ symmetry being maintained. Thus unlike a Hermiticity condition, an antilinear symmetry relation is  not basis dependent, and is thus far more powerful to work with.

While our discussion of $M(s)$ has only been made for finite-dimensional matrices, it immediately generalizes to infinite-dimensional ones since one can work in occupation number space. However, something unexpected can happen. If we evaluate a matrix element such as $\langle\Omega|\Omega\rangle$ where $|\Omega\rangle$ is the no-particle state, we can introduce a complete set of position eigenstates according to $\langle\Omega|\Omega\rangle=\int dx \langle\Omega|x\rangle\langle x|\Omega\rangle=\int dx \psi^*_0(x)\psi_0(x)$ where $x$ is real, and it can turn out that wave functions such as $\psi_0(x)$ might not be normalizable on the real $x$ axis. While they would be normalizable on the real $x$ axis in the standard self-adjoint Hermitian case, in the antilinear case one might need  to continue the coordinate $x$ into the complex plane in order to obtain a wave function that is normalizable, and it is only in such complex domains where $\int dx \psi^*_0(x)\psi_0(x)$ is finite that the Hamiltonian is then self adjoint. We thus turn now to a discussion of self-adjointness as it pertains to Hamiltonians with antilinear symmetry.

\subsection{Self-Adjointness}

In regard to self-adjointness, we note that to show that a quantum-mechanical operator such as the momentum operator  $\hat{p}=-i\partial_x$ (or the Hamiltonian that is built out of it) acts as a Hermitian operator in the space of the wave functions of the Hamiltonian, one has to integrate by parts and be able to throw away spatially asymptotic surface terms. ($[-i\int dx \psi_1^*\partial_x\psi_2]^*=-i\int dx \psi_2^*\partial_x\psi_1+i(\psi^*_2\psi_1)|^{\infty}_{-\infty}$.) In a $PT$ symmetric  or some general antilinearly symmetric situation  this procedure can be realized by allowing for the possibility that one may have to rotate into the complex $(x,p)$ plane in order to find so-called Stokes wedges in which one can throw surface terms away \cite{Bender2007} when it is not possible to do so on the real axis. A typical example is the divergent Gaussian $\exp(x^2)$. It is not normalizable on the real $x$-axis, but is normalizable on the imaginary $x$-axis, and would be of relevance if the momentum operator $\hat{p}$ were to be anti-Hermitian rather than Hermitian, and thus represented by $\partial_{x}$, with the $[\hat{x},\hat{p}]=i$ commutator being realized as $[-ix,\partial_x]=i$. The difference between the $-i\partial_x$ and $\partial_x$ representations of the momentum operator is only in a fully permissible commutation-relation-preserving similarity transformation into the complex plane through an angle $\theta=-\pi/2$, since the general angle  $\hat{S}=\exp(-\theta \hat{p}\hat{x})$ effects 
\begin{eqnarray}
\hat{S}\hat{p}\hat{S}^{-1}=\hat{p}\exp(-i\theta), \qquad \hat{S}\hat{x}\hat{S}^{-1}=\hat{x}\exp(i\theta),
\label{F19ab}
\end{eqnarray}
while preserving both the relation $[\hat{x},\hat{p}]=i$ and the eigenvalues of a Hamiltonian $\hat{H}(\hat{x},\hat{p})$ that is built out of $\hat{x}$ and $\hat{p}$. 

A commutation relation is actually not defined until one can specify a good test function on which it can act according to $[\hat{x},\hat{p}]\psi(x)=i\psi(x)$, as the commutation relation can be represented by $[\bar{x},-i\partial_{\bar{x}}]\psi(\bar{x})=i\psi(\bar{x})$ for any $\bar{x}=x\exp(i\theta)$, with wave functions potentially only being normalizable for specific, non-trivial domains in  $\theta$. It is the domain in the complex $x$ plane for which the test function is normalizable that determines the appropriate differential representation for an operator. Until one has looked at asymptotic boundary  conditions, one cannot determine whether an operator is self-adjoint or not, since such self-adjointness is determined not by the operator itself but by the space of states on which it acts. When acting on its own eigenstates according to $\hat{x}|x\rangle=x|x\rangle$, the position operator is self-adjoint and Hermitian.  When acting on the eigenstates of $\hat{H}(\hat{x},\hat{p})$ it may not be self-adjoint until it is continued into the complex plane according to $\hat{x}^{\prime}=\hat{S}\hat{x}\hat{S}^{-1}$. However now $\hat{x}^{\prime}$ would not be Hermitian. Since $\hat{p}^{\prime}=\hat{S}\hat{p}\hat{S}^{-1}$ would then not be Hermitian either, $\hat{H}^{\prime}(\hat{x}^{\prime},\hat{p}^{\prime})=\hat{S}\hat{H}(\hat{x},\hat{p})\hat{S}^{-1}$ would in general not be Hermitian as well. In securing self-adjointness one can thus lose Hermiticity. It is only when $\hat{x}$ is self-adjoint when acting on the eigenstates of $\hat{H}(\hat{x},\hat{p})$ without any continuation into the complex plane being needed (viz. $\theta=0$) that $\hat{H}(\hat{x},\hat{p})$ could be  Hermitian, with its wave functions $\psi(x)=\langle x|\psi\rangle$ then being normalizable on the real $x$ axis. 

A self-adjointness mismatch between the action of the  position and momentum operators on their own eigenstates and on those of the Hamiltonian is one of the key components of  the $PT$-symmetry program or of any general antilinear-symmetry program, with a continuation into the complex $(x,p)$ plane being required whenever there is  any such mismatch, something that is expressly found to be the case for $H=p^2+ix^3$. The art of the $PT$-symmetry program then is the art of determining in which domain in the relevant complex plane the wave functions of a Hamiltonian are well-behaved asymptotically, with many examples being provided in \cite{Bender2007,Special2012,Theme2013}.  In the following we will present examples in which manifestly non-Hermitian Hamiltonians that are either Jordan Block or have complex conjugate eigenvalues are nonetheless self-adjoint in appropriate Stokes wedges in the complex plane. Self-adjointness is thus more general than Hermiticity and encompasses it as a special case.

\subsection{Organization of the Paper}

The present paper is organized as follows. Given our above study of the properties of the particular matrix $M(s)$, in Sec. \ref{antilinearity} we extend our study of antilinearity of a Hamiltonian as an alternative to Hermiticity to the general case. And following \cite{Mostafazadeh2002} and \cite{Mannheim2013} we show that antilinearity is both necessary and sufficient to secure the time independence of the most general allowed Hilbert space inner products, and thus secure conservation of probability. While most of the results presented in Sec. \ref{antilinearity} are already in the literature, some of  our derivations are new. Also new is the centrality and emphasis we give to the time independence of inner products. In Secs. \ref{intro} and \ref {antilinearity} we study antilinearity in and of itself, while starting in Sec. \ref{cpt} we study how relativity constrains this analysis. The material presented in  Secs. \ref{intro} and \ref{antilinearity} is primarily preparatory, with the remainder of the paper then presenting new results that had not previously been reported in the literature. 

In  Sec. \ref{cpt} we show that the Lorentz group has a natural complex  extension, and then identify the linear component of a $CPT$ transformation as being a specific complex Lorentz transformation. With this property we can then show that once one imposes complex Lorentz invariance the antilinear symmetry associated with the time independence of inner products is uniquely prescribed to be $CPT$.  

In Sec. \ref{implications} we apply these results to some interesting $CPT$ theories such as the $H=p^2+ix^3$ theory and the fourth-order derivative Pais-Uhlenbeck two-oscillator model. We show that the Pais-Uhlenbeck model admits of explicit realizations in which the energy eigenvalues of the Hamiltonian come in complex conjugate pairs or in which the Hamiltonian is a Jordan-block  Hamiltonian that cannot be diagonalized at all. Both of these two realizations are shown to be $CPT$ symmetric, to thus provide explicit examples of manifestly non-Hermitian Hamiltonians that are $CPT$ invariant. 

One of the surprising results of our work is that we find that whether we use Hermiticity to derive the $CPT$ theorem or use complex Lorentz invariance and probability conservation to derive the $CPT$ theorem, in both the cases the allowed Hamiltonians that we obtain are always of exactly the same form, the same operator structure and the same reality pattern for coefficients. Despite this, it does not follow that the only allowed Hamiltonians are then Hermitian, since the Hermticity that is being appealed to here is that of the individual operators in the Hamiltonian and their coefficients and not that of the Hamiltonian itself. And we had noted above that when the generic $\hat{H}(\hat{x},\hat{p})$ acts on the eigenstates of $\hat{x}$ it might not be self-adjoint even though $\hat{x}$ itself is self-adjoint when acting on that very same basis. Moreover, as the $M(s)$ example given above shows, even if the secular equation $|H-\lambda I|=0$ is a real equation for any value of the parameters, it  can have real or complex solutions depending on the range of the parameters. As we discuss in Secs. \ref{cpt} and \ref{implications}, if we do start only from the requirements of time independence of inner products and complex Lorentz invariance, we may then obtain Hamiltonians that are Hermitian for certain ranges of parameters. For such cases though, we cannot immediately tell ahead of time what those ranges might be and need to solve the theory first, with Hermiticity not being determinable merely by inspection of the form of the Hamiltonian. Thus Hermiticity of a Hamiltonian never needs to be postulated, with it being output rather than input in those cases where it is found to occur.

In Sec. \ref{comparing} we show that the illustrative two-oscillator Pais-Uhlenbeck Hamiltonian is self-adjoint even when it is Jordan block or when energy eigenvalues come in complex conjugate pairs, to thus provide an explicit example in which a non-Hermitian Hamiltonian is self-adjoint. In this section we show that in general the connection between antilinearity and self-adjointness is very tight -- for any Hamiltonian antilinearity implies self-adjointness, and self-adjointness implies antilinearity. We should thus associate self-adjointness with antilinearity rather than with Hermiticity, with its association with Hermiticity being the special case.  

In deriving the $CPT$ theorem in Sec. \ref{cpt}, we find that a $CPT$-invariant Hamiltonian has to obey $H=H^*$. With the Euclidean time evolution operator being given by $\exp(-\tau H)$, it follows that for time-independent Hamiltonians the Euclidean time Green's functions and path integrals have to be real. In Sec. \ref{euclidean} we explore this aspect of the $CPT$ theorem in some Hermitian and non-Hermitian cases  and show that  $CPT$ symmetry is a necessary and sufficient condition for the reality of the field-theoretic Euclidean time Green's functions and path integrals, while Hermiticity is only a sufficient condition for such reality. As such, this result  generalizes to field theory a similar result found in \cite{Bender2002,Bender2010} for matrices.

In quantizing a physical system one can work directly with quantum operators acting on a Hilbert space and impose canonical commutation relations for the operators, a q-number approach, or one can quantize using Feynman path integrals, a purely c-number approach. In constructing the appropriate classical action needed for the path integral approach, one ordinarily builds the action out of real quantities, because real quantities are the eigenvalues of Hermitian  quantum operators. However, as we show in Sec. \ref{constraining}, this is inadequate in certain cases, and particularly so in minimally coupled electrodynamics (while $\partial_{\mu}-eA_{\mu}$  is real, it is only $i\partial_{\mu}-eA_{\mu}$ that can be Hermitian in the quantum case), with the correct $i\partial_{\mu}-eA_{\mu}$ based classical action being constructed by requiring that it be $CPT$ symmetric instead (classically $i\partial_{\mu}$ and $eA_{\mu}$ are both $CPT$ even, since classically the product $eA_{\mu}$ is $C$ even). 

Since the space of states needed for self-adjointness could be in the complex plane rather than on the real axis,  one has to ask what happens to the antilinear symmetry as one continues into the complex plane. In Sec. \ref{continuing} we show that despite the fact that the antilinear symmetry acts non-trivially on angles that are complex, in such a complex plane continuation both the antilinear operator and the Hamiltonian transform so that their commutation relation is preserved. 

A central theme of this paper is the primacy of antilinearity over Hermiticity. This is manifested in the canonical quantization approach to quantum mechanics, where c-number Poisson brackets  are replaced by q-number commutators, and one constructs a q-number Hamiltonian operator that acts on quantum-mechanical states in a quantum-mechanical Hilbert space. In and of itself nothing in the canonical  quantization procedure makes any reference to Hermiticity per se or forces the q-number Hamiltonian to necessarily be Hermitian (one usually just takes it to be so). However, as discussed in Sec. \ref{continuing}, there is, as with any symmetry, a correlation between an antilinear symmetry in the classical theory and one in the quantum theory that is derived from it by canonical quantization.  A quantum theory can thus inherit an antilinear symmetry from an underlying classical theory, and a quantum Hamiltonian can have an antilinear symmetry without being Hermitian, with antilinearity being  more far reaching than Hermiticity while encompassing it as a special case. 

The contrast between antilinearity and Hermiticity is even more sharp in path integral quantization, since path integral quantization  is a completely c-number approach in which no reference is made to any quantum-mechanical Hilbert space at all. Rather, path integral quantization enables one to construct quantum-mechanical matrix elements (viz. Green's functions such as $\langle \Omega|T[\phi(x_1)\phi(x_2)]|\Omega \rangle$ or the more general ones such as $\langle \Omega_L|T[\phi(x_1)\phi(x_2)]|\Omega_R \rangle$ that we introduce below in Sec. \ref{euclidean}) without one needing to construct the quantum operators and Hilbert space themselves. Once one has constructed these matrix elements one can construct  a quantum-mechanical Hamiltonian time evolution operator and  Hilbert space that would yield them. However, since there is no reference to any quantum-mechanical Hilbert space in the path integral itself (it being an integral over strictly classical paths alone), there is no immediate reason to presume that the resulting  quantum-mechanical system would be one in which the quantum Hamiltonian would be Hermitian. 

Path integral quantization thus raises the question \cite{Mannheim2013} of how quantum-mechanical Hermiticity ever comes into physics at all, and what there would be in any given c-number path integral that would indicate whether the associated quantum-mechanical Hamiltonian would or would not be Hermitian. In Sec. \ref{continuing} we address this question by showing that for any pair of canonical variables such as $q$ and $p$, there is a correspondence principle between complex similarity transformations on the q-number $\hat{q}$ and $\hat{p}$ in the quantum theory  and symplectic transformations through the selfsame complex angles on the c-number $q$ and $p$ in the classical  theory. Use of this  complex plane correspondence principle enables us to show that only if the path integral exists with a real measure and its Euclidean time continuation is real could the quantum-mechanical Hamiltonian be Hermitian, though even so, the results of this paper require that it would also possess an antilinear $CPT$ symmetry. However, if the path integral only exists with a complex measure, the Hamiltonian would be  $CPT$ symmetric but not Hermitian (though it could still be Hermitian in disguise). It is thus through the existence of a real measure path integral that Hermiticity can enter quantum theory.

In Sec. \ref{final} we make some final comments. In an Appendix we discuss the Majorana basis for the Dirac gamma matrices, a basis that is very convenient for discussing the relation between $CPT$ transformations and the complex Lorentz group. Also  in the Appendix we present a quantization scheme for fermion  fields in which complex conjugation acts non-trivially on the fermion fields. With this quantization scheme we find that all spin zero fermion multilinears are real, something that will prove central to the proof of the $CPT$ theorem that we give in this paper. In addition, we compare and contrast the charge conjugation operator $C$ with the ${\cal C}$ operator that appears \cite{Bender2007} in $PT$ studies. Finally in the Appendix we show how causality is maintained in all the various realizations (real, Jordan block, complex conjugate pair energy eigenvalues) of a non-Hermitian but $CPT$-symmetric fourth-order derivative scalar field theory.

\section{Antilinearity as a Basic Principle for Quantum Theory}
\label{antilinearity}

\subsection{Necessary Condition for the Reality of Eigenvalues}

In order to identify the specific role played by antilinearity, we consider  some generic discrete antilinear operator $A$ with $A^2=I$, an operator we shall write as $A=LK$ where $L$ is a linear operator,  $K$ is complex conjugation, $K^2=I$, $LL^*=I$, and $A^{-1}=KL^{-1}$. It is instructive to look first not at the eigenvector equation $H|\psi \rangle =E|\psi \rangle$ itself, but at the secular equation  $f(\lambda)=|H-\lambda I|=0$ that determines the eigenvalues of $H$. In \cite{Bender2002} it was noted that if $H$ has an antilinear symmetry, then the eigenvalues  obey 
\begin{eqnarray}
f(\lambda)=|H-\lambda I|=|AHA^{-1}-\lambda I|=|LKHK^{-1}L^{-1}-\lambda I|=|KHK^{-1}-\lambda I|=|H^*-\lambda I|=0.
\label{H20ab}
\end{eqnarray}
In consequence  $H$ and $H^*$ both have the same set of eigenvalues, with $f(\lambda)$ thus being a real function of $\lambda$ (viz. in an expansion $f(\lambda)=\sum a_n\lambda^n$ all $a_n$ are real). Then in \cite{Bender2010} the converse was shown, namely if $f(\lambda)$ is a real function of $\lambda$, $H$ must have an antilinear symmetry. If $f(\lambda)$ is a real function the eigenvalues can be real or appear in complex conjugate pairs (just as we found in our $M(s)$ example), while if  $f(\lambda)$ is not real the condition $f(\lambda)=0$ must have at least one complex solution. Antilinear symmetry is thus seen to be the necessary condition for the reality of  eigenvalues, while Hermiticity is only a sufficient condition. 

\subsection{Necessary and Sufficient Condition for the Reality of Eigenvalues}

As to a condition that is both necessary and sufficient, in $PT$ theory it was shown in \cite{Bender2010} that a non-Jordan-block,  $PT$-symmetric Hamiltonian will always possess an additional discrete linear symmetry, with there always being an  operator, called ${\cal C}$ in the $PT$ literature (see  \cite{Bender2007}), that obeys $[{\cal C},H]=0$, ${\cal C}^2=I$. In those cases in which this ${\cal C}$ operator can be constructed explicitly in closed form it is found to depend on the structure of the particular Hamiltonian of interest, and for our $M(s)$ example the ${\cal C}$ operator is given by  
\begin{eqnarray}
{{\cal C}}(s>1)=\frac{1}{\sin\alpha}\left(\sigma_1+ i\sigma_3\cos\alpha\right),\qquad
{{\cal C}}(s<1)=\frac{1}{i\sinh\beta}\left(\sigma_1+ i\sigma_3\cosh\beta\right),
\label{H21ab}
\end{eqnarray}
where $\sin \alpha=(s^2-1)^{1/2}/s$, $\sinh\beta=(1-s^2)^{1/2}/s$.  Given the existence of the ${\cal C}$ operator, in \cite{Bender2010} it was shown that if the $PT$ theory ${\cal C}$ commutes with $PT$ then all eigenvalues are real, while if it does not, then some of the eigenvalues must appear in complex conjugate pairs, with, as we elaborate on in the Appendix, no non-trivial such ${\cal C}$ existing in the Jordan-block case. Simultaneously satisfying the conditions $[PT,H]=0$, $[PT,{\cal C}]=0$ is thus both necessary and sufficient  for all the eigenvalues of a non-Jordan-block Hamiltonian to be real. In the Appendix we compare and contrast this ${\cal C}$ operator with the charge conjugation operator $C$.

\subsection{Antilinearity and Eigenvector Equations}

As well as the eigenvalue equation it is also instructive to look at the eigenvector equation 
\begin{eqnarray}
i\frac{\partial}{\partial t}|\psi(t)\rangle=H|\psi(t)\rangle=E|\psi(t)\rangle.
\label{H22ab}
\end{eqnarray}
On replacing the parameter $t$ by $-t$ and then multiplying by a general antilinear operator $A$ we obtain
\begin{eqnarray}
i\frac{\partial}{\partial t}A|\psi(-t)\rangle=AHA^{-1}A|\psi(-t)\rangle=E^*A|\psi(-t)\rangle.
\label{H23ab}
\end{eqnarray}
From (\ref{H23ab}) we see that if $H$ has an antilinear symmetry so that $AHA^{-1}=H$, then, as first noted by Wigner in his study of time reversal invariance,  energies can either be real and have eigenfunctions that obey $A|\psi(-t)\rangle=|\psi(t)\rangle$, or can appear in complex conjugate pairs that have conjugate eigenfunctions ($|\psi(t)\rangle \sim \exp(-iEt)$ and $A|\psi(-t)\rangle\sim \exp(-iE^*t)$).

To establish the converse, suppose we are given that the energy eigenvalues are real or appear in complex  conjugate pairs. In such a case not only would $E$ be an eigenvalue but $E^*$ would be too. Hence, we can set $HA|\psi(-t)\rangle=E^*A|\psi(-t)\rangle$ in (\ref{H23ab}), and obtain
\begin{eqnarray}
(AHA^{-1}-H)A|\psi(-t)\rangle=0.
\label{H24ab}
\end{eqnarray}
Then if the eigenstates of $H$ are complete, (\ref{H24ab}) must hold for every eigenstate, to yield $AHA^{-1}=H$ as an operator identity, with $H$ thus having an antilinear symmetry. 

An alternate argument is to note that if we are given that all energy eigenvalues of $H$ are real or in complex conjugate pairs, from  $H|\psi\rangle=E|\psi\rangle$, and thus $AHA^{-1}A|\psi\rangle=E^*A|\psi\rangle$, it follows that $H$ and $AHA^{-1}$ have the same set of energy eigenvalues and are thus isospectrally related via $H=SAHA^{-1}S^{-1}=SLKHK(SL)^{-1}$ with a linear $S$. Thus again $H$ has an antilinear symmetry (viz. $SLK$). Hence we see that if a Hamiltonian has an antilinear symmetry then its eigenvalues are either real or appear in complex conjugate pairs; while if all the energy eigenvalues are real or appear in complex conjugate pairs, the Hamiltonian must admit of an antilinear symmetry. 

\subsection{Antilinearity and the Time Independence of Inner Products}

While this analysis shows that $H$ will have an antilinear symmetry if its eigenvalues are real or appear in complex conjugate pairs, we still need a reason for why the eigenspectrum should in fact take this form.  To this end we look at the time evolution of inner products. Specifically, the eigenvector equation $i\partial_{t}|R \rangle=H|R \rangle =E|R \rangle$ only involves the kets and serves to identify right-eigenvectors. Since the bra states are not specified by an equation that only involves the kets, there is some freedom in choosing them. As discussed for instance in \cite{Mannheim2013}, in general one should not use the standard $\langle R|R \rangle$ Dirac inner product  associated with the Dirac conjugate $\langle R|$ of $|R \rangle$ since $\langle R(t)|R(t) \rangle=\langle R(0)|\exp(iH^{\dagger}t)\exp(-iHt)|R(0) \rangle$ is not equal to $\langle R(t=0)|R(t=0) \rangle$ when the Hamiltonian is not Hermitian, with this inner product  then not being preserved in time. Rather, one should introduce left-eigenvectors of the Hamiltonian according to $-i\partial_{t} \langle L|=\langle L|H= \langle L|E$, and use the more general inner product $\langle L|R \rangle$, since for it one does have 
\begin{eqnarray}
\langle L(t)|R(t) \rangle=\langle L(t=0)|\exp(iHt)\exp(-iHt)|R(t=0) \rangle=\langle L(t=0)|R(t=0) \rangle,
\label{H25ab}
\end{eqnarray}
with this inner product being preserved in time. While this inner product coincides with the Dirac inner product $\langle R|R \rangle$ for Hermitian $H$, for non-Hermitian $H$ one should use the $\langle L|R \rangle$ inner product  instead. Since a Hamiltonian cannot have eigenstates other than its left and right ones, the  $\langle L|R \rangle$ inner product  is the most general inner product  one could use.

\subsection{Time Independence of Inner Products and the $V$ operator}

In \cite{Mostafazadeh2002} and \cite{Mannheim2013} a procedure was given for constructing the left-eigenvectors from the right-eigenvectors. Since the norm $\langle R_j(t)|R_i(t)\rangle$ is not time independent when the Hamiltonian is not Hermitian, as long as the sets of all $\{|R_i(t)\rangle\}$ and all $\{\langle R_j(t)|V\}$ are both complete, the most general inner product one could introduce would be of the form $\langle R_j(t)|V|R_i(t)\rangle$, as written here in terms of some as yet to be determined operator $V$. On provisionally presupposing  $V$ to be time independent, we evaluate 
\begin{eqnarray}
i\frac{\partial}{\partial t} \langle R_j(t)|V|R_i(t)\rangle
=\langle R_j(t)|(VH-H^{\dagger}V)|R_i(t)\rangle.
\label{H26ab}
\end{eqnarray}
From (\ref{H26ab}) we see that the $V$-based inner products will be time independent if $V$ obeys the so-called pseudo-Hermitian condition $VH-H^{\dagger}V=0$. For time-independent Hamiltonians the operator $V$ then would indeed be time independent, just as we had presupposed. Since $\langle R|$ obeys $-i\partial_t \langle R| =\langle R| H^{\dagger}$, and thus obeys $-i\partial_t \langle R|V =\langle R| H^{\dagger}V$ if $VH-H^{\dagger}V=0$, we find that $\langle R|V$ then obeys $-i\partial_t \langle R| V=\langle R| VH$, and we can thus identify $\langle L|=\langle R|V$. Thus via the right-eigenvectors and the operator $V$ that obeys $VH-H^{\dagger}V=0$ one can construct the left-eigenvectors.\footnote{\label{F1} For our $M(s)$ example we presented the $V$ operator for the matrix $M(s<1)$ in  (\ref{H8ab}),  and in (\ref{H9ab}) showed that $V$-based $\langle R_j(t)|V|R_i(t) \rangle$ inner products are indeed time independent. For completeness we note that for $s>1$ the $V$ operator is given by $(\sigma_0+\sigma_2\cos\alpha)/\sin\alpha$, where $\sin \alpha=(s^2-1)^{1/2}/s$. Also we note that the associated $s>1$ ${\cal C}$ operator is given by ${{\cal C}}(s>1)=(\sigma_1+i\sigma_3\cos\alpha)/\sin\alpha$, with both it and the analogous $s<1$ ${\cal C}$ operator obeying ${{\cal C}}=PV$, a point we explore further in the Appendix.} 

From (\ref{H26ab}) we can also show that $VH-H^{\dagger}V=0$ if the $V$-based inner products are time independent \cite{Mostafazadeh2002}, \cite{Mannheim2013}. Specifically, from (\ref{H26ab}) we see that if we are given that all $V$-based inner products are time independent, then if the set of all $|R_i(t)\rangle$ is complete, the right-hand side of (\ref{H26ab}) must vanish for all states, with the condition  $VH-H^{\dagger}V=0$ then emerging as an operator identity. The conditions that all $V$-based inner products are time independent and the condition that $VH-H^{\dagger}V=0$ are thus equivalent.

Now the operator $V$ may or may not be not be invertible ($V$ will not be invertible if the eigenvectors are complete but do not form a Reisz basis \cite{Siegl2012}), and so we need to discuss both invertible and non-invertible cases. With $H$ and $H^{\dagger}$ being related by $H^{\dagger}=VHV^{-1}$ when $V$ is invertible, it follows that in the invertible case $H$ and $H^{\dagger}$ both have the same set of eigenvalues. In consequence, the eigenvalues of $H$ are either real or appear in complex conjugate pairs. Thus, as we noted above, $H$ must have an antilinear symmetry. Hence if all $\langle R_j(t)|V|R_i(t)\rangle$ inner products are time independent and $V$ is invertible, the Hamiltonian must have an antilinear symmetry. Now if the Hamiltonian has an antilinear symmetry, its eigenvalues are then real or in complex conjugate pairs, and $H$ and $H^{\dagger}$ must thus be isospectrally related by some operator $V$ according to $H^{\dagger}=VHV^{-1}$. Thus, as noted in \cite{Mostafazadeh2002} and \cite{Mannheim2013}, pseudo-Hermiticity implies antilinearity and antilinearity implies pseudo-Hermiticity.

Regardless of whether or not $V$ is invertible, we note that if  $|R_i(t) \rangle$ is a right-eigenstate of $H$ with energy eigenvalue $E_i=E_i^R+iE_i^I$, in general we can write
\begin{eqnarray}
&&\langle R_j(t)|V|R_i(t) \rangle
=\langle R_j(0)|V|R_i(0) \rangle e^{-i(E_i^R+iE_i^I)t+i(E_j^R-iE_j^I)t}
\label{H27ab}
\end{eqnarray}
Since $V$ has been chosen so that the $\langle R_j(t)|V|R_i(t) \rangle$ inner products are to be time independent,  the only allowed non-zero inner product are those that obey
\begin{eqnarray}
&&E_i^R=E_j^R,\qquad E_i^I=-E_j^I,
\label{H28ab}
\end{eqnarray}
with all other $V$-based inner products having to obey $\langle R_j(0)|V|R_i(0) \rangle=0$. We recognize (\ref{H28ab}) as being precisely none other than the requirement that eigenvalues be real or appear in complex conjugate pairs, just as required of antilinear symmetry. Since this analysis does not require the invertibility of $V$, the time independence of the $V$-based inner products thus implies that the Hamiltonian must have an antilinear symmetry regardless of whether or not  $V$ is invertible. As had been noted above, in the presence of complex energy eigenvalues the time independence of inner products is maintained because the only non-zero overlap of any given right-eigenvector with a complex energy eigenvalue is that with the appropriate left-eigenvector with the eigenvalue needed to satisfy (\ref{H28ab}), i.e. precisely between decaying and growing modes.

Thus regardless of whether or not $V$ is invertible, if all $V$-based inner products are time independent it follows that  the energy eigenvalues are either real or appear in complex conjugate pairs. Thus, as had been noted above, $H$ must have an antilinear symmetry. While construction of the needed $V$ operator is not a straightforward task, the $V$ operator must exist if the Hamiltonian has an antilinear symmetry, with a symmetry condition, even an antilinear one, being  something that is much easier to identify, and thus more powerful since it guarantees that such a $V$ must exist even if one cannot explicitly construct it in closed form. With the operator $V$ we note that the time evolution operator $U=\exp(-iHt)$ obeys $VU^{-1}=U^{\dagger}V$ (and thus $U^{-1}=V^{-1}U^{\dagger}V$ if $V$ is invertible), to thus generalize the standard unitarity condition $U^{-1}=U^{\dagger}$ that holds for Hermitian Hamiltonians (where $V=I$).

Time independence of inner products under the evolution of a Hamiltonian and antilinearity of that Hamiltonian thus complement each other, with the validity of either one ensuring the validity of the other. Since on physical grounds one must require time independence of inner products if one is to construct a quantum theory with probability conservation, that requirement entails not that the Hamiltonian be Hermitian, but that it instead possess an antilinear symmetry. Since it in addition requires that  $VH-H^{\dagger}V=0$ and thus that $\langle L|=\langle R|V$, the resulting left-right $\langle R|V|R\rangle=\langle L|R \rangle$ norm is thus the most general time-independent inner product that one could write down. Antilinearity thus emerges as a basic requirement of quantum theory, to thus supplant the standard requirement of Hermiticity.

\section{Antilinearity and the $CPT$ Theorem}
\label{cpt}

\subsection{Complex Lorentz Invariance for Coordinates}

While our above remarks apply to any discrete antilinear symmetry, it is of interest to ask whether there might be any specially chosen or preferred one, and in this section we show that once we impose Lorentz invariance (as extended to include complex transformations) there is such a choice, namely $CPT$. We thus extend the $CPT$ theorem to non-Hermitian Hamiltonians, and through the presence  of complex conjugate pairs of energy eigenvalues to unstable states, a result we announced in  \cite{Mannheim2016}. (The familiar standard proofs always involved Hermiticity -- see e.g. \cite{Streater1964,Weinberg1995}, with the axiomatic field theory proof  \cite{Streater1964} involving complex Lorentz invariance as well.) With the Hamiltonian being the generator of time translations we can anticipate a connection to the Lorentz group and to spacetime operators, and with time reversal being a spacetime-based antilinear operator we can anticipate that the discrete symmetry would involve $T$. The possible antilinear options that have a spacetime connection are thus $T$, $PT$, $CT$ and $CPT$. As we will see, of the four it will be $CPT$ that will be automatically selected. (Some alternate discussion of the $CPT$ theorem in the presence of unstable states may be found in \cite{Selover2013}.)

While Lorentz invariance is ordinarily thought of as involving real transformations only, so that $x^{\prime\mu}=\Lambda^{\mu}_{\phantom{\mu}\nu}x^{\nu}$ is real, the line element $\eta_{\mu\nu}x^{\mu}x^{\nu}$ is left invariant even if $\Lambda^{\mu}_{\phantom{\mu}\nu}$ is complex.  Specifically, if we introduce a set of six antisymmetric Lorentz generators $M_{\mu\nu}$ that obey 
\begin{eqnarray}
[M_{\mu\nu},M_{\rho\sigma}]&=&i(-\eta_{\mu\rho}M_{\nu\sigma}+\eta_{\nu\rho}M_{\mu\sigma}
-\eta_{\mu\sigma}M_{\rho\nu}+\eta_{\nu\sigma}M_{\rho\mu}),
\label{H29ab}
\end{eqnarray}
as written here with ${\rm diag}[\eta_{\mu\nu}]=(1,-1,-1,-1)$, and introduce six antisymmetric angles $w^{\mu\nu}$, the Lorentz transformation $\exp(iw^{\mu\nu}M_{\mu\nu})$ will not only leave the $\tilde{x}^{\mu}x_{\mu}$ line element invariant with real $w^{\mu\nu}$, it will do so with complex $w^{\mu\nu}$ as well since the reality or otherwise of $w^{\mu\nu}$ plays no role in the analysis. To see this in  detail it is instructive to ignore metric and dimension issues and consider invariance of the two-dimensional line element $s^2=\tilde{x}x=x_1^2+x_2^2$. If we introduce a rotation matrix 
\begin{eqnarray}
R=\left(\matrix{\phantom{-} \cos \alpha&\sin \alpha\cr -\sin\alpha&\cos \alpha \cr}\right),
\label{H30ab}
\end{eqnarray}
because this matrix is orthogonal, the line element is preserved ($\tilde{x}x\rightarrow \tilde{x}\tilde{R}Rx=\tilde{x}R^{-1}Rx=\tilde{x}x$). Since a product of two rotations obeys $\widetilde{R_1R_2}=\tilde{R_2}\tilde{R_1}=R_2^{-1}R_1^{-1}=(R_1R_2)^{-1}$, the product is also orthogonal, with rotation matrices thus forming a group. Suppose we now make $\alpha$ complex. Then even with complex angle $R$ remains orthogonal, the line element is still preserved, and the class of all real and complex rotations forms a group. Since this analysis immediately generalizes to the coordinate representation of $SO(4)$ and consequently to that of the Lorentz $SO(3,1)$, we see that the $SO(3,1)$ length $\tilde{x}^{\mu}x_{\mu}$ is left invariant under real and complex Lorentz transformations, with the group structure remaining intact. 

\subsection{Complex Lorentz Invariance for Fields}

For field theories similar remarks apply to the action $I=\int d^4x L(x)$. With $L(x)$ having spin zero, this action is invariant under real Lorentz transformations of the form $\exp(iw^{\mu\nu}M_{\mu\nu})$ where the six $w^{\mu\nu}=-w^{\nu\mu}$ are real parameters and the six $M_{\mu\nu}=-M_{\nu\mu}$ are the generators of the Lorentz group. Specifically, with $M_{\mu\nu}$ acting on the Lorentz spin zero  $L(x)$ as $x_{\mu}p_{\nu}-x_{\nu}p_{\mu}$, under an infinitesimal Lorentz transformation the change in the action is  given by $\delta I=2w^{\mu\nu}\int d^4x x_{\mu}\partial_{\nu}L(x)$, and thus by $\delta I=2w^{\mu\nu}\int d^4x \partial_{\nu}[x_{\mu}L(x)]$. Since the change in the action is a total divergence, the familiar invariance of the action under real Lorentz transformations is secured. However, we now note that nothing in this argument depended on $w^{\mu\nu}$ being real, with the change in the action still being a total divergence even if $w^{\mu\nu}$ is complex. The action $I=\int d^4x L(x)$ is thus actually invariant under complex Lorentz transformations as well and not just  under real ones, with complex Lorentz invariance thus being  just as natural to physics as real Lorentz invariance.

\subsection{Majorana Spinors}

In extending the discussion to spinors there is a subtlety since Dirac spinors reside not in $SO(3,1)$ but in its complex covering group. While this immediately implies the potential relevance of complex transformations,  if one were to work with unitary transformations they would not remain unitary if $w^{\mu\nu}$ is complexified. (For transformations of the form $R=\exp(i\alpha J)$ with generic generator $J$, under a complexification of $\alpha$ the relation $R^{-1}=R^{\dagger}$ is not preserved if $J$ is Hermitian,  while the relation  $R^{-1}=\tilde{R}$ is preserved if $J$ is antisymmetric.) However, Dirac spinors are reducible under the Lorentz group, with it being Majorana and Weyl spinors that are irreducible, with a Dirac spinor being writable as a sum of two Majorana spinors or two Weyl spinors. Now these two spinors are related since a Majorana spinor can be written as a Weyl spinor plus its charge conjugate (see e.g. \cite{Mannheim1984}), and we shall thus work with Majorana spinors in the following. As such, Majorana spinors are the natural counterparts of the coordinates, since unlike $SO(4)$, which only has one real four-dimensional irreducible representation (the vector), because of the Minkowski nature of the spacetime metric the group $SO(3,1)$ has two inequivalent real four-dimensional representations, the vector representation and the Majorana spinor representation. This is most easily seen in the Majorana  basis for the Dirac matrices (see e.g. \cite{Mannheim1984}), with the two irreducible representations being  reproduced in the Appendix.

Now while $SO(3,1)$ possesses a real four-dimensional irreducible Majorana spinor representation, this is not the case for the $SO(4,2)$ conformal group of which $SO(3,1)$ is a subgroup, since the four-dimensional spinor representation of the conformal group is complex, not real.\footnote{\label{F5} In terms of generators $M_{\mu\nu}$, $P_{\mu}$, $D$, and $K_{\mu}$, together with (\ref{H29ab}) the conformal algebra takes the form $[M_{\mu\nu},P_{\sigma}]=i(\eta_{\nu\sigma}P_{\mu}-\eta_{\mu\sigma}P_{\nu})$, $[P_{\mu},P_{\nu}]=0$, 
$[M_{\mu\nu},K_{\sigma}]=i(\eta_{\nu\sigma}K_{\mu}-\eta_{\mu\sigma}K_{\nu})$,
$[M_{\mu\nu},D]=0$,  $[K_{\mu},K_{\nu}]=0$, $[K_{\mu},P_{\nu}]=2i(\eta_{\mu\nu}D-M_{\mu\nu})$,
$[D,P_{\mu}]=iP_{\mu}$, $[D,K_{\mu}]=-iK_{\mu}$. It admits of a four-dimensional spinor representation of the form $M_{\mu\nu}=(i/4)[\gamma_{\mu},\gamma_{\nu}]$, $K_{\mu}+P_{\mu}=\gamma_{\mu}$, $K_{\mu}-P_{\mu}=\gamma_{\mu}\gamma^5$, $D=i\gamma^5/2$. In the Majorana basis of the gamma matrices $(i/4)[\gamma_{\mu},\gamma_{\nu}]$ and $\gamma_{\mu}$ are pure imaginary, while $\gamma_{\mu}\gamma^5$ and  $i\gamma^5/2$ are real. Thus unlike the $SO(3,1)$ Majorana spinor representation, the $SO(4,2)$ spinor representation is complex. (In passing we note that with $\exp(i\alpha D)$ being equal to $i\gamma^5$ when $D= i\gamma^5/2$ and $\alpha=-i\pi$, when acting on Majorana spinors a dilatation acts in precisely the same way as the linear part of the $CPT$ operator is shown to behave below.)} However, since $SO(4,2)$ is an orthogonal group,  its group structure will remain intact under complex conformal transformations, just as we had found to be the case for $SO(3,1)$. Now conformal invariance is the full symmetry of the light cone, and if all elementary particle masses are to arise though vacuum breaking, the fermion and gauge boson sector of the fundamental action that is to describe their dynamics would then be conformal invariant, just as is indeed the case in the standard $SU(3)\times SU(2)\times U(1)$ theory of strong, electromagnetic and weak interactions. With the spinor representation of the conformal group being complex, it is then natural that the spinor representation of its $SO(3,1)$ subgroup would be complex too, with its two separate Majorana spinor components being combined into a single irreducible representation of the conformal group. Thus with a Dirac spinor being irreducible under the conformal group even as it is reducible under $SO(3,1)$, through the conformal group we are again led to complex Lorentz invariance. 

\subsection{Complex Lorentz Invariance for Majorana Spinors}

With Majorana spinors living in $SO(3,1)$ itself rather than its covering group, the extension to complex Lorentz transformations parallels that for the coordinates. With spinors being Grassmann variables, to implement such a parallel treatment we work in the Majorana basis of the Dirac gamma matrices where the Dirac space matrix $C$ that transposes according to $C\gamma^{\mu}C^{-1}=-\widetilde{\gamma^{\mu}}$ coincides with $\gamma^0$. Following e.g. \cite{Mannheim1985}, we introduce a ``line element" in Grassmann space, viz. $\tilde{\psi} C\psi$ (the tilde here denotes transposition in the Dirac gamma matrix space alone and not in the field space of $\psi$). In the Majorana basis $C$ is antisymmetric, just as needed since the Grassmann $\psi$ and $\tilde{\psi}$ obey an anticommutation algebra. With the Lorentz generators behaving as $M^{\mu\nu}=i[\gamma^{\mu},\gamma^{\nu}]/4$ in the Dirac gamma matrix space, under a Lorentz transformation we find that 
\begin{eqnarray}
\tilde{\psi} C\psi \rightarrow \tilde{\psi} \exp(iw^{\mu\nu}\widetilde{M_{\mu\nu}})C\exp(iw^{\mu\nu}M_{\mu\nu})\psi. 
\label{F31ab}
\end{eqnarray}
Then, with $\widetilde{M_{\mu\nu}}=-CM_{\mu\nu}C^{-1}$, the invariance of $\tilde{\psi} C\psi$ is secured. Moreover, since this analysis is independent of whether $w^{\mu\nu}$ is real or complex, the invariance of   $\tilde{\psi} C\psi$ is secured not just for real $w^{\mu\nu}$ but for complex $w^{\mu\nu}$ as well. Because of the signature of the spacetime metric, the three Lorentz $M_{0i}$ boosts are symmetric in the Majorana  basis for the Dirac gamma matrices while the three $M_{ij}$ rotations are antisymmetric. Since this same pattern is found for the vector representation, on recalling that $\tilde{x}^{\mu}x_{\mu}$ is invariant under complex Lorentz transformations, we see that in the Majorana spinor space the Lorentz group structure also remains intact under complex transformations, with  the Majorana spinor line element being left invariant under the complex Lorentz group. Using Majorana spinors we can thus extend complex Lorentz invariance to the spinor sector.

To make an explicit connection between Majorana spinors and Dirac spinors at the quantum field theory level, we introduce a unitary charge conjugation operator which in quantum field space transforms a general Dirac spinor into its charge conjugate according to  $\hat{C}\psi\hat{C}^{-1}=\psi^c$.\footnote{\label{F6} While for our purposes here it  will suffice to take $\hat{C}$ to be a linear operator, in \cite{Mannheim2016} we actually explored a non-standard but occasionally studied (see e.g. \cite{Selover2013}) antilinear option for $\hat{C}$.} On introducing $\psi_M=(\psi+\psi^c)/2$, $\psi_A=(\psi-\psi^c)/2$, we can write $\psi=\psi_M+\psi_A$, where $\psi_M$ and $\psi_A$ obey $\hat{C}\psi_M\hat{C}^{-1}=\psi_M$, $\hat{C}\psi_A\hat{C}^{-1}=-\psi_A$, with $\psi_M$ being self conjugate (just like the $x^{\mu}$) and $\psi_A$ being anti-self-conjugate.  For convenience in the following we set $\psi_M=\psi_1$, $\psi_A=i\psi_2$ where $\hat{C}\psi_1\hat{C}^{-1}=\psi_1$, $\hat{C}\psi_2\hat{C}^{-1}=-\psi_2$. The utility of this particular  $\psi=\psi_M+\psi_A=\psi_1+i\psi_2$ decomposition is that it is preserved under an arbitrary similarity transformation $S$, with the transformed $\psi_1$ and $\psi_2$ respectively being self-conjugate and anti-self-conjugate under the transformed charge conjugation operator $\hat{C}^{\prime}=S\hat{C}S^{-1}$. As we discussed in Sec. \ref{intro}, the Hermiticity condition  $H_{ij}=H^*_{ji}$ is not preserved under a general similarity transformation, with self-conjugacy having a basis-independent  status that Hermiticity does not possess. While the Hermiticity condition $H_{ij}=H^*_{ji}$ for an operator is not basis independent, we note that in the Majorana basis  of the Dirac gamma matrices charge conjugation is the same as Hermitian conjugation. Thus in that basis we can take $\psi_1$ and $\psi_2$ to be Hermitian fields, and in the following we shall work in the Majorana basis and use the $\psi=\psi_1+i\psi_2$ decomposition of a general Dirac spinor where $\hat{C}\psi_1\hat{C}^{-1}=\psi_1=\psi^{\dagger}$, $\hat{C}\psi_2\hat{C}^{-1}=-\psi_1=-\psi_1^{\dagger}$. In the Majorana  basis for the Dirac gamma matrices $\hat{P}$ and  $\hat{T}$ implement 
\begin{eqnarray}
\hat{P}\psi(\vec{x},t)\hat{P}^{-1}=\gamma^0\psi(-\vec{x},t),\qquad \hat{T}\psi(\vec{x},t)\hat{T}^{-1}=\gamma^1\gamma^2\gamma^3\psi(\vec{x},-t)
\label{F32ab}
\end{eqnarray}
as it is these transformations that leave the action for a free Dirac field invariant. In terms of the $\psi_1$, $\psi_2$ basis $\hat{C}\hat{P}\hat{T}$ itself thus implements 
\begin{eqnarray}
\hat{C}\hat{P}\hat{T}[\psi_1(x)+i\psi_2(x)]\hat{T}^{-1}\hat{P}^{-1}\hat{C}^{-1} =i\gamma^5 [\psi_1(-x)-i\psi_2(-x)],
\label{F33ab}
\end{eqnarray}
a relation  that will prove central in  the following.

As regards complex Lorentz transformations, we note that for Dirac spinors quantities such as $\bar{\psi}\psi=\psi^{\dagger}\gamma^0\psi$ would not be invariant under a complex Lorentz transformation if it is applied to both $\psi$ and $\psi^{\dagger}$ as is. However, with $\psi_1$ and $\psi_2$ both being taken to be Hermitian Majorana spinors, we should write $\psi^{\dagger}\gamma^0\psi$ as $(\tilde{\psi}_1-i\tilde{\psi}_2)\gamma^0(\psi_1+i\psi_2)$ (in constructing $\tilde{\psi}_i$ the transposition acts only on their four components in  the Dirac gamma matrix space and not on quantum fields themselves), and then implement the transformation on the separate $\psi_1$ and $\psi_2$, since they transform as $\psi_i\rightarrow \exp(iw^{\mu\nu}M_{\mu\nu})\psi_i$, $\tilde{\psi}_i \rightarrow \tilde{\psi}_i \exp(iw^{\mu\nu}\widetilde{M_{\mu\nu}})$.

Given that $\psi$ transforms as  $\psi\rightarrow \exp(iw^{\mu\nu}M_{\mu\nu})\psi$ under a real or a complex Lorentz transformation, we might initially expect that $\psi^{\dagger}$ transforms as $\psi^{\dagger}\rightarrow \psi^{\dagger}\exp(-i[w^{\mu\nu}]^*M_{\mu\nu}^{\dagger})$, rather than as the relation $\psi^{\dagger}\rightarrow \psi^{\dagger}\exp(iw^{\mu\nu}\widetilde{M_{\mu\nu}})$ that we have found. To appreciate the distinction we need to introduce the quantum field-theoretic Lorentz generators $\hat{\Lambda}=\exp (iw^{\mu\nu}\hat{M}_{\mu\nu})$, which generate $\hat{\Lambda}^{-1}\psi\hat{\Lambda}= \exp(iw^{\mu\nu}M_{\mu\nu})\psi$ and thus  $\hat{\Lambda}^{\dagger}\psi^{\dagger}[\hat{\Lambda}^{-1}]^{\dagger}= \psi^{\dagger}\exp(-i[w^{\mu\nu}]^*M_{\mu\nu}^{\dagger})$, where $\hat{M}^{\mu\nu}=\int d^3x (x^{\mu}\hat{T}^{0\nu}-x^{\nu}\hat{T}^{0\mu})$ and $\hat{T}^{\mu\nu}$ is the quantum field energy-momentum tensor.   Even if we were to take $\hat{M}^{\mu\nu}$  to be Hermitian (which it would not be if $\hat{H}=\int d^3x \hat{T}^{00}$ is not Hermitian), with complex $w^{\mu\nu}$ the operator $\hat{\Lambda}$ would not be unitary, and there is thus no otherwise troublesome relation of the form $\hat{\Lambda}^{-1}\psi^{\dagger}\hat{\Lambda}= \psi^{\dagger}\exp(-i[w^{\mu\nu}]^*M_{\mu\nu}^{\dagger})$.  
In this way we can extend complex Lorentz invariance to $\bar{\psi}\psi$.
 
To determine what happens to a general matrix element under a complex Lorentz transformation, we recall that in Sec. \ref{antilinearity} we had introduced a $V$ operator that effects $VH=H^{\dagger}V$. Given this $V$, for a Lorentz transformation $\hat{\Lambda}=\exp (iw^{\mu\nu}\hat{M}_{\mu\nu})$ first with real $w^{\mu\nu}$, we can set  
\begin{eqnarray}
\hat{\Lambda}^{\dagger}V=\exp (-iw^{\mu\nu}\hat{M}_{\mu\nu}^{\dagger})V=V\exp (-iw^{\mu\nu}\hat{M}_{\mu\nu})=V\hat{\Lambda}^{-1}. 
\label{F34ab}
\end{eqnarray}
With the matrix element $\langle R|V|R\rangle$ transforming into $\langle R|\hat{\Lambda}^{\dagger}V\hat{\Lambda}|R\rangle$ under a Lorentz transformation on the states, $\langle R|V|R\rangle$ transforms into $\langle R|V\hat{\Lambda}^{-1}\hat{\Lambda}|R\rangle$, to thus be invariant. However, this procedure will not work as is if $w^{\mu\nu}$ is complex, and so in the complex Lorentz case we will need to find an alternate matrix element. This alternate is provided by the $\hat{C}\hat{P}\hat{T}$ operator. Specifically, we note that given a quantum field-theoretic action that is $CPT$ even, its variation with respect to the $C$ even, $P$ even, $T$ even metric $g_{\mu\nu}$ yields an energy-momentum tensor $\hat{T}^{\mu\nu}$ that is $CPT$ even too. In consequence $\hat{H}$ is $CPT$ even, while the $\hat{M}^{\mu\nu}=\int d^3x (x^{\mu}\hat{T}^{0\nu}-x^{\nu}\hat{T}^{0\mu})$ generators that are constructed from it are $CPT$ 
odd.\footnote{\label{F7} Setting $\hat{C}\hat{P}\hat{T}[x^{\mu}\hat{T}^{0\nu}(x)-x^{\nu}\hat{T}^{0\mu}(x)][\hat{C}\hat{P}\hat{T}]^{-1}=(x^{\mu}\hat{T}^{0\nu}(-x)-x^{\nu}\hat{T}^{0\mu}(-x))=-(-x^{\mu}\hat{T}^{0\nu}(-x)-(-x)^{\nu}\hat{T}^{0\mu}(-x))$ yields a net minus sign when $CPT$ acts on the time independent $\hat{M}^{\mu\nu}=\int d^3x (x^{\mu}\hat{T}^{0\nu}(x)-x^{\nu}\hat{T}^{0\mu}(x))=\int d^3x (-x^{\mu}\hat{T}^{0\nu}(-x)+x^{\nu}\hat{T}^{0\mu}(-x))$ at $t=0$.} Thus if we now apply $CPT$ to a complex Lorentz transformation generator we obtain 
\begin{eqnarray}
\hat{C}\hat{P}\hat{T}\exp(iw^{\mu\nu}\hat{M}_{\mu\nu})[\hat{C}\hat{P}\hat{T}]^{-1}=\exp(i[w^{\mu\nu}]^*\hat{M}_{\mu\nu}), 
\label{F35ab}
\end{eqnarray}
and thus obtain $V\hat{C}\hat{P}\hat{T}\hat{\Lambda}^{-1}= \hat{\Lambda}^{\dagger}V\hat{C}\hat{P}\hat{T}$. On defining the more general matrix element $\langle R|V\hat{C}\hat{P}\hat{T}|R\rangle$, we find that it transforms into $\langle R|\hat{\Lambda}^{\dagger}V\hat{C}\hat{P}\hat{T}\hat{\Lambda}|R\rangle$ under a complex Lorentz transformation on the states. It thus transforms into $\langle R|V\hat{C}\hat{P}\hat{T}\hat{\Lambda}^{-1}\hat{\Lambda}|R\rangle$, to thus be invariant.  Finally we note that even if the $\hat{M}_{\mu\nu}$ are Hermitian (so $V=I$), it is $\langle R|\hat{C}\hat{P}\hat{T}|R\rangle$ that is invariant under complex Lorentz transformations  and not the standard Dirac norm $\langle R|R\rangle$. This then is how one constructs matrix elements that are invariant under complex Lorentz transformations.

\subsection{Connection Between Complex Lorentz Transformations and $PT$ and $CPT$ Transformations}

The utility of complex Lorentz invariance is that it has a natural connection to both $PT$ and $CPT$ transformations. For coordinates $PT$ implements $x^{\mu}\rightarrow -x^{\mu}$, and thus so does $CPT$ since the coordinates are charge conjugation even (i.e. unaffected by a charge conjugation transformation).  With a boost in the $x_1$-direction implementing $x_1^{\prime}=x_1\cosh\xi +t\sinh \xi $, $t^{\prime}=t\cosh\xi +x_1\sinh \xi $, the complex $\Lambda^{0}_{\phantom{0}1}(i\pi)$ boost with $\xi=i\pi$ implements $x_1\rightarrow -x_1$, $t\rightarrow -t $. With the $\Lambda^{0}_{\phantom{0}2}(i\pi)$ boost implementing $x_2\rightarrow -x_2$, $t\rightarrow -t $, and with the $\Lambda^{0}_{\phantom{0}3}(i\pi)$ boost implementing $x_3\rightarrow -x_3$, $t\rightarrow -t $, the sequence $\pi\tau=\Lambda^{0}_{\phantom{0}3}(i\pi)\Lambda^{0}_{\phantom{0}2}(i\pi)\Lambda^{0}_{\phantom{0}1}(i\pi)$ implements $\pi\tau:x^{\mu}\rightarrow -x^{\mu}$, just as required of a $PT$ or $CPT$ transformation on the coordinates. 

With Lorentz transformations on real coordinates obeying $(\Lambda^0_{\phantom {0}0})^2-(\Lambda^1_{\phantom{0}0})^2-(\Lambda^2_{\phantom{0}0})^2-(\Lambda^3_{\phantom{0}0})^2=1$, there are four disconnected $L^{{\rm det}}_{{\rm sgn}}$ domains, classified according to ${\rm det}\Lambda=\pm 1$ and ${\rm sgn} \Lambda^0_{\phantom {0}0}=\pm 1$. The domains $L^+_+$ and $L^+_-$ are then connected by a $PT$ transformation on the coordinates. Complex Lorentz transformations thus cover the otherwise disconnected $L^+_+$ and $L^+_-$ domains, with this thus being an interesting geometrical aspect of $PT$ transformations.

With $\Lambda^{0}_{\phantom {0}i}(i\pi)$ implementing $\exp(-i\pi\gamma^0\gamma_i/2)=-i\gamma^0\gamma_i$ in the Dirac gamma matrix space, quite remarkably, we find that as an operator in quantum field space $\hat{\pi}\hat{\tau}=\hat{\Lambda}^{0}_{\phantom{0}3}(i\pi)\hat{\Lambda}^{0}_{\phantom{0}2}(i\pi)\hat{\Lambda}^{0}_{\phantom{0}1}(i\pi)$ implements 
\begin{eqnarray}
\hat{\pi}\hat{\tau}\psi_1(x)\hat{\tau}^{-1}\hat{\pi}^{-1}=\gamma^5 \psi_1(-x),\qquad \hat{\pi}\hat{\tau}\psi_2(x)\hat{\tau}^{-1}\hat{\pi}^{-1}=\gamma^5 \psi_2(-x).
\label{F36ab}
\end{eqnarray}
Thus up to an overall complex phase, we recognize this transformation as acting as none other than (the linear part of) a $CPT$ transformation, and thus see that $CPT$ is naturally associated with the complex Lorentz group, even having a Lorentz invariant structure since $\gamma^5$ commutes with all of the $M^{\mu\nu}=i[\gamma^{\mu},\gamma^{\nu}]/4$ Lorentz generators. 

In general then, we can implement a $CPT$ transformation as $K\hat{\pi}\hat{\tau}$ where the complex conjugation $K$ serves as the antilinear component of $CPT$. Because of the factor $i$ that is present in  $\hat{C}\hat{P}\hat{T}\psi_1(x)\hat{T}^{-1}\hat{P}^{-1}\hat{C}^{-1} =i\gamma^5 \psi_1(-x)$ but not in  $\hat{\pi}\hat{\tau}\psi_1(x)\hat{\tau}^{-1}\hat{\pi}^{-1}=\gamma^5 \psi_1(-x)$, the effect of $K\hat{\pi}\hat{\tau}$ on a fermion bilinear can at most differ from the effect of $\hat{C}\hat{P}\hat{T}$ on the bilinear by a phase that is real. In the Appendix we construct an explicit anticommutation quantization scheme for Majorana fields in which the phase is found to be equal to one in all combinations of fermion bilinears and quadrilinears that have spin zero, a property that will prove central to our derivation of the $CPT$ theorem. With the fermions being in the fundamental representation of the Lorentz group from which all other representations can be constructed,  this result then generalizes to the arbitrary spin zero fermion multilinear. Since the Hamiltonian is constructed from the Lagrangian by first forming the energy-momentum tensor from it and then setting $H=\int d^3x T_{00}$, the only terms of interest for exploring properties of the Hamiltonian are those that are associated with  spin zero terms present in the Lagrangian. With the $K \hat{\pi}\hat{\tau}$ phase of all such spin zero terms being real, none of these terms is affected by $K$ at all. Thus given complex Lorentz invariance, and given the fact that the individual spin zero terms themselves are $K$ invariant even if they contain factors $i$ (which some are shown in  the Appendix to do), to establish $CPT$ invariance we now only need to be able to monitor any other factors of $i$ that might appear in the Lagrangian, such as in combinations of fields or in any numerical coefficients that might be present in the Lagrangian.

\subsection{Discrete Transformations on Fermion Spin Zero Multilinears}

To see first how such a monitoring is achieved in the Hermitian case, we recall that, as noted for instance in \cite{Weinberg1995}, every representation of the Lorentz group transforms under $\hat{C}\hat{P}\hat{T}$  as $\hat{C}\hat{P}\hat{T}\phi(x)\hat{T}^{-1}\hat{P}^{-1}\hat{C}^{-1}= \eta(\phi)\phi(-x)$, with a  $\phi$-dependent  intrinsic $CPT$ phase  $\eta(\phi)$ that depends on the spin of each $\phi$, and for integer spin systems (bosons or fermion multilinears (bilinears, quadrilinears, etc.)) obeys $\eta^2(\phi)=1$. Moreover, all spin zero fields (both scalar and pseudoscalar) expressly have $\eta(\phi)=1$. Since the most general Lorentz invariant Lagrangian density must be built out of sums of appropriately contracted spin zero products of fields with arbitrary numerical coefficients, and since it is only spin zero fields that can multiply any given net spin zero product an arbitrary number of times and still yield net spin zero, all net spin zero products of fields must have a net $\eta(\phi)$ equal to one. Generically, such products could involve $\phi_+\phi_+$, $\phi_+\phi_-$, or $\phi_-\phi_-$ type contractions where $\phi_{\pm}=\phi_1\pm i\phi_2$. Establishing $CPT$ invariance of the Lagrangian density (and thus that of the Hamiltonian) requires showing that the numerical coefficients are all real and that only $\phi_+\phi_-$ (or $\phi_+\phi_++\phi_-\phi_-$) type contractions appear. As noted in \cite{Weinberg1995}, this will precisely be the case if the Lagrangian density is Hermitian, with the $CPT$ invariance of the Hamiltonian then following.

\begin{table}[htbp]
\centering
\begin{tabular}{c|ccccccc}
&C&P&T&CP&CT&PT&CPT\\ 
\hline
$\bar{\psi}\psi$&+&+&+&+&+&+&+\\ 
$\bar{\psi}i\gamma^5\psi$&+&-&-&-&-&+&+\\
$\bar{\psi}\gamma^{0}\psi$&-&+&+&-&-&+&-\\
$\bar{\psi}\gamma^{i}\psi$&-&-&-&+&+&+&-\\
$\bar{\psi}\gamma^{0}\gamma^5\psi$&+&-&+&-&+&-&-\\
$\bar{\psi}\gamma^{i}\gamma^5\psi$&+&+&-&+&-&-&-\\
$\bar{\psi}i[\gamma^{0},\gamma^{i}]\psi$&-&-&+&+&-&-&+\\
$\bar{\psi}i[\gamma^{i},\gamma^{j}]\psi$&-&+&-&-&+&-&+\\
$\bar{\psi}[\gamma^{0},\gamma^{i}]\gamma^5\psi$&-&+&-&-&+&-&+\\
$\bar{\psi}[\gamma^{i},\gamma^{j}]\gamma^5\psi$&-&-&+&+&-&-&+\\
\hline
\end{tabular}
\caption{C, P, and T assignments for fermion bilinears}
\end{table}
\begin{table}[htbp]
\centering
\begin{tabular}{c|ccccccc}
&C&P&T&CP&CT&PT&CPT\\ 
\hline
$\bar{\psi}\psi$&+&+&+&+&+&+&+\\ 
$\bar{\psi}i\gamma^5\psi$&+&-&-&-&-&+&+\\
$\bar{\psi}\psi\bar{\psi}\psi$&+&+&+&+&+&+&+\\
$\bar{\psi}\psi\bar{\psi}i\gamma^5\psi$&+&-&-&-&-&+&+\\
$\bar{\psi}i\gamma^5\psi\bar{\psi}i\gamma^5\psi$&+&+&+&+&+&+&+\\
$\bar{\psi}\gamma^{\mu}\psi\bar{\psi}\gamma_{\mu}\psi$&+&+&+&+&+&+&+\\
$\bar{\psi}\gamma^{\mu}\psi\bar{\psi}\gamma_{\mu}\gamma^5\psi$&-&-&+&+&-&-&+\\
$\bar{\psi}\gamma^{\mu}\gamma^5\psi\bar{\psi}\gamma_{\mu}\gamma^5\psi$&+&+&+&+&+&+&+\\
$\bar{\psi}i[\gamma^{\mu},\gamma^{\nu}]\psi\bar{\psi}i[\gamma_{\mu},\gamma_{\nu}]\psi$&+&+&+&+&+&+&+\\
$\bar{\psi}i[\gamma^{\mu},\gamma^{\nu}]\psi\bar{\psi}[\gamma_{\mu},\gamma_{\nu}]\gamma^5\psi$&+&-&-&-&-&+&+\\
$\bar{\psi}i[\gamma^{\mu},\gamma^{\nu}]\gamma^5\psi\bar{\psi}i[\gamma_{\mu},\gamma_{\nu}]\gamma^5\psi$&+&+&+&+&+&+&+\\
\hline
\end{tabular}
\caption{C, P, and T assignments for fermion bilinears and quadrilinears that have spin zero}
\end{table}

To appreciate the $\eta(\phi)$ pattern, it is instructive to look at the intrinsic $C$, $P$ and $T$ parities of  fermion bilinears  as given in Table I, and for the moment we take the bilinears to be Hermitian. (In Table I associated changes in the signs of $\vec{x}$ and $t$ are implicit.) Even though it is not independent of the other fermion bilinears we have included the spin two, parity minus $\bar{\psi}[\gamma^{\mu},\gamma^{\nu}]\gamma^5\psi$, so that we can contract it into a spin zero combination with $\bar{\psi}i[\gamma^{\mu},\gamma^{\nu}]\psi$. In constructing spin zero combinations from these fermions we can use $\bar{\psi}\psi$ and $\bar{\psi}i\gamma^5\psi$ themselves or contract $\bar{\psi}\psi$ and $\bar{\psi}i\gamma^5\psi$ with themselves  or with each other an arbitrary number of times. Similarly, we can contract $\bar{\psi}\gamma^{\mu}\psi$ and $\bar{\psi}\gamma^{\mu}\gamma^5\psi$ with themselves or with each other, and we can contract $\bar{\psi}i[\gamma^{\mu},\gamma^{\nu}]\psi$ and $\bar{\psi}[\gamma^{\mu},\gamma^{\nu}]\gamma^5\psi$ with themselves  or with each other. As we see from Table I, it is only for $CPT$ that the net intrinsic phase shows any universal behavior, being correlated \cite{Weinberg1995} with the spin of the bilinear by being even or odd according to whether the spin is even or odd. Initially the factors of $i$ in $\bar{\psi}i\gamma^5\psi$ and $\bar{\psi}i[\gamma^{\mu},\gamma^{\nu}]\psi$ were introduced to make the bilinears be Hermitian. Now we see that the very same factors of $i$ can be introduced in order to make the intrinsic $CPT$ parity of the bilinears alternate with spin, and in consequence we do not need to impose Hermiticity on the fermion bilinears at all, and can define the bilinears as being of the form $\psi^{\dagger}\gamma^0\psi=(\tilde{\psi}_1-i\tilde{\psi}_2)\gamma^0(\psi_1+i\psi_2)$ etc., where $\psi_1$ and $\psi_2$ are Majorana spinors that transform as $\hat{C}\psi_1\hat{C}^{-1}=\psi_1$, $\hat{C}\psi_2\hat{C}^{-1}=-\psi_2$. 

Given the correlation between intrinsic $CPT$ parity and spin, from Table II we see that for the fermion bilinears and quadrilinears every contraction that has spin zero has even intrinsic  $CPT$ parity. Moreover, as we also see from Table II, $CPT$ is the only transformation that produces the same positive sign for every one of the spin zero contractions. ($PT$ almost has this property, failing to meet it only for $\bar{\psi}\gamma^{\mu}\psi\bar{\psi}\gamma_{\mu}\gamma^5\psi$.) Thus in a spin zero Lagrangian density, it is only under $CPT$ that every term in it has the same net intrinsic parity. $CPT$ is thus singled out as being different from all the other spacetime transformations. 

\subsection{Derivation of the $CPT$ Theorem}

To now derive a $CPT$ theorem for non-Hermitian Hamiltonians, we note first that, as shown in the Appendix, every single one of the  spin zero fermion combinations that is listed in Table II is unchanged under complex conjugation Since the action of $\hat{\pi}\hat{\tau}=\hat{\Lambda}^{0}_{\phantom{0}3}(i\pi)\hat{\Lambda}^{0}_{\phantom{0}2}(i\pi)\hat{\Lambda}^{0}_{\phantom{0}1}(i\pi)$ on a general spin zero combination will leave it invariant while reversing the signs of all four components of $x^{\mu}$, the action of  $K\hat{\pi}\hat{\tau}$ on any spin zero combination will do so too.   $K\hat{\pi}\hat{\tau}$ thus has precisely the same effect on the spin zero terms as $\hat{C}\hat{P}\hat{T}$, to thus lead to the same positive intrinsic $CPT$ parities as  listed in the last column in Table II. Thus to implement CPT we only need to implement $K\hat{\pi}\hat{\tau}$. On now applying the Lorentz transformation $\hat{\pi}\hat{\tau}$ to a general spin zero action, every single spin zero combination in it will transform the same way, to give $I=\int d^4x {{\cal L}}(x) \rightarrow \int d^4x{{\cal L}}(-x)$. However since $\int d^4x{{\cal L}}(-x)=\int d^4x{{\cal L}}(x)$ we see that $I$ is left invariant. The full $CPT$ transformation on the action thus reduces to $I\rightarrow KIK=\int d^4x K{{\cal L}}(x)K$. Finally, since we had shown in Sec. \ref{antilinearity} that a Hamiltonian must admit of an antilinear symmetry if it is to effect time-independent evolution of inner products, with this probability conservation requirement we then infer that  $K{{\cal L}}(x)K={{\cal L}}(x)$. The Lagrangian density and thus the Hamiltonian are thereby $CPT$ symmetric, and we thus obtain our desired $CPT$ theorem for non-Hermitian Hamiltonians.

In addition, we note that since $K$ complex conjugates all factors of $i$, even, as noted in the Appendix,  including those in the matrix representations of the quantum fields, we see that the Hamiltonian obeys $H=H^*$, to thus be real. While this condition is somewhat analogous to $H=H^{\dagger}$, in the standard approach to the $CPT$ theorem the $H=H^{\dagger}$ condition is input, while in our approach $H=H^*$ is output. With the use of complex conjugation under $K$, we see that the action of $K$ entails that in ${{\cal L}}(x)$ all numerical coefficients are real, with only general bosonic or fermionic $(\phi_1-i\phi_2)(\phi_1+i\phi_2)$ or $(\phi_1-i\phi_2)(\phi_1-i\phi_2)+(\phi_1+i\phi_2)(\phi_1+i\phi_2)$ type contractions being allowed. 

As we see, quite remarkably we finish up with the same allowed generic structure for ${{\cal L}}(x)$ as in the Hermitian case, except that now no restriction to Hermiticity has been imposed. In our approach we do not require the fields in ${{\cal L}}(x)$  to be Hermitian, we only require that they have a well-defined behavior under $CPT$, so that now we obtain $CPT$ symmetry of a Hamiltonian even if the Hamiltonian is Jordan-block or its energy eigenvalues appear in complex conjugate pairs. In the standard Hermiticity-based approach to the $CPT$ theorem one requires the fields in ${{\cal L}}(x)$ to be Hermitian and requires the coefficients in the action to be real. However, as we had noted in  our discussion in Sec. \ref{intro}, this is not sufficient to secure the Hermiticity of a Hamiltonian that is built out of the fields in the action, since when the Hamiltonian acts on the eigenstates of the field operators themselves the Hamiltonian may not be self-adjoint. $CPT$ symmetry thus goes beyond Hermiticity, and under only two requirements, viz.  conservation of  probability (for the antilinear part of the $CPT$ transformation) and invariance under the complex Lorentz group (for the linear part of the $CPT$ transformation), $CPT$ invariance of the Hamiltonian then follows, with no restriction to Hermiticity being needed. 

\subsection{No Vacuum Breaking of $CPT$ Symmetry}

While we have shown that the Hamiltonian is $CPT$ invariant, there is still the possibility that $CPT$ might be broken in the vacuum. However with every spin zero combination of fields being $CPT$ even as per Table II, then since Lorentz invariance only permits spin zero field configurations to acquire a non-vanishing vacuum expectation value, $CPT$ symmetry could not be broken spontaneously. Lorentz invariance thus plays a double role as it is central to making both the Hamiltonian and the vacuum be $CPT$ symmetric. 

\subsection{How to Distinguish Hermiticity from $CPT$ Invariance}

Since we obtain exactly the same generic form for the Hamiltonian whether we use Hermiticity or  invariance under complex conjugation times complex Lorentz invariance, and thus obtain Hamiltonians that on the face of it always appear to be Hermitian, we will need some criterion to determine which case we are in. As we will see, just as in the example given in (\ref{H1ab}), it depends on the values of the parameters. As regards the behavior in time, we note that if we have a real wave equation that does not mean that the associated frequencies are necessarily real, since solutions to real equations could come in complex conjugate pairs. As regards the behavior in space, that depends on asymptotic boundary conditions (viz. self-adjointness), since a real wave equation can have non-normalizable solutions that diverge asymptotically, and in Sec. \ref{comparing} we discuss this issue in detail.

For the time dependence issue, consider the neutral scalar field with action $I_{\rm S}=\int d^4x [\partial_{\mu}\phi\partial^{\mu}\phi-m^2\phi^2]/2$ and Hamiltonian $H=\int d^3x[\dot{\phi}^2+\vec{\nabla}\phi\cdot \vec{\nabla}\phi+m^2\phi^2]/2$. Solutions to the  wave equation $-\ddot{\phi}+\nabla^2\phi-m^2\phi=0$  obey $\omega^2(\vec{k})=\vec{k}^2+m^2$. Thus the poles in the scalar field propagator are at $\omega(\vec{k})=\pm[\vec{k}^2+m^2]^{1/2}$, the field can be expanded as $\phi(\vec{x},t)=\sum [a(\vec{k})\exp(-i\omega(\vec{k}) t+i\vec{k}\cdot\vec{x})+a^{\dagger}(\vec{k})\exp(+i\omega(\vec{k}) t-i\vec{k}\cdot\vec{x})]$,  and the Hamiltonian is given by $H=\sum [\vec{k}^2+m^2]^{1/2}[a^{\dagger}(\vec{k})a(\vec{k})+a(\vec{k})a^{\dagger}(\vec{k})]/2$.  

For either sign of $m^2$ the $I_{\rm S}$ action is CPT symmetric, and for both signs $I_{\rm S}$ appears to be Hermitian. For $m^2>0$, $H$ and $\phi(\vec{x},t)$ are indeed Hermitian and all frequencies are real. However, for $m^2<0$, frequencies become complex when $\vec{k}^2<-m^2$. The poles in the propagator move into the complex plane, the field $\phi(\vec{x},t)$ then contains modes that grow or decay exponentially in time, while $H$ contains energies that are complex. Thus now $H\neq H^{\dagger}$ and  $\phi\neq \phi^{\dagger}$. As we see,  whether or not an action is $CPT$ symmetric is an intrinsic property of the unconstrained action itself prior to any stationary variation, but whether or not a Hamiltonian is Hermitian is a property of the stationary solution.\footnote{\label{F8} While one can construct the Hamiltonian from the energy-momentum tensor, the energy-momentum tensor is only conserved in solutions to the equations of motion. Hermiticity is thus tied to the solutions to the theory in a way that $CPT$ is not.} Hermiticity of a Hamiltonian or of the fields that it is composed of cannot be assigned a priori, and can only be determined after the theory has been solved. However, the $CPT$ properties of Hamiltonians or fields can be assigned a priori, and thus that is how Hamiltonians and fields should be characterized.  One never needs to postulate Hermiticity at all.

\section{Some Implications of the $CPT$ Theorem in the Non-Hermitian Case}
\label{implications}

\subsection{$CPT$ Symmetry and Unstable States}

In the classic application of the $CPT$ theorem, the theorem was used to establish the equality of the lifetimes of unstable particles and their antiparticles, with the most familiar application being in $K$ meson decays. However, such use of the theorem  was made via a $CPT$ theorem whose derivation had only been obtained for Hamiltonians that are Hermitian, and for such Hamiltonians states should not decay at all. To get round this one by hand adds a non-Hermitian term to the Hamiltonian, with the added term being the same one in both the particle and the antiparticle decay channels. In addition, one also by hand imposes a non-$CPT$-invariant boundary condition that only allows for decaying modes and forbids growing ones. In our approach we have no need to do this since the time-independent inner products that we use precisely provide for time-independent transitions between decaying states and the growing states into which they decay without any need to add in any terms by hand. $CPT$ invariance then requires that  the transition rates for the decays of particles and their antiparticles be equal.

\subsection{$CPT$ Symmetry and $PT$ Symmetry}

Our derivation of the $CPT$ theorem for non-Hermitian Hamiltonians provides a fundamental justification for the $PT$ studies of Bender and collaborators. These studies are mainly quantum-mechanical ones in which the field-theoretic charge conjugation operator plays no role (i.e. $[\hat{C},\hat{H}]=0$). The $CPT$ symmetry of any given relativistic theory thus ensures the $PT$ symmetry of any charge conjugation invariant quantum-mechanical theory that descends from it, doing so regardless of whether or not the  Hamiltonian is Hermitian, and independent of whether or not $P$ or $T$ themselves are conserved.   

\subsection{The $H=p^2+ix^3$ Theory and $CPT$ Symmetry}

To appreciate the above points within a specific context, we recall that it was the $H=p^2+ix^3$ theory that first engendered interest in $PT$ symmetry, since despite not being Hermitian but instead being $PT$ symmetric, it had an entirely real set of energy eigenvalues \cite{Bender1998,Bender1999}, \cite{Bender2007} (and is actually Hermitian in disguise). Now the presence of the factor $i$ initially suggests that $H$ might not have descended from a $CPT$-invariant theory since our derivation of the $CPT$ theorem led us to numerical coefficients that are all real. However, in this particular case the factor of $i$ arises because the $H=p^2+ix^3$ theory does not descend directly from a $CPT$-invariant Hamiltonian but from a similarity transformation of one that does, an allowable  transformation since it does not affect energy eigenvalues. 

To be specific, consider an initial $CPT$-symmetric, time-independent Hamiltonian 
\begin{eqnarray}
H=\int d^3x[-\Pi^2(\vec{x},t=0)+\Phi^3(\vec{x},t=0)]
\label{F37ab}
\end{eqnarray}
with real coefficients, and the $C$, $P$, and $T$ assignments for $\Phi$, $\Pi$, and $-\Pi^2+\Phi^3$ as indicated in Table III as per the pseudoscalar $\bar{\psi}i\gamma^5\psi$ assignments listed in Table I.\footnote{\label{F9} With generic canonical field-theoretic commutator having the form $[\Phi(\bar{x}^{\prime},t=0),\Pi(\bar{x},t=0)]=i\delta^3(\bar{x}-\bar{x}^{\prime})$, a neutral field and its conjugate always have the same $C$ and $P$, and have opposite $T$, $PT$, and  $CPT$.} Since $H$ is time independent, we only need to evaluate the fields in it at $t=0$. The similarity transformation $S=\exp[(\pi/2)\int d^3x \Pi(\vec{x},t=0)\Phi(\vec{x},t=0)]$ effects 
\begin{eqnarray}
S\Phi(\vec{x},t=0) S^{-1} &=& -i\Phi(\vec{x},t=0)\equiv -ix,\qquad S\Pi(\vec{x},t=0) S^{-1}= i\Pi(\vec{x},t=0)\equiv ip,
\nonumber\\
SHS^{-1}&=&\int d^3x[\Pi^2(\vec{x},t=0)+i\Phi^3(\vec{x},t=0)]\equiv p^2+ix^3, 
\label{F38ab}
\end{eqnarray}
where we have introduced the compact notation $x$, $p$ and $p^2+ix^3$. The similarity transformation also leads to the $C$, $P$, and $T$ assignments for $x$ and $p$ as indicated in Table III, and a thus $CPT$ even $SHS^{-1}=p^2+ix^3$.\footnote{\label{F10} With $\Pi(\vec{x},t=0)$ being a time derivative of $\Phi(\vec{x},t=0)$,  the integral $\int d^3x \Pi(\vec{x},t=0)\Phi(\vec{x},t=0)$ is a Lorentz scalar. With the integral being composed of self-conjugate fields, its $CPT$ properties are  fixed by $K$ alone. Because of the presence of the factor $i$ in $[\Phi(\bar{x}^{\prime},t=0),\Pi(\bar{x},t=0)]=i\delta^3(\bar{x}-\bar{x}^{\prime})$, the integral is $T$ odd, and thus $CPT$ odd. Given the structure of this canonical commutator, a similarity transformation with $S=\exp[(\pi/2)\int d^3x \Pi(\vec{x},t=0)\Phi(\vec{x},t=0)]$  generates factors of $i$, to thus change the $T$, $PT$, and  $CPT$ properties of the fields. 
} Then, with both $\Phi$ and $\Pi$ being charge conjugation even neutral fields, the $PT$ symmetry of $H=p^2+ix^3$ directly follows. 

\begin{table}[ht]
\centering
\begin{tabular}{c|ccccc}
&~~C~~&~~P~~&~~T~~&~~PT~~&~~CPT~~\\ 
\hline
$\Phi$&+&-&-&+&+\\ 
$\Pi$&+&-&+&-&-\\
$-\Pi^2+\Phi^3$&+&&&+&+\\
$x$&+&-&+&-&-\\
$p$&+&-&-&+&+\\
$p^2+ix^3$&+&&&+&+\\
\hline
\end{tabular}
\caption{C, P, and T assignments for $\Phi$, $\Pi$, $x$, and $p$}
\end{table}

\subsection{The Pais-Uhlenbeck Two-Oscillator Theory and $CPT$ Symmetry}

Given our derivation of the $CPT$ theorem without assuming Hermiticity, it would be of interest to find an explicit $CPT$-invariant Hamiltonian whose energy eigenvalues come in complex conjugate pairs or whose Hamiltonian is not diagonalizable. To this end we consider the fourth-order Pais-Uhlenbeck two-oscillator ($[z,p_z]=i$ and  $[x,p]=i$) model studied in \cite{Bender2008a,Bender2008b}. Its action and Hamiltonian are given by 
\begin{eqnarray}
I_{\rm PU}=\frac{\gamma}{2}\int dt\left[{\ddot z}^2-\left(\omega_1^2
+\omega_2^2\right){\dot z}^2+\omega_1^2\omega_2^2z^2\right],
\label{H39ab}
\end{eqnarray}
\begin{eqnarray}
H_{\rm PU}=\frac{p^2}{2\gamma}+p_zx+\frac{\gamma}{2}\left(\omega_1^2+\omega_2^2 \right)x^2-\frac{\gamma}{2}\omega_1^2\omega_2^2z^2,
\label{H40ab}
\end{eqnarray}
where initially $\omega_1$ and $\omega_2$ are taken to be real (and positive for definitiveness). Once one sets  $\omega_1=(\bar{k}^2+M_1^2)^{1/2}$, $\omega_2=(\bar{k}^2+M_2^2)^{1/2}$ and drops the spatial dependence, this Hamiltonian becomes the quantum-mechanical limit of  a covariant fourth-order derivative neutral scalar field theory \cite{Bender2008b}, with action and propagator 
\begin{eqnarray}
I_S&=&\frac{1}{2}\int d^4x\bigg{[}\partial_{\mu}\partial_{\nu}\phi\partial^{\mu}
\partial^{\nu}\phi-(M_1^2+M_2^2)\partial_{\mu}\phi\partial^{\mu}\phi
+M_1^2M_2^2\phi^2\bigg{]},
\nonumber\\
D(k^2)&=&\frac{1}{(k^2-M_1^2)(k^2-M_2^2)}
=\frac{1}{M_1^2-M_2^2}\bigg{(}\frac{1}{k^2-M_1^2}-\frac{1}{k^2-M_2^2}\bigg{)},
\label{H41ab}
\end{eqnarray}
and Hamiltonian $H=\int d^3x T_{00}$ where
\begin{eqnarray}
T_{00}&=&\pi_{0}\dot{\phi}+\frac{1}{2}\pi_{00}^2+\frac{1}{2}(M_1^2+M_2^2)\dot{
\phi}^2-\frac{1}{2}M_1^2M_2^2\phi^2
-\frac{1}{2}\pi_{ij}\pi^{ij}+\frac{1}{2}(
M_1^2+M_2^2)\phi_{,i}\phi^{,i},
\nonumber \\
\pi^{\mu}&=&\frac{\partial{\cal L}}{\partial \phi_{,\mu}}-\partial_{\lambda
}\left(\frac{\partial {\cal L}}{\partial\phi_{,\mu,\lambda}}\right),\qquad \pi_{\mu\lambda}=\frac{\partial {\cal L}}{\partial \phi_{,\mu,\lambda}}.
\label{H42ab}
\end{eqnarray}
The $H_{\rm PU}$ Hamiltonian  turns out not to be Hermitian but to instead be $PT$ symmetric \cite{Bender2008a,Bender2008b}, with all energy eigenvalues nonetheless being given by $E(n_1,n_2)=(n_1+1/2)\omega_1+(n_2+1/2)\omega_2$, an expression that is real when $\omega_1$ and $\omega_2$ are both real. (When the frequencies are real all the poles of the propagator are on the real axis.) In addition, $H_{\rm PU}$ is $CPT$ symmetric since $H_{PU}$ is separately charge conjugation invariant ($[C,H_{PU}]=0$), while thus descending from a neutral scalar field theory with an action $I_S$ that is $CPT$ invariant itself. The theory is also free of ghost states of negative norm, since when one uses the needed positive definite $PT$ theory norm (viz. the one constructed  via $\langle \psi|{\cal C}PT|\psi\rangle$ \cite{Bender2007} where ${\cal C}$ this time is the $PT$ theory ${\cal C}$ operator described earlier -- a norm that, as we show in the Appendix,  is equivalent to the $\langle L|R\rangle$ norm introduced earlier), the relative minus sign in the partial fraction decomposition of the propagator given in (\ref{H41ab})  is generated not by the structure of the Hilbert space itself but  by the ${\cal C}$ operator \cite{Bender2008a,Bender2008b}, since with it obeying ${\cal C}^2=I$, it has eigenvalues equal to plus and minus one.  The negative residue of the pole in the $1/(k^2-M_2^2)$ term in  (\ref{H41ab}) is not due to a negative Dirac norm. Rather it means that one should not be using the Dirac norm at all. 

With the eigenvectors of $H_{\rm PU}$ being complete if $\omega_1$ and $\omega_2$ are real and unequal \cite{Bender2008a}, for real and unequal $\omega_1$ and $\omega_2$, $H_{\rm PU}$ while not Hermitian is Hermitian in disguise, with the explicit similarity transformation needed to bring it to a Hermitian form being given by  \cite{Bender2008a}  
\begin{eqnarray}
Q&=&\left( \frac{pq+\gamma^2\omega_1^2\omega_2^2xy}{\gamma
\omega_1\omega_2}\right)\log\left(\frac{\omega_1+\omega_2}{\omega_1-\omega_2}\right), 
\nonumber\\
e^{-Q/2}H_{\rm PU}e^{Q/2}&=&{\tilde H}_{\rm PU}=\frac{{p}^2}{2\gamma}+\frac{{q}^2}{2\gamma\omega_1^2}+\frac{\gamma}{
2}\omega_1^2 x^2+\frac{\gamma}{2}\omega_1^2\omega_2^2{y}^2={\tilde H}_{\rm PU}^{\dagger},
\label{F43ab}
\end{eqnarray}
where in terms of the operators given in (\ref{H40ab}), $y=-iz$, $q=ip_z$, and $[y,q]=i$. In this particular case $x=x^{\dagger}$, $p=p^{\dagger}$, $y=y^{\dagger}$, $q=q^{\dagger}$, $Q=Q^{\dagger}$, $V=e^{-Q}$, and ${{\cal C}}=PV$.  As we see,  ${\tilde H}_{\rm PU}$ is a perfectly well-behaved, standard Hermitian two-oscillator system that manifestly cannot have any states of negative norm. Thus for the two oscillator frequencies being real and unequal, while not Hermitian, $H_{\rm PU}$ is nonetheless Hermitian in disguise. As we now show, when we take the two frequencies to be equal or be in a complex conjugate pair this will no longer be the case.

\subsection{$CPT$ Symmetry when Energies are in Complex Conjugate Pairs}

If we set  $\omega_1=\alpha+i\beta$, $\omega_2=\alpha-i\beta$ with real $\alpha$ and $\beta$, we see that despite the fact that $\omega_1$ and $\omega_2$ are now complex, quite remarkably, the quantities $(\omega_1^2+\omega_2^2)/2=\alpha^2-\beta^2$ and  $\omega_1^2\omega_2^2=(\alpha^2+\beta^2)^2$ both remain real. In consequence $H_{\rm PU}$ remains $CPT$ invariant, but now the energies come in complex conjugate pairs as per $E(n_1,n_2)=(n_1+1/2)(\alpha+i\beta)+(n_2+1/2)(\alpha-i\beta)$.  With all the terms in the $I_{\rm PU}$ action still being real, the theory looks very much like a Hermitian theory, but it is not since energy eigenvalues come in complex conjugate pairs. The Pais-Uhlenbeck two-oscillator theory with complex conjugate frequencies thus provides an example of a theory that looks Hermitian but is not. The Pais-Uhlenbeck two-oscillator model with frequencies that come in complex conjugate pairs thus serves as an explicit example of a $CPT$-invariant but non-Hermitian Hamiltonian in which energy eigenvalues come in complex conjugate pairs, while showing that one can indeed write down theories of this type. (This example also shows that one can have dissipation despite the absence any odd-time-derivative dissipative terms in (\ref{H39ab}).)

\subsection{$CPT$ Symmetry in the Jordan-Block Case}

It is also of interest to note that when $\omega_1=\omega_2=\alpha$  with $\alpha$ real, the seemingly Hermitian $H_{\rm PU}$  becomes of non-diagonalizable, and thus of manifestly non-Hermitian,  Jordan-block form \cite{Bender2008b} (the  similarity transformation in (\ref{F43ab}) that effects $e^{-Q/2}H_{\rm PU}e^{Q/2}={\tilde H}_{\rm PU}$ becomes undefined when $\omega_1=\omega_2$), with its $CPT$ symmetry not being impaired. (In \cite{Bender2008b} the emergence of a Jordan-block Hamiltonian in the equal frequency limit was associated with the fact that the partial fraction decomposition of the propagator given in (\ref{H41ab}) becomes undefined when $M_1^2=M_2^2$, i.e. when $\omega_1^2=\omega_2^2$, since the $1/(M_1^2-M_2^2)$ prefactor becomes singular.) Thus for $\omega_1$ and $\omega_2$ both real and unequal, both real and equal, or being in a complex conjugates of each other, in all cases one has a non-Hermitian but $CPT$-invariant Hamiltonian that descends from a quantum field theory whose Hamiltonian while not Hermitian is nonetheless $CPT$ symmetric.\footnote{\label{F12} In \cite{Bender2008b} we carried out the construction of the energy eigenvalue spectrum and Hilbert space for the relativistic scalar field theory action $I_S$ of (\ref{H41ab}) itself, enabling us to show that there are no states of negative norm when $M_1^2\neq 0$, $M_2^2\neq 0$, and to identify the zero norm states that appear when $M_1^2=M_2^2=0$. Just as with the Pais-Uhlenbeck oscillator, these results carry over directly to the complex conjugate case where $M_1^2=M^2+iN^ 2$, $M_2^2=M^2-iN^ 2$. In \cite{Mannheim2011} we showed that these same results apply to the conformal gravity action $I_{\rm W}=-\alpha_g\int d^4x (-g)^{1/2}C_{\lambda\mu\nu\kappa} C^{\lambda\mu\nu\kappa}$ (equivalent to the $M_1^2=M_2^2=0$ case) when linearized around a flat background. In the Appendix we show how causality is maintained for all the various field-theoretic choices for $M_1^2$ and $M_2^2$ that are of interest to us here.} 

Even though the work of \cite{Bender2008a,Bender2008b} shows explicitly that $H_{\rm PU}$ is not Hermitian (being quadratic $H_{\rm PU}$ is exactly solvable),  it nonetheless appears to be so. However, while not Hermitian, $H_{\rm PU}$ is self-adjoint, and so we turn now to a discussion of the distinction between Hermiticity and self-adjointness. This will involve the introduction of Stokes wedges in the complex plane, regions where wave functions are asymptotically bounded, with such wedges playing a key role in $PT$ or any general antilinear symmetry studies \cite{Bender2007}.

\section{Comparing Antilinearity, Self-adjointness, and Hermiticity}
\label{comparing}

\subsection{Self-Adjointness and the Pais-Uhlenbeck Hamiltonian}

To understand the issue of self-adjointness we again consider the Pais-Uhlenbeck Hamiltonian, and make a standard wave-mechanics representation of the Schr\"odinger equation $H_{\rm PU}\psi_n=E_n\psi_n$ by setting
$p_z=-i\partial/\partial z$, $p_x=-i\partial/\partial x$. In this representation we find two classes of eigenstates, one a potentially physical class with positive energy eigenvalues when $\omega_1$ and $\omega_2$ are both  real and positive, and the other, an unphysical class with negative energy eigenvalues. The state whose energy is $(\omega_1+\omega_2)/2$, the lowest energy state in the positive energy sector, has an eigenfunction of the form \cite{Mannheim2007}
\begin{eqnarray}
\psi_+(z,x)&=&\exp\bigg[\frac{\gamma}{2}(\omega_1+\omega_2)\omega_1\omega_2z^2
+i\gamma\omega_1\omega_2zx-\frac{\gamma}{2}(\omega_1+\omega_2)x^2\bigg],
\label{H44ab}
\end{eqnarray}
while the state whose energy is $-(\omega_1+\omega_2)/2$, the highest energy state in an unbounded from below negative energy sector, has an eigenfunction of the form 
\begin{eqnarray}
\psi_-(z,x)&=&\exp\bigg[-\frac{\gamma}{2}(\omega_1+\omega_2)\omega_1\omega_2z^2
+i\gamma\omega_1\omega_2zx+\frac{\gamma}{2}(\omega_1+\omega_2)x^2\bigg].
\label{H45ab}
\end{eqnarray}
With $\psi_+(z,x)$ diverging at large $z$ and $\psi_-(z,x)$ diverging at large $x$, neither of theses two states is normalizable. Thus in trying to show that $H_{\rm PU}$  obeys $\int\psi^*_1 H\psi_2 =[\int \psi^*_2H\psi_1]^*$, we are unable to drop the surface terms that are generated in an integration by parts, and have to conclude \cite{Bender2008a,Bender2008b} that in the basis of wave functions associated with the positive energy eigenfunctions (or  negative for that matter) $H_{\rm PU}$ is not self-adjoint. Self-adjointness of a differential operator in a given basis means that one can throw away surface terms. Moreover, without actually looking at asymptotic boundary conditions, one cannot in fact determine if a differential operator is self-adjoint from the form of the operator itself, since such self-adjointness is determined not by the operator but by the space of states on which it acts.

Since there is only a sensible physical interpretation of a theory if the energy spectrum is bounded from below, we thus seek a viable interpretation of the  $\psi_+(z,x)$ sector of the Pais-Uhlenbeck model. Inspection of $\psi_+(z,x)$ shows that $\psi_+(z,x)$ would be normalizable if we were to replace $z$ by $iz$, and thus replace $p_z$ by $-\partial_z$ (so to maintain $[z,p_z]=i$). In other words we cannot presume a priori that $p_z$ is Hermitian in the basis of eigenfunctions of $H_{\rm PU}$, and thus cannot presume a priori that $H_{PU}$ is Hermitian either. The complete domain in the complex $z$ plane in which the wave function is normalizable is known as a Stokes wedge. If we draw a letter $X$ in the complex $z$ plane and also draw a letter $X$ in the complex $x$ plane, then $\psi_+(z,x)$ is normalizable if $z$ is in the north or south quadrant of its letter $X$, and $x$ is in the east or west quadrant of its letter $X$. The needed Stokes wedges contain purely imaginary $z$ and purely real $x$. And in these particular wedges we can construct normalizable wave functions whose energy eigenvalues are strictly bounded from below. Since the wave functions of the excited states are just polynomials functions of $z$ and $x$ times the ground state wave function \cite{Mannheim2007,Bender2008b}, in the same Stokes wedges these wave functions are normalizable too. While $H_{\rm PU}$ is not Hermitian, in these particular Stokes wedges we see that $H_{\rm PU}$ is nonetheless self-adjoint.  

Inspection of $\psi_+(z,x)$ shows that in these particular Stokes wedges the asymptotic behavior is not modified if we set $\omega_1=\omega_2=\alpha$, with $\alpha>0$. With this being true also for the excited states \cite{Mannheim2007,Bender2008b}, the Jordan-block limit of the Pais-Uhlenbeck Hamiltonian is thus self-adjoint even though it is manifestly not Hermitian. Moreover, if we set $\omega_1=\alpha+i\beta$, $\omega_2=\alpha-i\beta$ ($\alpha$ still positive and $\beta$ real) we obtain $\omega_1+\omega_2=2\alpha$, $\omega_1\omega_2=\alpha^2+\beta^2$. Thus, quite remarkably,  all the terms in $\psi_+(z,x)$ not only remain real, they undergo no sign change, with the wave functions thus still being  normalizable in the selfsame Stokes wedges. With this also being the case for the excited states, even in the complex energy sector, an again manifestly non-Hermitian situation, $H_{\rm PU}$ is still self-adjoint.  

\subsection{Self-Adjointness and Antilinearity}

While of course many operators are both Hermitian and self-adjoint, as we see from the Pais-Uhlenbeck example self-adjointness should not in general be associated with Hermiticity. The Pais-Uhlenbeck model shows that there is  instead  a connection between antilinearity and self-adjointness, and this turns out to be general. Specifically, below in Sec. \ref{euclidean} we will show that if a Hamiltonian has an antilinear symmetry the Euclidean time path integral is real. Moreover, if the real parts of the energy eigenvalues of the Hamiltonian are bounded from below and all are positive, the Euclidean time path integral is well-behaved and finite. In consequence, the Minkowski time path integral is finite too. Then, because of the complex plane correspondence principle that we derive below in Sec. \ref{continuing}, the quantum Hamiltonian must be self-adjoint in some domain in the complex plane. In general then, antilinearity implies self-adjointness. As to the converse, we note that if a Hamiltonian is self-adjoint in some direction in the complex plane, in that direction asymptotic surface terms would vanish and left-right inner products would be time independent. 

While we can show that $i\partial_t\langle L(t)|R(t)\rangle=\langle L(t)|\overrightarrow{H}|R(t)\rangle-\langle L(t)|\overleftarrow{H}|R(t)\rangle$ is immediately zero when $\hat{H}$ is represented as an infinite-dimensional matrix in Hilbert space, when $\hat{H}$ is represented as a differential operator, it acts to the right on $|R(t)\rangle$ and to the left on $\langle L(t)|$. To then show that $i\partial_t\langle L(t)|R(t)\rangle$ is zero requires the vanishing of the asymptotic surface term generated in an integration by parts. With such surface terms vanishing when $\hat{H}$ is self-adjoint, self-adjointness thus leads to  probability conservation. In addition, we note that if in matrix elements of the form $\langle R|\hat{H}|R \rangle=\int dx dy\psi^*_{R}(x)\langle x|\hat{H}|y\rangle\psi_R(y)$ we can drop surface terms in an integration by parts, we would have both self-adjointness and Hermiticity. However, when we need to distinguish between left- and right-eigenstates and introduce matrix elements of the form $\langle L|\hat{H}|R \rangle=\int dx dy\psi^*_{L}(x)\langle x|\hat{H}|y\rangle\psi_R(y)$, this time if we can drop surface terms in an integration by parts, we would still have self-adjointness but would not have Hermiticity (i.e. not have $\hat{H}_{LR}=(\hat{H}_{RL})^*$) since $\psi^*_L(x)$ is not the same as $\psi^*_R(x)$. Self-adjointness is thus distinct  from Hermiticity while encompassing it as the special case in which self-adjointness is secured without the need to continue into the complex plane. Probability would then be conserved and, as shown  in Sec. \ref{antilinearity}, the Hamiltonian would then have an antilinear symmetry. Thus antilinearity implies self-adjointness, and self-adjointness implies antiinearity.

\subsection{Connection Between the $CPT$ Norm and Left-Right Norm}

Now that we have identified $CPT$ as the basic antilinear symmetry for quantum theory, we see that the overlap of a state with its $CPT$ conjugate is  time independent  since the Hamiltonian is itself $CPT$ symmetric, with this norm thus being preserved in time. Now in Sec. \ref{antilinearity} we introduced a different time-independent  norm, the overlap of a right-eigenvector with a left-eigenvector. Thus up to a phase  we can now identify the left-eigenvector as the $CPT$ conjugate of the right-eigenvector. 

The issue of the phase is of relevance since the utility of the $CPT$ norm or of the left-right norm is not just in the time independence. The sign of the norm is also of significance. Since non-Hermitian Hamiltonians that have a real and complete eigenspectrum can be brought to a Hermitian form by a similarity transformation (cf. (\ref{H14ab}) above), and since the signs and magnitudes of inner products do not change under a similarity transformation, prior to making the transformation one must be able to define  a positive definite norm for such non-Hermitian Hamiltonian. The norm in question is not actually the overlap of a state with its $CPT$ conjugate, but is instead the left-right norm $\langle L|R\rangle=\langle R|V|R\rangle$. However, as we discuss in more detail in the Appendix, in many cases the $V$ operator can be written as $V=P{{\cal C}}$ where ${{\cal C}}={{\cal C}}^{-1}$  is the $PT$ theory  ${\cal C}$ operator. The $V$ norm is thus equivalent to a $P{{\cal C}}$ norm. With both of these norms being positive definite, their interpretation as probabilities is secured. 

The issue of the sign is also of significance for a different reason. For the unequal frequency fourth-order derivative Pais-Uhlenbeck Hamiltonian it is found that if one quantizes the theory using the Dirac norm, these norms turn out to be negative (see e.g. \cite{Bender2008a}), causing one to think that such theories are not unitary or of physical relevance. However, the fact that the Dirac norm is found to be negative is actually a signal that one is quantizing in the wrong Hilbert space and that the Hamiltonian is not Hermitian. When quantized with the ${{\cal C}}CPT$ norm used in $(C)PT$ theories (${\cal C}$ added to $(C)PT$), the norms are then positive definite \cite{Bender2008a}, with the theory then being fully acceptable. 

By same token, conformal gravity, equally a fourth-order derivative theory, is actually free of any negative Dirac norm ghost states \cite{Mannheim2011,Mannheim2012}, to thus be a fully acceptable quantum gravity theory. Moreover, it turns out that the Hamiltonian of (linearized) conformal gravity is actually Jordan block \cite{Mannheim2011,Mannheim2012} (analog of the equal frequency Pais-Uhlenbeck model), to thus manifestly not be Hermitian but to instead possess an antilinear CPT symmetry.

\subsection{$CPT$ symmetry and the Construction of Field-Theoretic Lagrangians}

As regards the difference between Hermiticity and antilinearity, we note additionally that in constructing field-theoretic Lagrangian densities it is standard practice, particularly when spinors are involved, to add on to the chosen Lagrangian density its Hermitian conjugate. This is done in order to make the ensuing Hamiltonian be Hermitian, since one simply postulates as a priori input that it should be. However, as we have seen, this is too restrictive a condition, with quantum theory being richer. Moreover, it is anyway unnecessary and one never actually needs to impose Hermiticity at all, since one should instead add on the $CPT$ conjugate (if one had initially chosen a Lagrangian density that was not $CPT$ invariant). Not only does this encompass Hermiticity while allowing more general possibilities, $CPT$ symmetry does not even need to be postulated as it is an output requirement for any quantum theory that has probability conservation and complex Lorentz invariance.

\section{Antilinearity and Euclidean Time Green's Functions and Path Integrals}
\label{euclidean}

\subsection{Hermitian Case}

To explore the interplay between antilinear symmetry and path integrals it suffices to discuss self-conjugate fields, and so we assume $C$ invariance and reduce $CPT$ symmetry to $PT$ symmetry. So consider now the generic two-point path integral  $\int {{\cal D}}[\phi]\phi(0,t)\phi(0,0)\exp(iS)$ with classical action $S=\int d^4x {{\cal L}}(x)$, as integrated over the paths of some generic self-conjugate field $\phi(\vec{x},t)$, with $\vec{x}$ conveniently taken to be zero. In theories in which the Hamiltonian is Hermitian, the left and right vacua needed for the two-point function are Hermitian conjugates of each other, and we can represent the associated time-ordered two-point function as a path integral
\begin{eqnarray}
\theta(t)\langle \Omega|\phi(0,t)\phi(0,0)|\Omega\rangle e^{-iE_0t}
&+&\theta(-t)\langle \Omega|\phi(0,0)\phi(0,t)|\Omega\rangle e^{+iE_0t}
\nonumber\\
&=&\int_{-\infty}^{\infty} {{\cal D}}[\phi]\phi(0,t)\phi(0,0)\exp(iS),
\label{H46ab}
\end{eqnarray}
where $E_0$ is the energy of the state $|\Omega\rangle$. Since the treatment of the $t>0$ and $t<0$ parts of the two point function are analogous, we shall only discuss the $t>0$ part in the following.
On introducing the time evolution operator, using the completeness relation  $H=\sum_n|n\rangle E_{n}\langle n|$,  and taking $\phi(\vec{x},t)$ to be Hermitian, evaluation of the $t>0$ part of the two-point function yields 
\begin{eqnarray}
\langle \Omega|\phi(0,t)\phi(0,0)|\Omega\rangle e^{-iE_0t}
&=&\langle \Omega|e^{iHt}\phi(0,0)e^{-iHt}\phi(0,0)|\Omega\rangle e^{-iE_0t}
\nonumber\\
&=&\sum_n\langle \Omega |\phi(0,0)|n\rangle \langle n|\phi(0,0) |\Omega\rangle e^{-iE_nt}
=\sum_n|\langle \Omega |\phi(0,0)|n\rangle |^2 e^{-iE_nt}.
\label{H47ab}
\end{eqnarray}
In arriving at this result we have identified $\langle n|\phi(0,0) |\Omega\rangle$ as the complex conjugate of $\langle \Omega |\phi(0,0)|n\rangle$. Such an identification can immediately be made if the states $|n\rangle$ are also eigenstates of a Hermitian $\phi(0,0)$, except for the fact that they actually cannot be since $[\phi,H]=i\partial_t\phi$ is not equal to zero. Nonetheless, in its own eigenbasis we can set  $\phi=\sum_{\alpha}|\alpha\rangle \phi_{\alpha}\langle\alpha|$, where the $\phi_{\alpha}$ are real. Consequently, we can set
\begin{eqnarray}
\langle \Omega|\phi(0,0)|n\rangle &=&\sum_{\alpha}\langle \Omega |\alpha\rangle \phi_{\alpha}\langle\alpha |n\rangle,
\nonumber\\
\langle n|\phi(0,0)|\Omega\rangle &=&\sum_{\alpha}\langle n |\alpha\rangle \phi_{\alpha}\langle\alpha |\Omega\rangle
=\sum_{\alpha}\langle\alpha |n\rangle^*\phi_{\alpha} \langle \Omega |\alpha\rangle^* =\langle \Omega|\phi(0,0)|n\rangle^*,
\label{H48ab}
\end{eqnarray}
from which the last equality in (\ref{H47ab}) then follows after all.

If we now substitute the Euclidean time $\tau=it$ in  (\ref{H47ab}) we obtain 
\begin{eqnarray}
&&\langle \Omega|\phi(0,t)\phi(0,0)|\Omega\rangle e^{-iE_0t}
=\sum_n|\langle \Omega |\phi(0,0)|n\rangle |^2 e^{-E_n\tau}.
\label{H49ab}
\end{eqnarray}
In Euclidean time this expression is completely real since all the eigenvalues of a Hermitian Hamiltonian are real, to thus confirm that in this case the Euclidean time two-point function and the Euclidean time path integral are completely real. The Euclidean time two-point function is convergent at large positive $\tau$ if all the $E_n$ are greater or equal to zero. (The complex $t$ plane Wick rotation is such that $t>0$ corresponds to $\tau>0$.\footnote{\label{F16} With $t$-plane singularities having $t_I>0$ (the typical oscillator path integral behaves as $1/\sin[(\omega-i\epsilon)t]$), and with circle at infinity terms vanishing in the lower half plane (cf. $\exp[-i\omega(t_R+it_I)]$), with $\tau=it$ a lower right quadrant Wick rotation yields $i\int_{0}^{\infty} dt =-i\int_{-i\infty}^{0} dt =\int_{0}^{\infty} d\tau$.}) Also, its expansion at large $\tau$ is dominated by $E_0$, with the next to leading term being given by the next lowest energy $E_1$ and so on. Finally, in order for the time-ordered two-point function given in (\ref{H46ab})  to be describable by a Euclidean time path integral with convergent exponentials, as per continuing in time according to $\tau=it$, we would need $iS=i\int dt d^3x {{\cal L}}(\vec{x},t)=\int d\tau d^3x {{\cal L}}(\vec{x},-i\tau)$ to be real and negative definite on every path. 

\subsection{$CPT$ Symmetric Case with All Energies Real}

We can obtain an analogous outcome when the Hamiltonian is not Hermitian, and as we now show, it will precisely be $PT$ symmetry (i.e. $CPT$ symmetry) that will achieve it for us. As  described earlier, in general we must distinguish between left- and right-eigenvectors, and so in general  the $t>0$ two-point function will represent $\langle \Omega_{L}|\phi(0,t)\phi(0,0)|\Omega_{R}\rangle e^{-iE_0t}$. Now in the event that the left-eigenvectors are not the Dirac conjugates of the right-eigenvectors of $H$, the general completeness and orthogonality relations (in the non-Jordan-block case) are given by \cite{Mannheim2013} $\sum_n|R_{n}\rangle\langle L_{n}|=\sum_n|L_{n}\rangle\langle R_{n}|=I$, $\langle L_{n}|R_{m}\rangle=\langle R_{m}|L_{n}\rangle=\delta (n,m)$, while the spectral decomposition of the Hamiltonian is given by $H=\sum_n|R_{n}\rangle E_{n}\langle L_{n}|$. Consequently, we can set
\begin{eqnarray}
&&\langle \Omega_{L}|\phi(0,t)\phi(0,0)|\Omega_{R}\rangle e^{-iE_0t}
=\sum_n\langle \Omega_{L}|\phi(0,0)|R_{n}\rangle e^{-iE_{n}t}\langle L_{n}|\phi(0,0)|\Omega_{R}\rangle.
\label{H50ab}
\end{eqnarray}

To analyze this expression we will need to determine the matrix elements of $\phi(0,0)$. To use Hermiticity for $\phi(0,0)$ is complicated and potentially not fruitful. Specifically, if we insert  $\phi=\sum_{\alpha}|\alpha\rangle \phi_{\alpha}\langle\alpha|$ in the various matrix elements  of interest, on recalling that  $\langle L|=\langle R|V$, we obtain 
\begin{eqnarray}
&&\langle \Omega_{L}|\phi(0,0)|R_{n}\rangle =\sum_{\alpha}
\langle \Omega_R |V|\alpha\rangle \phi_{\alpha}\langle\alpha |R_n\rangle,
\nonumber\\
&&\langle L_{n}|\phi(0,0)|\Omega_{R}\rangle =\sum_{\alpha}\langle R_n |V|\alpha\rangle \phi_{\alpha}\langle\alpha |\Omega_R\rangle
=\sum_{\alpha}\langle\alpha |V^{\dagger}|R_n\rangle^*\phi_{\alpha} \langle \Omega_R |\alpha\rangle^*. 
\label{H51ab}
\end{eqnarray}
This last expression is not only not necessarily equal to $\langle \Omega_L|\phi(0,0)|R_n\rangle^*$, it does not even appear to be related to it.

To be able to obtain a quantity that does involve the needed complex conjugate, we note that  as well as being Hermitian, as a self-conjugate neutral scalar field, $\phi(0,0)$ is $PT$ even. Its $PT $ transformation properties are straightforward since we can write everything in the left-right energy eigenvector  basis (as noted in Sec. \ref{intro} relations such as  $[PT,\phi]=0$ and thus $PT\phi T^{-1}P^{-1}=\phi$ are  basis independent). On applying a $PT$ transformation and recalling that $P^2=1$, $T^2=1$, we obtain  
\begin{eqnarray}
\phi&=&\sum_{i,j}|R_i \rangle \phi_{ij}\langle L_j|=PT\phi T^{-1}P^{-1}
=PT\phi TP=\sum_{i,j}PT|R_i \rangle \phi^*_{ij}\langle L_j|TP.
\label{H52ab}
\end{eqnarray}
As per (\ref{H23ab}), for energy eigenvalues that are real we have $PT|R_i \rangle=|R_i \rangle$, $\langle L_j|TP=\langle L_j|$, with $PT\phi TP=\phi$  thus yielding
\begin{eqnarray}
\phi_{ij}=\phi^*_{ij},\qquad \langle L_i|\phi |R_j \rangle =\phi_{ij}.
\label{H53ab}
\end{eqnarray}
Thus we can set
\begin{eqnarray}
\langle \Omega_{L}|\phi(0,t)\phi(0,0)|\Omega_{R}\rangle e^{-iE_0t}=\sum_n \phi_{0n}\phi_{n0} e^{-iE_{n}t}.
\label{H54ab}
\end{eqnarray}
With $\phi_{0n}$ and $\phi_{n0}$ both being real, with real $E_n$ this expression is completely real when the time is Euclidean. Thus in the real eigenvalue sector of a  $PT$-symmetric theory, the Euclidean time two-point function and the Euclidean time path integral are completely real. Since they both are completely real, we confirm that the form $\langle \Omega_{L}|\phi(0,t)\phi(0,0)|\Omega_{R}\rangle$ is indeed the correct $PT$-symmetry generalization of the Hermitian theory form $\langle \Omega|\phi(0,t)\phi(0,0)|\Omega\rangle$ used in (\ref{H46ab}) above.

\subsection{$CPT$ Symmetric Case with Some Energies in Complex Pairs}

In the event that energy eigenvalues appear in complex conjugate pairs, we have two cases to consider, namely cases in which there are also real eigenvalues, and cases in which all eigenvalues are in complex conjugate pairs. In both the cases we shall sequence the energy eigenvalues in order of increasing real parts of the energy eigenvalues. Moreover, in cases where there are both real and complex energy eigenvalues we shall take the one with the lowest real part to have a purely real energy. 

For energy eigenvalues that are in complex conjugate pairs according to $E_{\pm}=E_R\pm iE_I$, as per (\ref{H23ab}) we have 
\begin{eqnarray}
PT|R_{\pm} \rangle=|R_{\mp} \rangle,\quad
\langle L_{\pm}|TP=\langle L_{\mp}|,
\label{H55ab}
\end{eqnarray}
with time dependencies $|R_{\pm} \rangle\sim \exp(-iE_{\pm}t)=\exp(-iE_Rt\pm E_It)$, $\langle L_{\pm}|=\langle R_{\pm}|V \sim \exp(iE_{\mp}t)=\exp(iE_Rt \pm E_It)$.  Given (\ref{H27ab}) and (\ref{H28ab}), we see that these eigenvectors have no overlap with the eigenvectors associated with purely real eigenvalues. In the complex conjugate energy eigenvalue sector we can set  $\sum_n[|R^{+}_{n}\rangle\langle L^{-}_{n}|+|R^{-}_{n}\rangle\langle L^{+}_{n}|]=I$ as summed over however many complex conjugate pairs there are. Also we can set  $\langle L^{-}_{n}|R^{+}_{m}\rangle=\langle L^{+}_{n}|R^{-}_{m}\rangle=\delta (n,m)$, while the previous spectral decomposition of the Hamiltonian given by $H=\sum_n|R_{n}\rangle E_{n}\langle L_{n}|$ is augmented with $H=\sum_n[|R^{+}_{n}\rangle E^{+}_{n}\langle L^{-}_{n}|+|R^{-}_{n}\rangle E^{-}_{n}\langle L^{+}_{n}|]$. Thus just as in  our discussion of transition matrix elements in Secs. \ref{intro} and \ref{antilinearity}, the non-trivial overlaps are always between states with exponentially decaying and exponentially growing  behavior in time.

Now while the Hamiltonian does not link the real and complex conjugate energy sectors the scalar field can. In this mixed sector, with summations being suppressed, the decomposition of the scalar field is given by
\begin{eqnarray}
\phi&=&|R_{i} \rangle \phi_{i-}\langle L_{-}|+|R_{i} \rangle \phi_{i+}\langle L_{+}|
+|R_{-} \rangle \phi_{-i}\langle L_{i}|+|R_{+} \rangle \phi_{+i}\langle L_{i}|,
\nonumber\\
PT\phi TP&=&|R_{i} \rangle \phi^*_{i-}\langle L_{+}|+|R_{i} \rangle \phi^*_{i+}\langle L_{-}|
+|R_{+} \rangle \phi^*_{-i}\langle L_{i}|+|R_{-} \rangle \phi^*_{+i}\langle L_{i}|,
\label{H56ab}
\end{eqnarray}
with $PT\phi TP=\phi$  thus yielding
\begin{eqnarray}
&&\phi_{i-}=\phi^*_{i+},~~ \phi_{i+}=\phi^*_{i-},~~ \phi_{-i}=\phi^*_{+i},~~ \phi_{+i}=\phi^*_{-i},
\nonumber\\
&&\langle L_{i}|\phi |R_{+}\rangle =\phi_{i-},~~  \langle L_{i}|\phi |R_{-}\rangle =\phi_{i+},~~
\langle L_{+}|\phi |R_{i}\rangle =\phi_{-i},~~  \langle L_{-}|\phi |R_{i}\rangle =\phi_{+i}.
\label{H57ab}
\end{eqnarray}
The contribution of this sector to the two-point function is given by 
\begin{eqnarray}
&&\langle \Omega_{L}|\phi(0,t)\phi(0,0)|\Omega_{R}\rangle  e^{-iE_0t}
=\phi_{0-}\phi_{+0} e^{-iE_{R}t+E_It}+\phi_{0+}\phi_{-0} e^{-iE_{R}t-E_It}.
\label{H58ab}
\end{eqnarray}
Via (\ref{H57ab}) we see that the Euclidean time Green's function and path integral are completely real, just as desired.  
 
On comparing (\ref{H58ab}) with (\ref{H54ab}), we see that (\ref{H58ab})  is a direct continuation of (\ref{H54ab}),  with pairs of states with real energy eigenvalues in (\ref{H54ab}) continuing into pairs of states with complex conjugate energy eigenvalues in (\ref{H58ab}). This pattern is identical to the one exhibited by the two-dimensional matrix example given in (\ref{H1ab}). Since we have to go through a Jordan-block phase in order to make the continuation from real to complex energy eigenvalues, we can infer that also in the $PT$-symmetric Jordan-Block case the Euclidean time Green's function and path integral will be real. In fact this very situation has already been encountered in a specific model, the real frequency realization of the fourth-order Pais-Uhlenbeck two-oscillator model. The Hamiltonian of the theory is $PT$ symmetric, and in the equal-frequency limit becomes Jordan block. For both the real and unequal frequency case and the real and equal frequency case the Euclidean time path integral is found to be real \cite{Mannheim2007}, with the unequal-frequency path integral continuing into the equal-frequency path integral in the limit,  while nicely generating none other than the Euclidean time continuation of the non-stationary $t\exp(-iE t)$ wave function  described in Sec. \ref{intro}.

\subsection{$CPT$ Symmetric Case with All Energies in Complex Pairs}

In the event that all the energy eigenvalues of the theory are in complex conjugate pairs, we need to  evaluate two-point function matrix elements taken in these states. Since the Hamiltonian does not  induce  transitions  between differing pairs we only need to consider one such pair. In this sector we can expand $\phi$ according to 
\begin{eqnarray}
\phi&=&|R_{+} \rangle \phi_{+-}\langle L_{-}|+|R_{-} \rangle \phi_{-+}\langle L_{+}|,
\qquad
PT\phi TP=|R_{-} \rangle \phi^*_{+-}\langle L_{+}|+|R_{+} \rangle \phi^*_{-+}\langle L_{-}|,
\label{H59ab}
\end{eqnarray}
with $PT\phi TP=\phi$  thus yielding
\begin{eqnarray}
\phi_{+-}=\phi^*_{-+},\qquad \phi_{-+}=\phi^*_{+-},
\qquad
\langle L_{-}|\phi |R_{+}\rangle =\phi_{+-},\qquad  \langle L_{+}|\phi |R_{-}\rangle =\phi_{-+}.
\label{H60ab}
\end{eqnarray}
In this sector we can thus set
\begin{eqnarray}
\langle \Omega_{+}|\phi(0,t)\phi(0,0)|\Omega_{-}\rangle  
&=&\phi_{-+}\phi_{-+} e^{-iE_{R}t-E_It},
\nonumber\\
 \langle \Omega_{-}|\phi(0,t)\phi(0,0)|\Omega_{+}\rangle  
&=&\phi_{+-}\phi_{+-} e^{-iE_{R}t+E_It}.
\label{H61ab}
\end{eqnarray}
From (\ref{H60ab}) we see that  the Euclidean time Green's function and path integral associated with the sum $\langle \Omega_{+}|\phi(0,t)\phi(0,0)|\Omega_{-}\rangle +\langle \Omega_{-}|\phi(0,t)\phi(0,0)|\Omega_{+}\rangle$ are completely real. (The difference would be purely imaginary.) Thus, as indicated in Sec. \ref{intro}, in all possible cases we find that if the Hamiltonian is $PT$ symmetric the Euclidean time Green's functions and path integrals are real.\footnote{\label{F17} We should however add a caveat. Given the fact that in Sec. \ref{cpt} we showed that $H=KHK=H^*$, we would initially conclude that for Euclidean times $\tau=it$ and a time-independent Hamiltonian, the time evolution operator $\exp(-iHt)=\exp(-H\tau)$ would automatically be real. Consequently, the associated Euclidean time path integrals and Green's functions would be real too.  However, like the condition $H=H^{\dagger}$, the condition $H=H^*$ is not preserved under a similarity transformation. Thus initially we could only establish reality of the  Euclidean time Green's functions and path integrals in a restricted class of bases. As the analysis of Sec. \ref{cpt} shows, when $C\phi(\vec{x},t)C^{-1}=\phi(\vec{x},t)$ those bases include the ones in which $PT\phi(\vec{x},t)[PT]^{-1}=\phi(-\vec{x},-t)$. However, while the operator identity $H=H^*$ would transform non-trivially under a similarity transform, with the Green's functions being matrix elements of the fields as per $\langle \Omega_{L}|\phi(0,t)\phi(0,0)|\Omega_{R}\rangle$, the Euclidean time Green's functions  and path integrals would be left invariant under the similarity transform and thus always take the real values obtained in the basis in which $CPT\phi(\vec{x},t)[CPT]^{-1}=\phi(-\vec{x},-t)$. That this must be the case is because the terms in the Euclidean time path integral behave as $\exp(-E_i\tau)$ times left-right matrix elements of the field operators where the $E_i$ are energy eigenvalues, and energy eigenvalues and field operator matrix elements  are left invariant under similarity transformations.}

To prove the converse, we note  that when we continue the path integral to Euclidean time and take the  large $\tau=it$ limit,  the leading term is of the form $\exp(-E_0 \tau)$ where $E_0$ is the energy of the ground state. The next to leading term is the first excited state and so on (as sequenced according to the real parts of the energy eigenvalues, all taken to be positive). If the Euclidean time path integral is real, it is not possible for there to be any single  isolated complex energy eigenvalue. Rather, any such complex eigenvalues must come in complex conjugate pairs, and likewise the left-right overlap matrix elements of the fields (the coefficients of the $\exp(-E \tau)$ terms) must equally come in complex conjugate pairs. Thus if the Euclidean time path integral is real we can conclude that all the energies and matrix elements are real or appear in complex conjugate pairs. Moreover, if the energies are all real but one obtains some matrix elements that are not stationary (i.e. $\sim \tau\exp(-E\tau)$), we can conclude that the Hamiltonian is Jordan block. Hence, according to our previous discussion, in all cases the Hamiltonian of the theory must be $PT$ symmetric. We thus establish that  $PT$ (i.e. $CPT$) symmetry is a both necessary and sufficient condition for the reality of the Euclidean time path integral, and  generalize to field theory the analogous result for $|H-\lambda I|$ that was obtained in \cite{Bender2010} for matrix mechanics.

\section{Constraining the Path Integral Action via $CPT$ Symmetry}
\label{constraining}

The discussion given above regarding path integrals was based on starting with matrix elements of products of quantum fields and rewriting them as path integrals. Thus we begin with the q-number theory in which the quantum-mechanical Hilbert space is already specified and construct a c-number path integral representation of its Green's functions from it. However, if one wants to use path integrals to quantize a theory in the first place one must integrate the exponential of $i$ times the classical action over classical paths. Thus we start with the classical action, and if we have no knowledge beforehand of the structure of the quantum action, we cannot construct the classical action by taking the quantum action and replacing each q-number quantity in it by a c-number (i.e. by replacing q-number operators that obey non-trivial $\hbar$-dependent commutation relations by c-number quantities for which all commutators are zero.) Moreover, while a quantum field theory may be based on Hermitian operators, such Hermiticity is an intrinsically quantum-mechanical concept that cannot even be defined until a quantum-mechanical Hilbert space has been constructed on which the quantum operators can then act. Or stated differently, since path integration is an entirely classical procedure involving integration of a purely classical action over classical paths there is no reference to any Hermiticity of operators in it at all. And even if one writes the Lagrangian in the classical action as the Legendre transform of the classical Hamiltonian, one cannot attach any notion of Hermiticity to the classical  Hamiltonian either.

To try to get round this problem one could argue that since the eigenvalues of Hermitian operators are real, and since such eigenvalues are c-numbers, one should build the classical action out of these eigenvalues, with the classical action then being a real c-number. And if the classical action is real, in Euclidean time $i$ times the action would be real too. The simplest example of a real classical action  is the one inferred from the quantum Lagrangian $m\dot{x}^2/2$ for a free, non-relativistic quantum particle with a q-number position operator that obeys $[\hat{x},\hat{p}]=i\hbar$. On setting $\hbar=0$ one constructs  the classical Lagrangian as the same  $m\dot{x}^2/2$ except that now $x$ is a c-number that obeys $[x,p]=0$. Another familiar example is the neutral scalar field Lagrangian $\partial_{\mu}\phi\partial^{\mu}\phi$, with the same form serving in both the q-number and c-number cases. If we take the fields to be charged, while we could use a Lagrangian of the form  $\partial_{\mu}\phi\partial^{\mu}\phi^*$ in the c-number case, in the q-number case we would have to use $\partial_{\mu}\phi\partial^{\mu}\phi^{\dagger}$.

\subsection{Gauge Field and Fermion Field Considerations}

Despite this, this prescription fails as soon as one couples to a gauge field or introduces a fermion field. For a gauge field one can take the quantum-mechanical $A_{\mu}$ to be Hermitian and the  classical-mechanical $A_{\mu}$ to be real. With such a real $A_{\mu}$ one could introduce a classical Lagrangian density of the form $(\partial_{\mu}\phi-A_{\mu}\phi)(\partial^{\mu}\phi^*-A^{\mu}\phi^*)$. Now while this particular classical Lagrangian  density would be locally invariant under $\phi\rightarrow e^{\alpha(x)}\phi$, $A_{\mu}\rightarrow A_{\mu}+\partial_{\mu}\alpha(x)$, it would not be acceptable since a path integration based on it would not produce conventional quantum electrodynamics. Rather, to generate  conventional quantum electrodynamics via path integration one must take the classical Lagrangian density to be of the form  $(\partial_{\mu}\phi-iA_{\mu}\phi)(\partial^{\mu}\phi^*+iA^{\mu}\phi^*)$. Now in this particular case we already know the answer since the $(\partial_{\mu}\phi-iA_{\mu}\phi)(\partial^{\mu}\phi^{\dagger}+iA^{\mu}\phi^{\dagger})$ form (or equivalently $(i\partial_{\mu}\phi+A_{\mu}\phi)(-i\partial^{\mu}\phi^{\dagger}+A^{\mu}\phi^{\dagger})$) is the form of the quantum-mechanical Lagrangian density. However, that does not tell us what classical action to use for other theories for which the quantum-mechanical action is not known ahead of time. 

To address this issue we need to ask why one should include the factor of $i$ in the quantum Lagrangian in the first place. The answer is that in quantum mechanics it is not $\partial_{\mu}$ that is Hermitian. Rather, it is $i\partial_{\mu}$. Then since  $\partial_{\mu}$ is anti-Hermitian one must combine it with some anti-Hermitian function of the Hermitian $A_{\mu}$, hence $iA_{\mu}$. We thus have a mismatch between the quantum and classical theories, since while $\partial_{\mu}$ is real it is not Hermitian. We must thus seek some entirely different rule for determining the classical action needed for path integration, one that does not rely on any notion of Hermiticity at all. That needed different rule is $CPT$ symmetry.

Because of the structure of the Lorentz force $\vec{F}=e\vec{E}+e\vec{v}\times \vec{B}$, in classical electromagnetism one should not be able to distinguish between a charge $e$ moving in given $\vec{E}$ and $\vec{B}$ fields and the oppositely signed charge moving in $-\vec{E}$ and $- \vec{B}$ fields (opposite since these $\vec{E}$ and $\vec{B}$ fields are themselves set up by charges). In consequence both $e$ and $A_{\mu}$ are taken to be charge conjugation odd so that the combination $eA_{\mu}$ is charge conjugation even. Thus in order to implement $CPT$ invariance for classical electromagnetic couplings where $A_{\mu}$ always appears multiplied by $e$, one only needs to implement $PT$ invariance. Now under a $PT$ transformation $A_{\mu}$ is $PT$ even. Thus with $\partial_{\mu}$ being $PT$ odd,\footnote{\label{F18} In a $PT$ transformation on the coordinates, $\partial_{\mu}$ transforms into $-\partial_{\mu}$. In a $PT$ transformation on the fields $\partial_{\mu}\phi(x^{\lambda})$ transforms into $ \partial_{\mu}\phi(-x^{\lambda})$, i.e. into $-[\partial/\partial (-x^{\mu})]\phi(-x^{\lambda})$.  Thus, under a $d^4x $ integration the $PT$ transform of $\partial_{\mu}\phi(x^{\lambda})$ acts as $-\partial_{\mu}\phi(x^{\lambda})$. Thus, under a transformation on coordinates or fields, in the action  $\partial_{\mu}$ acts as a $PT$ odd operator.} we see that we must always have $\partial_{\mu}$ be accompanied by $ieA_{\mu}$ and not by $eA_{\mu}$ itself, since then both $\partial_{\mu}$ and $ieA_{\mu}$ would have the same negative sign under $PT$. To then construct a coupling term that has zero Lorentz spin, is $PT$ (and thus $CPT$) even, and obeys $K{{\cal L}}(x)K={{\cal L}}(x)$ (cf. the discussion in Sec. \ref{cpt}), we must take ${{\cal L}}(x)=(\partial_{\mu}\phi-ieA_{\mu}\phi)(\partial^{\mu}\phi^*+ieA^{\mu}\phi^*)$, with $PT$ and $CPT$ symmetry thus readily being implementable at the level of the classical action. We must thus use $CPT$ symmetry at the classical level in order to fix the structure of the classical path integral action. And moreover, $CPT$ symmetry can be implemented not just on one classical path such as the stationary one, it can be implemented on every classical path, stationary or  non-stationary alike. When this is done, the resulting quantum theory obtained via path integral quantization will also be $CPT$ symmetric, with the associated quantum Hamiltonian being $CPT$ symmetric too, and being so regardless of whether or not it might be Hermitian.

The situation for fermion fields is analogous. Specifically, for fermion fields we could introduce Grassmann fermions and take the path integral action to be $\int d^4x \bar{\psi}\gamma^{\mu}\partial_{\mu}\psi$. However, this expression is not $CPT$ invariant, and it is $CPT$ symmetry that tells us to introduce a factor of $i$ and use the standard $\int d^4x \bar{\psi}i\gamma^{\mu}\partial_{\mu}\psi$ instead.

\subsection{Gravity Considerations}

Similar considerations apply to path integral actions that involve gravity, and again there is a simplification, since just like the classical $eA_{\mu}$, the metric $g_{\mu\nu}$ is charge conjugation even. Thus if we take a relativistic flat spacetime theory that is already $CPT$ invariant and replace $\eta_{\mu\nu}$ by $g_{\mu\nu}$, replace ordinary derivatives by covariant ones,  and couple to gravity via the standard Levi-Civita connection 
\begin{eqnarray}
\Lambda^{\lambda}_{\phantom{\alpha}\mu\nu}=\frac{1}{2}g^{\lambda\alpha}(\partial_{\mu}g_{\nu\alpha} +\partial_{\nu}g_{\mu\alpha}-\partial_{\alpha}g_{\nu\mu}),
\label{F62ab}
\end{eqnarray}
$CPT$ invariance would not be impaired.  Now in coupling to gravity one can use a geometric connection $\Gamma^{\lambda}_{\phantom{\alpha}\mu\nu}$ that is more general than the standard Levi-Civita connection. One could for instance introduce a torsion-dependent connection of the form
\begin{eqnarray}
K^{\lambda}_{\phantom{\alpha}\mu\nu}=\frac{1}{2}g^{\lambda\alpha}(Q_{\mu\nu\alpha}+Q_{\nu\mu\alpha}-Q_{\alpha\nu\mu}),
\label{F63ab}
\end{eqnarray} 
where $Q^{\lambda}_{\phantom{\alpha}\mu\nu}=\Gamma^{\lambda}_{\phantom{\alpha}\mu\nu}-\Gamma^{\lambda}_{\phantom{\alpha}\nu\mu}$ is the antisymmetric part of the connection. Or one could use the modified Weyl connection introduced in \cite{Mannheim2014,Mannheim2015b,Mannheim2016b}, viz. 
\begin{eqnarray}
V^{\lambda}_{\phantom{\alpha}\mu\nu}=-\frac{2ie}{3}g^{\lambda\alpha}\left(g_{\nu\alpha}A_{\mu} +g_{\mu\alpha}A_{\nu}-g_{\nu\mu}A_{\alpha}\right),
\label{F64ab}
\end{eqnarray}
where $A_{\mu}$ is the electromagnetic vector potential. As shown in \cite{Mannheim2014}, both $K^{\lambda}_{\phantom{\alpha}\mu\nu}$ and $V^{\lambda}_{\phantom{\alpha}\mu\nu}$ transform in the same $CPT$ way (viz. $CPT$ odd) as $\Lambda^{\lambda}_{\phantom{\alpha}\mu\nu}$ (with $V^{\lambda}_{\phantom{\alpha}\mu\nu}$ doing so precisely because of the factor of $i$),  and thus neither of them modifies the $PT$ or $CPT$ structure of the theory in any way, with the theory remaining $CPT$ invariant. 

Our use of the modified  $V^{\lambda}_{\phantom{\alpha}\mu\nu}$ connection is of interest for another reason. When first introduced by Weyl in an attempt to metricate (geometrize) electromagnetism and give gravity a conformal structure, the connection was taken to be of the form 
\begin{eqnarray}
W^{\lambda}_{\phantom{\alpha}\mu\nu}=-eg^{\lambda\alpha}\left(g_{\nu\alpha}A_{\mu} +g_{\mu\alpha}A_{\nu}-g_{\nu\mu}A_{\alpha}\right).
\label{F65ab}
\end{eqnarray}
Apart from an overall normalization factor, this connection differs from the modified one by not possessing the factor of $i$. Since Weyl was working in  classical gravity, everything was taken to be real, with the  $\partial_{\mu}$ derivative in the Levi-Civita connection  being replaced by $\partial_{\mu}-2eA_{\mu}$ in order to generate $W^{\lambda}_{\phantom{\alpha}\mu\nu}$. From the perspective of classical physics the Weyl prescription was the natural one to introduce. However, it turns out that this prescription does not work for fermions, since if the Weyl connection is inserted into the curved space Dirac action as is, it is found to drop out identically \cite{Mannheim2014}, with Weyl's attempt to metricate electromagnetism thus failing for fermions. However, when instead the modified  $V^{\lambda}_{\phantom{\alpha}\mu\nu}$ is inserted into the curved space Dirac action, it is found \cite{Mannheim2014} to precisely lead to minimally coupled electromagnetism with action $\int d^4x (-g)^{1/2}i\bar{\psi}\gamma^{\mu}(\partial_{\mu}+\Gamma_{\mu}-ieA_{\mu})\psi$ (the $2/3$ factor in $V^{\lambda}_{\phantom{\alpha}\mu\nu}$ serves to give $A_{\mu}$ the standard minimally coupled weight), where $\Gamma_{\mu}$ is the fermion spin connection as evaluated with the Levi-Civita connection alone. Thus the geometric prescription that leads to the correct coupling of fermions to the vector potential is not to replace $\partial_{\mu}$ by $\partial_{\mu}-2eA_{\mu}$ in the Levi-Civita connection, but to replace it by $\partial_{\mu}-(4ie/3)A_{\mu}$ instead. We note that it is this latter form that respects $CPT$ symmetry, and in so doing it leads to  a geometrically-generated electromagnetic Dirac action that is automatically $CPT$ symmetric. Hence even in the presence of gravity we can establish a $CPT$ theorem. Now as we had noted above,  the conformal gravity theory possesses a non-diagonalizable Jordan-block Hamiltonian. It thus provides an explicit field-theoretic model in which the $CPT$ theorem holds in a non-Hermitian gravitational theory.

Beyond being an example of a non-Hermitian but $CPT$-invariant theory, conformal gravity is of interest in its own right, with the case for local conformal gravity having been made in \cite{Mannheim2012,Mannheim2015}, and the case for local conformal symmetry having been made in \cite{tHooft2014,tHooft2015}. Moreover, if we introduce a fermion Dirac action $I_{\rm D}=\int d^4x (-g)^{1/2}i\bar{\psi}\gamma^{\mu}(\partial_{\mu}+\Gamma_{\mu}-ieA_{\mu})\psi$, then as noted in \cite{tHooft2010a}, if we perform a path integration over the fermions of the form
\begin{eqnarray}
I_{\rm path}=\int D[\psi]D[\bar{\psi}]\exp\left(i\int d^4x (-g)^{1/2}i\bar{\psi}\gamma^{\mu}(\partial_{\mu}+\Gamma_{\mu}-ieA_{\mu})\psi\right), 
\label{F66ab}
\end{eqnarray}
we obtain an effective action of the form 
\begin{eqnarray}
I_{\rm EFF}=\int d^4x (-g)^{1/2}[a C_{\lambda\mu\nu\kappa} C^{\lambda\mu\nu\kappa}+b F_{\mu\nu}F^{\mu\nu}]
\label{F67ab}
\end{eqnarray}
($a$ and $b$ are numerical coefficients), i.e. we obtain none other than the conformal gravity action (as evaluated with the standard Levi-Civita connection) plus the Maxwell action.  Since the $I_{\rm D}$ fermion action is the completely standard one that is used for fermions coupled to gravity and electromagnetism all the time, we see that the emergence of the conformal gravity action is unavoidable in any conventional standard theory. (In a study of quantum gravity 't Hooft  \cite{tHooft2015} has commented  that the inclusion of the conformal gravity action seems to be inevitable.) Since we have seen that the conformal gravity action is not Hermitian but nonetheless $CPT$ symmetric, in any fundamental theory of physics one would at some point have to deal with the issues raised in this paper.

\section{Continuing the $CPT$ and $PT$ Operators and Path Integrals into the Complex Plane}
\label{continuing}

As we have seen, there are two different ways to obtain a real  Euclidean time path integral in which all energy eigenvalues are real -- the Hamiltonian could be Hermitian, or the theory could be in the real eigenvalue realization of  a $CPT$ symmetric but non-Hermitian (and possibly even Jordan-block) Hamiltonian. Thus one needs to ask how is one to determine which case is which. In \cite{Mannheim2013}  a candidate resolution of this issue was suggested. Specifically, the real time (i.e. Minkowski not Euclidean) path integral was studied in some specific models that were charge conjugation invariant (as we discussed in Sec. \ref{constraining}, charge conjugation essentially plays no role at the classical level anyway since at the classical level $eA_{\mu}$ is charge conjugation invariant). In these studies it was found that in the Hermitian case the path integral existed with a real measure, while in the $CPT$ and thus $PT$ case the fields in the path integral measure (but not the coordinates on which they depend) needed to be continued into the complex plane.\footnote{\label{F19} Since we are continuing operators into the complex plane and not the coordinates on which they depend,  for a field $\phi(\vec{x},t)$ we continue the dependence of $\phi$ on $\vec{x}$ and $t$, but not $\vec{x}$ or $t$  themselves (i.e. in  $\phi(x)=\sum a_n(x_{\mu}x^{\mu})^n$ we continue the $a_n$). If we descend to quantum mechanics $\vec{x}$ serves as a non-relativistic stand in for $\phi(\vec{x},t)$ and becomes the operator while $t$ remains a parameter. Then it is the operator that is continued, with its eigenvectors being continued along with it, while its eigenvalues are unaffected. Also, while we continue into the complex plane, for each field component we are restricting to one-dimensional contours in the complex plane (just like the one-dimensional contour on the real axis that we use if we do not continue into the complex plane at all). We are not doubling the number of degrees of freedom by giving the field independent real and imaginary components and then integrating over both of them.} (Continuing the path integral measure into the complex plane is also encountered in 't Hooft's study of quantum gravity \cite{tHooft2011}.) Moreover, should this pattern of behavior prove to be the general rule, it would then explain how quantum Hermiticity arises in a purely c-number based path integral quantization procedure in the first place, since the path integral itself makes no reference to any  Hilbert space whatsoever. Specifically, the general rule would then be that  only if the real time path integral exists with a real measure, and its Euclidean time continuation is real, would the quantum matrix elements that the path integral describes then be associated with a Hermitian Hamiltonian acting on a Hilbert space with a standard Dirac norm. In the section we provide a proof of this proposition.

\subsection{The Pais-Uhlenbeck Two-Oscillator Theory Path Integral}

To see what specifically happens to the path integral in the non-Hermitian case, it is instructive to begin by considering  the path integral associated with the illustrative Pais-Uhlenbeck two-oscillator model that we discussed in Secs.  \ref{implications} and \ref{comparing}. With charge conjugation playing no role in the path integral,  it suffices to discuss the path integral from the perspective of  $PT$ symmetry. For real Minkowski time the path integral is given by 
\begin{eqnarray}
&&G(z_f,x_f,t_f;z_i,x_i,t_i)=\int_i^f {{\cal D}}[z]{{\cal D}}[x] 
 \exp\left[\frac{i\gamma}{2}\int_i^fdt\bigg{(}
\dot{x}^2-(\omega_1^2+\omega^2_2)x^2+\omega_1^2\omega_2^2 z^2\bigg{)}\right].
\label{H68ab}
\end{eqnarray}
Here the path integration is  over independent $z(t)$ and $x(t)$ paths since the equations of motion are fourth-order derivative equations, and thus have twice the number of degrees of freedom as second-order ones, with $x(t)$ replacing $\dot{z}(t)$ and $\dot{x}(t)$ replacing $\ddot{z}(t)$ in the $I_{\rm PU}$ action given in (\ref{H39ab})  \cite{Mannheim2007}.  To enable the path integration  to be asymptotically damped we use the Feynman prescription and replace $\omega_1^2$ and $\omega_2^2$ by $\omega_1^2-i\epsilon$ and $\omega_2^2-i\epsilon$. This then generates an additional contribution to the path integral action of the form 
\begin{eqnarray}
&&i\Delta S =\frac{\gamma}{2}\int_i^f dt \bigg{(}
-2\epsilon x^2+\epsilon(\omega_1^2+\omega_2^2) z^2\bigg{)}.
\label{H69ab}
\end{eqnarray}
While this term provides damping for real $x$ if $\omega_1^2+\omega_2^2$ is positive, it does not do so for real $z$. Thus just as we had discussed in Sec. \ref{comparing} in regard to normalizable wave functions,  to obtain the required damping $z$ needs to be continued into the Stokes wedges associated with the north and south quadrants of a letter $X$ drawn in the complex $z$ plane. In these particular wedges the path integration converges, and is then well-defined. Moreover, since $\omega_1^2+\omega_2^2$ is real and positive for $\omega_1$ and $\omega_2$ both real and unequal, for $\omega_1$ and $\omega_2$ both real and equal, and for $\omega_1$ and $\omega_2$ complex conjugates of each other, the damping is achieved in all three of the possible realizations  of the Pais-Uhlenbeck oscillator, with the path integral existing in all of these three cases, and existing in the self-same Stokes wedge in the three cases. 

The boundaries between Stokes wedges are known as Stokes lines, with it being necessary to continue $z$ into the complex plane until it crosses a Stokes line (the arms of the letter $X$ in the Pais-Uhlenbeck case) in order to get a well-defined real time path integral. For the Pais-Uhlenbeck oscillator with real and unequal $\omega_1$ and $\omega_2$ the well-defined path integral that then ensues is associated with a $PT$-symmetric Hamiltonian, which while not Hermitian is Hermitian in disguise, with all energy eigenvalues being real and bounded from below \cite{Bender2008a}, and with the Euclidean time path integral being real and finite.\footnote{\label{F20} In Euclidean time the Pais-Uhlenbeck Lagrangian given in (\ref{H68ab}) takes the form ${{\cal L}}=(\gamma/2)[-(dx/d\tau)^2-(\omega_1^2+\omega_2^2)x^2+\omega_1^2\omega_2^2z^2]$. On putting $z$ on the imaginary axis as required by (\ref{H69ab}), with positive $\gamma$ the Lagrangian is then negative definite in every Euclidean path. With the needed Euclidean action being given by $\int d\tau{{\cal L}}$,  the needed action is negative definite on every Euclidean path, and the Euclidean time path integral is finite.} And even if $\omega_1$ and $\omega_2$ are complex conjugates of each other, the Euclidean time path integral is still real and finite. The need to continue the path integral measure into the complex plane thus reflects the fact the Pais-Uhlenbeck Hamiltonian is not self-adjoint on the real $z$ axis but is instead $PT$  (and thus $CPT$) symmetric. 

\subsection{Continuing Classical Symplectic Transformations into the Complex Plane}

In order to generalize this result, below we will establish a general complex plane correspondence principle for Poisson brackets and commutators, and then use it to show that in general whenever a continuation of the path integral measure into the complex plane is required, the associated quantum Hamiltonian could not be self-adjoint on the real axis. Moreover, since the discussion depends on the $PT$ symmetry of the Hamiltonian (here we leave out $C$ for simplicity), in a continuation into the complex plane we also need to ask what happens to the $PT$ symmetry. As we now show, it too is continued so that the $[PT,H]=0$ commutator remains intact.  We give the discussion for particle mechanics, with the generalization to fields being direct.

In classical mechanics one can make symplectic transformations that preserve Poisson brackets. A general discussion may for instance be found in \cite{Mannheim2013}, and we adapt that discussion here and consider the simplest case, namely that of a phase space consisting of just one $q$ and one $p$. In terms of the two-dimensional column vector $\eta=\widetilde{(q,p)}$ (the tilde denotes transpose) and an operator $J=i\sigma_2$ we can write a general Poisson bracket as  
\begin{eqnarray}
\{u,v\}=\frac{\partial u}{\partial q}\frac{\partial v}{\partial p}-\frac{\partial u}{\partial p}\frac{\partial v}{\partial q}
=\widetilde{\frac{\partial u}{\partial \eta}}J\frac{\partial v}{\partial \eta}.
\label{H70ab}
\end{eqnarray}
If we now make a phase space transformation to a new two-dimensional vector $\eta^{\prime}=\widetilde{(q^{\prime},p^{\prime})}$ according to 
\begin{eqnarray}
M_{ij}=\frac{\partial \eta^{\prime}_i}{\partial \eta_j},\qquad \frac{\partial v}{\partial \eta}=\tilde{M}\frac{\partial v}{\partial \eta^{\prime}},\qquad \widetilde{\frac{\partial u}{\partial \eta}}=\widetilde{\frac{\partial u}{\partial \eta^{\prime}}}M,
\label{H71ab}
\end{eqnarray}
the Poisson bracket then takes the form
\begin{eqnarray}
\{u,v\}=\widetilde{\frac{\partial u}{\partial \eta^{\prime}}}MJ\tilde{M}\frac{\partial v}{\partial \eta^{\prime}}.
\label{H72ab}
\end{eqnarray}
The Poisson bracket will thus be left invariant for any  $M$ that obeys the symplectic symmetry relation $MJ\tilde{M}=J$.

In the two-dimensional case the relation $MJ\tilde{M}=J$ has a simple solution, viz. $M=\exp(-i\omega \sigma_3)$, and thus for any $\omega$ the Poisson bracket algebra is left invariant. With $q$ and $p$  transforming as 
\begin{eqnarray}
\eta^{\prime}=e^{-i\omega \sigma_3}\eta,~~ q\rightarrow q^{\prime}=e^{-i\omega}q,~~ p\rightarrow p^{\prime}=e^{i\omega}p,
\label{H73ab}
\end{eqnarray}
the $qp$ product and the phase space measure $dqdp$ respectively transform into $q^{\prime}p^{\prime}$ and $dq^{\prime}dp^{\prime}$. With the classical action $\int dt (p\dot{q}-H(q,p))$ transforming into $\int dt (p^{\prime}\dot{q}^{\prime}-H(q^{\prime},p^{\prime}))$, under a symplectic transformation the path integral of the theory is left invariant  too.

Now though it is not always stressed in classical mechanics studies, since $i\omega$  is just a number the Poisson bracket algebra is left invariant even if, in our notation, $\omega$ is not pure imaginary. This then permits us to invariantly continue the path integral into the complex $(q,p)$ plane. Now one ordinarily does not do this because one ordinarily works with (phase space) path integrals that are already well-defined with real $q$ and $p$. However, in the $PT$ case  the path integral is often not well-defined for real $q$ and $p$ but can become so in a suitable Stokes wedge region in the complex $(q,p)$ plane. This means that as one makes the continuation one crosses a Stokes line, with the theories on the two sides of the Stokes line being inequivalent. 

As regards what happens to a $PT$ transformation when we continue into the complex plane, we first need to discuss the effect of $PT$ when $q$ and $p$ are real. When they are real, $P$ effects $q \rightarrow -q$, $p \rightarrow -p$, and $T$ effects $q \rightarrow q$, $p \rightarrow -p$. We can thus set  $PT=-\sigma_3K$ where $K$ effects complex conjugation on anything other than the real $q$ and $p$ that may stand to the right, and set 
\begin{eqnarray}
PT\left(\matrix{q \cr p}\right)=PT\eta=-\sigma_3\left(\matrix{q \cr p}\right)=-\sigma_3\eta.
\label{H74ab}
\end{eqnarray}
Let us now make a symplectic transformation to a new $PT$ operator $(PT)^{\prime}=MPTM^{-1}$. With $i\omega$ being complex the transformation takes the form
\begin{eqnarray}
MPTM^{-1}=e^{-i\omega\sigma_3}(-\sigma_3)e^{-i\omega^*\sigma_3}K.
\label{H75ab}
\end{eqnarray}
With $\eta$ being real, we thus obtain  
\begin{eqnarray}
(PT)^{\prime}\eta^{\prime}=e^{-i\omega\sigma_3}(-\sigma_3)e^{-i\omega^*\sigma_3}e^{i\omega^*\sigma_3}\eta=-\sigma_3\eta^{\prime}.~
\label{H76ab}
\end{eqnarray}
Thus the primed variables transform the same way under the transformed PT operator as the unprimed variables do under the unprimed PT operator. With the Hamiltonian transforming as $H^{\prime}(q^{\prime},p^{\prime})=MH(q,p)M^{-1}$, the classical $\{PT,H\}=\{(PT)^{\prime},H^{\prime}\}=0$ Poisson bracket  is left invariant, in much the same manner as discussed for quantum commutators in Sec. \ref{intro}. The utility of this remark is that once the path integral is shown to be $PT$ symmetric for all real paths, the $PT$ operator will transform in just the right way to enable the path integral to be $PT$ symmetric for complex paths as well. $PT$ symmetry can thus be used to constrain complex plane path integrals in exactly the same way as it can be used to constrain real ones, and to test for $PT$ symmetry one only needs to do so for the real measure case.

\subsection{Continuing Quantum Similarity Transformations into the Complex Plane}

It is also instructive to discuss the quantum analog. Consider a pair of quantum operators $\hat{q}$ and $\hat{p}$ that obey $[\hat{q},\hat{p}]=i$. Apply a similarity transformation of the form $\exp(\omega \hat{p}\hat{q})$ where $\omega$ is a complex number. This yields 
\begin{eqnarray}
\hat{q}^{\prime}&=&e^{\omega \hat{p}\hat{q}}\hat{q} e^{-\omega \hat{p}\hat{q}}=e^{-i\omega }\hat{q},
\qquad
\hat{p}^{\prime}=e^{\omega \hat{p}\hat{q}}\hat{p} e^{-\omega \hat{p}\hat{q}}=e^{i\omega }\hat{p},
\label{H77ab}
\end{eqnarray}
and preserves the commutation relation according to $[\hat{q}^{\prime},\hat{p}^{\prime}]=i$. Now introduce quantum operators $P$ and $T$ that obey $P^2=I$, $T^2=I$, $[P,T]=0$, and effect 
\begin{eqnarray}
P\hat{q}P&=&-\hat{q},~~ T\hat{q}T=\hat{q},~~ PT\hat{q}TP=-\hat{q},~~
P\hat{p}P=-\hat{p},~~T\hat{p}T=-\hat{p},~~ PT\hat{p}TP=\hat{p}.
\label{H78ab}
\end{eqnarray}
Under the similarity transformation the $PT$ and $TP$ operators transform according to 
\begin{eqnarray}
(PT)^{\prime}&=&e^{\omega \hat{p}\hat{q}}PT e^{-\omega \hat{p}\hat{q}}=e^{\omega \hat{p}\hat{q}} e^{\omega^* \hat{p}\hat{q}}PT,
\qquad
(TP)^{\prime}=e^{\omega \hat{p}\hat{q}}TP e^{-\omega \hat{p}\hat{q}}=TPe^{-\omega^* \hat{p}\hat{q}} e^{-\omega \hat{p}\hat{q}}.
\label{H79ab}
\end{eqnarray}
From (\ref{H78ab}) and (\ref{H79ab}) we thus obtain
\begin{eqnarray}
&&(PT)^{\prime}\hat{q}^{\prime}(TP)^{\prime}=e^{\omega \hat{p}\hat{q}} e^{\omega^* \hat{p}\hat{q}}PTe^{-i\omega }\hat{q}TPe^{-\omega^* \hat{p}\hat{q}} e^{-\omega \hat{p}\hat{q}}
\nonumber\\
&&=e^{\omega \hat{p}\hat{q}} e^{\omega^* \hat{p}\hat{q}}e^{i\omega^* }(-\hat{q})e^{-\omega^* \hat{p}\hat{q}} e^{-\omega \hat{p}\hat{q}}
=e^{\omega \hat{p}\hat{q}} e^{i\omega^* }e^{-i\omega^*} (-\hat{q})e^{-\omega \hat{p}\hat{q}}=- e^{-i\omega }\hat{q}=-\hat{q}^{\prime},
\label{H80ab}
\end{eqnarray}
\begin{eqnarray}
(PT)^{\prime}\hat{p}^{\prime}(TP)^{\prime}&=&e^{\omega \hat{p}\hat{q}} e^{\omega^* \hat{p}\hat{q}}PTe^{i\omega }\hat{p}TPe^{-\omega^* \hat{p}\hat{q}} e^{-\omega \hat{p}\hat{q}}
\nonumber\\
&=&e^{\omega \hat{p}\hat{q}} e^{\omega^* \hat{p}\hat{q}}e^{-i\omega^* }\hat{p}e^{-\omega^* \hat{p}\hat{q}} e^{-\omega \hat{p}\hat{q}}
=e^{\omega \hat{p}\hat{q}} e^{-i\omega^* }e^{i\omega^*}\hat{p} e^{-\omega \hat{p}\hat{q}}=e^{i\omega }\hat{p}=\hat{p}^{\prime}.
\label{H81ab}
\end{eqnarray}
Thus the primed variables transform the same way under the transformed PT operator as the unprimed variables do under the unprimed PT operator. With the Hamiltonian being a function of $\hat{q}$ and $\hat{p}$, the $[PT,\hat{H}]=[(PT)^{\prime},\hat{H}^{\prime}]=0$ commutator is left invariant.

As we see, the classical and quantum cases track into each other as we continue into the complex plane, with both the Poisson bracket and commutator algebras being maintained for every $\omega$. We can thus quantize the theory canonically by replacing Poisson brackets by commutators along any direction in the complex $(q,p)$ plane, and in any such direction there will be a correspondence principle for that direction. We  thus generalize the notion of correspondence principle to the complex plane. And in so doing we see that even if the untransformed $\hat{q}$ and $\hat{p}$ are Hermitian, as noted earlier, the transformed $\hat{q}^{\prime}$ and $\hat{p}^{\prime}$ will in general not be since the transformations are not unitary ($(\hat{q}^{\prime})^{\dagger}=e^{i\omega^*}\hat{q}^{\dagger}=e^{i\omega^*}\hat{q}\neq e^{-i\omega}\hat{q}$). However, what will be preserved is their $PT$ structure, with operators thus having well-defined  transformation properties under a $PT$ (i.e. $CPT$) transformation.  

\subsection{Continuing Path Integrals into the Complex Plane}

In order to apply this complex plane correspondence principle to path integrals, we need to compare the path integral and canonical quantization determinations of Green's functions. To this end we look at the matrix element $iG(i,f)=\langle q_i|\exp(-i\hat{H}t)|q_f\rangle$. If one introduces left- and right-eigenstates of the quantum Hamiltonian, then, as we had noted in Sec. \ref{euclidean},  the completeness and orthogonality relations take the form  
\begin{eqnarray}
\sum_n|R_{n}\rangle\langle L_{n}|=\sum_n|L_{n}\rangle\langle R_{n}|=I,\qquad  \langle L_{n}|R_{m}\rangle=\langle R_{m}|L_{n}\rangle=\delta (n,m), 
\label{F82ab}
\end{eqnarray}
while the spectral decomposition of the Hamiltonian is given by $\hat{H}=\sum_n|R_{n}\rangle E_{n}\langle L_{n}|$. Inserting complete sets of states into $G(i,f)$ yields 
\begin{eqnarray}
iG(i,f)=\sum \langle q_i|R_{n}\rangle\exp(-iE_nt)\langle L_{n}|q_f\rangle. 
\label{F83ab}
\end{eqnarray}
In terms of wave functions we thus have 
\begin{eqnarray}
iG(i,f)=\sum \psi_{R_n}(q_i)\exp(-iE_nt)\psi^*_{L_n}(q_f), 
\label{F84ab}
\end{eqnarray}
and can thus express $G(i,f)$ in terms of the eigenfunctions of $\hat{H}$. 

Similarly, if we introduce eigenstates of the position and momentum operators $\hat{q}$ and $\hat{p}$, and insert them into time slices of $\langle q_i|\exp(-i\hat{H}t)|q_f\rangle$, we obtain the path integral representation $iG(i,f)=\int {\cal D}[q]{\cal D}[p]\exp[iS_{\rm CL}(q,p)]$ where $S_{\rm CL}(q,p)=\int dt[p\dot{q}-H(p,q)]$ is the value taken by the classical action on each classical path that connects $q_i$ at $t=0$ with $q_f$ at $t$. Now even in the non-Hermitian Hamiltonian case this expression is the standard path integral representation of $iG(i,f)$ since it only involves the eigenstates of $\hat{q}$ and $\hat{p}$ and makes no reference to the eigenstates of $\hat{H}$. Even if neither $\hat{q}$ nor $\hat{p}$ is self-adjoint when acting on the space of eigenstates of $\hat{H}$, they are always self-adjoint and Hermitian when acting on their own position and momentum eigenstates. As had been noted in Sec. \ref{intro} such a self-adjointness mismatch between the action of the  position and momentum operators on their own eigenstates and on those of the Hamiltonian is central to the $PT$-symmetry program, with a continuation into the complex $(q,p)$ plane being required whenever there is any such mismatch. Thus while there are various ways to represent $\langle q_i|\exp(-i\hat{H}t)|q_f\rangle$, even though it was not originally intended when path integrals were first introduced, we see that writing $iG(i,f)$  as 
$iG(i,f)=\int {\cal D}[q]{\cal D}[p]\exp[iS_{\rm CL}(q,p)]$ provides us with an ideal platform to effect a continuation of $q$ and $p$  into the complex plane.

From the perspective of  path integrals it initially appears that the path integral representation is not sensitive to the domain in the complex $q$ plane in which the wave functions of the quantum Hamiltonian might be normalizable and in which the Hamiltonian acts on them as a self-adjoint operator. However, there is sensitivity to the Hamiltonian, not in writing the path integral down, but in determining the appropriate domain to use for the path integral measure. Specifically, since we may need to continue the coordinates through some complex angle in the complex plane in order to make the quantum Hamiltonian be self-adjoint, the complex plane correspondence principle requires that we would then have to continue the path integral measure through exactly the self-same complex angle. As we show below, when we do need to make such a continuation, it will be the very continuation that will enable the path integral to actually be well-defined and exist.

To implement this continuation we make a similarity transformation $\hat{S}=\exp(-\theta \hat{p}\hat{ q})$ on $\hat{q}$ to obtain $\hat{S}\hat{q}\hat{S}^{-1}=\exp(i\theta) \hat{q}$. With the eigenstates of $\hat{q}$ obeying $\hat{q}|q\rangle=q|q\rangle$, we obtain $\hat{S}\hat{q}\hat{S}^{-1}\hat{S}|q \rangle=\exp(i\theta)\hat{q}\hat{S}|q\rangle=q\hat{S}|q\rangle$, and can thus identify $\hat{S}|q\rangle=|\exp(-i\theta)q\rangle$. Applying a similar analysis to $\langle q|\hat{q}=\langle q|q$ yields $\langle q|\hat{S}^{-1}\hat{S}\hat{q}\hat{S}^{-1}=\langle q|\hat{S}^{-1}\hat{q}\exp(i\theta)=\langle q|\hat{S}^{-1}q$, and can thus identify $\langle q|\hat{S}^{-1}=\langle q\exp(-i\theta)|$. Then with the eigenstates of $\hat{q}$ obeying $\int dq|q\rangle\langle q|=I$ and thus $\int dq\hat{S}|q\rangle\langle q|\hat{S}^{-1}=I$, on setting $q^{\prime}=\exp(-i\theta)q$, we obtain $\exp(i\theta)\int dq^{\prime}|q^{\prime}\rangle\langle q^{\prime}|=I$. The presence of the factor $\exp(-i\theta)$ reflects the fact that $\hat{S}\hat{q}\hat{S}^{-1}$ is not Hermitian since $\hat{S}$ is not unitary. By the same token, with $\hat{S}\hat{p}\hat{S}^{-1}=\exp(-i\theta) \hat{p}$, we obtain $\exp(-i\theta)\int dp^{\prime}|p^{\prime}\rangle\langle p^{\prime}|=I$ where $p^{\prime}=\exp(i\theta)p$.

On introducing the matrix elements $\langle q|R\rangle =\psi_R(q)$, $\langle L|q\rangle =\psi^*_L(q)$, the matrix element  $\langle L|R\rangle$ is given by 
\begin{eqnarray}
\langle L|R\rangle =\int dq \langle L|q\rangle\langle q|R\rangle=\int dq \psi^*_L(q)\psi_R(q). 
\label{F85ab}
\end{eqnarray}
If the wave functions are not normalizable when $q$ is real, we must transform the coordinates into the complex plane to obtain 
\begin{eqnarray}
\langle L|R\rangle =\exp(i\theta)\int dq^{\prime} \langle L|q^{\prime}\rangle\langle q^{\prime}|R\rangle=\exp(i\theta)\int dq^{\prime} \psi^*_L(q^{\prime})\psi_R(q^{\prime}). 
\label{F86ab}
\end{eqnarray}
The theory is well-defined and the $\langle L|R\rangle$ norm  is finite (i.e. probability is finite) if there exists some domain in the complex $q^{\prime}$ plane in which $\int dq^{\prime} \psi^*_L(q^{\prime})\psi_R(q^{\prime})$ is finite. 

In such a domain we must consider Green's functions of the form $iG^{\prime}(i,f)=\langle q^{\prime}_i|\exp(-iHt)|q^{\prime}_f\rangle$. They can be represented by both matrix elements and path integrals of respective form
\begin{eqnarray}
iG^{\prime}(i,f)=\sum \psi_{R_n}(q^{\prime}_i)\exp(-iE_nt)\psi^*_{L_n}(q^{\prime}_f),\qquad iG^{\prime}(i,f)=\int {\cal D}[q^{\prime}]{\cal D}[p^{\prime}]\exp[iS_{\rm CL}(q^{\prime},p^{\prime})]. 
\label{F87ab}
\end{eqnarray}
Since the domain of $q$ and $p$ is chosen so that wave functions are normalizable, on normalizing them to one we obtain 
\begin{eqnarray}
\int dq_iiG(i,i)=\sum \exp(-iE_nt). 
\label{F88ab}
\end{eqnarray}
If all the energy eigenvalues have real parts that are positive (i.e. real parts of the energies  bounded from below), then on sequencing the sum on $n$ so that $Re[E_{n+1}] > Re[E_n]$ and setting $\tau=it$, we find that the modulus of $\exp(-E_{n+1}\tau)/\exp(-E_n\tau$ is less than one for all $n$ if $\tau>0$, with the sum $\sum \exp(-E_n\tau)$ thus being convergent when $\tau$ is positive. In consequence the associated Euclidean time path integral must also be convergent in the same complex $q$, $p$ domain. The complex plane correspondence principle thus translates into the equivalence of the two representations of the Green's function, with the domain in which the quantum Hamiltonian is self-adjoint being associated with the classical domain for which the path integral exists. 

We can thus associate a real path integral measure with real self-adjoint quantum fields, and can associate a complex path integral measure with quantum fields that are only self-adjoint in Stokes wedges that do not include the real axis. Self-adjointness of the quantum Hamiltonian thus correlates with finiteness of the path integral. In consequence, only if the path integral is convergent with a real measure and its Euclidean time continuation is real (i.e. every term in $ \sum \exp(-E_n\tau)$ is real) could the Hamiltonian be Hermitian, though even so the Hamiltonian  would still be $PT$  (i.e. $CPT$) symmetric. However, if the path integral is only convergent if the measure is  complex,  the Hamiltonian would be $PT$ (i.e.  $CPT$) symmetric but not Hermitian (though still possibly Hermitian in  disguise of course). It is thus through the existence of path integrals that are convergent when the measure is real  that Hermiticity can enter quantum theory. However, as noted earlier in our comparison of $CPT$ symmetry and Hermiticity, the emergence of Hermiticity would be output rather than input, with it being dependent on what appropriate path integral measure would be needed in order for the path integral to actually be convergent. Thus, in quantizing physical theories via path integral quantization, Hermiticity of a Hamiltonian never needs to postulated at all, with its presence or absence being determined by the domain of convergence of the path integral of the problem.

\section{Final Comments}
\label{final}

In this paper we have studied the implications for quantum theory of antilinearity of a Hamiltonian and have presented various theorems. We have seen that if a Hamiltonian has an antilinear symmetry, then its eigenvalues are either or real or appear in complex conjugate pairs; while if its eigenvalues are either or real or appear in complex conjugate pairs, then the Hamiltonian must possess an antilinear symmetry. Similarly, we have seen that if a Hamiltonian has an antilinear symmetry, then its left-right inner products are time independent and probability is conserved; while if its left-right inner products are time independent and probability is conserved, then the Hamiltonian must possess an antilinear symmetry. In addition, we have discussed the distinction between Hermiticity and self-adjointness, and have shown that if a Hamiltonian is self-adjoint it must have an antilinear symmetry, and if it has an  antilinear symmetry it must be self-adjoint. Such self-adjointness has primacy over Hermiticity since non-Hermitian Hamiltonians can be self-adjoint. When complex Lorentz invariance is imposed we have shown that the antilinear symmetry is then uniquely specified to be $CPT$. Since no restriction to Hermiticity is required, we thus extend the $CPT$ theorem to non-Hermitian Hamiltonians, and through the presence  of complex conjugate pairs of energy eigenvalues to unstable states. 

As our discussion of the various Levi-Civita, generalized Weyl, and torsion connections given in Sec. \ref{constraining} shows, we  even extend the $CPT$ theorem to include gravity, with its extension to the conformal gravity theory showing that one can have a $CPT$ theorem when a gravitational Hamiltonian (as defined via a linearization about flat spacetime) is not only not Hermitian, one can even have a $CPT$ theorem when a gravitational Hamiltonian is not even diagonalizable. $CPT$ symmetry is thus seen to be altogether more far reaching than Hermiticity, and in general Hamiltonians should be taken to be $CPT$ symmetric  rather than Hermitian.  With Hermiticity of a Hamiltonian when it is in fact found to occur being a property of the solution to a $CPT$-invariant theory and not an input requirement, Hermiticity never needs to be postulated at all.

In comparing $CPT$ symmetry with Hermiticity we note that $C$, $P$, and $T$ symmetries all have a natural connection to spacetime, since $P$ affects spatial coordinates, $T$ affects the time coordinate, and $C$ relates particles propagating forward in time to antiparticles propagating backward in time. As stressed in \cite{Bender2007}, Hermiticity has no such physical association, being instead a purely mathematical requirement. While one can use such a mathematical requirement to derive the $CPT$ theorem, our point here is that one can derive the $CPT$ theorem entirely from physical considerations, namely conservation of probability and invariance under complex Lorentz transformations.

A further distinction between antilinearity and Hermiticity is to be found in Feynman path integral quantization, with Feynman path integral quantization being a purely c-number approach to quantization, while Hermiticity of a Hamiltonian is only definable at the q-number level. Moreover, we have shown that in order to construct the correct classical action needed for a path integral quantization one must impose $CPT$ symmetry on each classical path. Such a requirement has no counterpart in any Hermiticity condition since Hermiticity of a Hamiltonian is only definable after the quantization has been performed and the quantum Hilbert space has been constructed. Hermiticity is thus quite foreign to c-number path integrals while $CPT$ symmetry is perfectly compatible with them.

When Hermiticity was first introduced into quantum mechanics its was done so because in experiments one measures real quantities, and one would like to associate them with real eigenvalues of quantum-mechanical operators, with the operators then being observables. However, one does not need to impose Hermiticity in order to obtain real eigenvalues since Hermiticity is only a sufficient condition for obtaining real eigenvalues, with it being antilinearity that is the necessary condition. In addition, we note that since the eigenvectors of a Hermitian Hamiltonian are stationary, they cannot describe decays. Now while decays would require energy eigenvalues to be complex, the imaginary part of a complex energy is real, and is thus also an observable. Specifically, in a scattering experiment one measures a cross section as a function of energy, and on observing a resonance one identifies the position of peak of the resonance as the real part of the energy of the state and the value of its width as its imaginary part, i.e. one measures two real numbers, the position of the peak and the width.  Thus both the position of the peak and the value of the width are real observable quantities even though the resonance state is described by a complex energy. While such complex energies are foreign to Hermitian Hamiltonians they are perfectly natural for antilinearly symmetric ones, since the presence of complex conjugate pairs of energy eigenvectors and energy eigenvalues ensures the time independence of the appropriate inner products and conservation of probability, just as discussed in Secs.  \ref{intro} and \ref{antilinearity}. Antilinearity thus outperforms Hermiticity. To conclude we note that  $CPT$ symmetry is more far reaching than Hermiticity and can supplant it as a fundamental requirement for physical theories, with it being antilinearity (as realized as  $CPT$) rather than Hermiticity that should be taken to be a guiding principle for quantum theory.

\section{Appendix}

\setcounter{equation}{0}
\def\theequation{A\arabic{equation}}
\subsection{The Majorana Basis for the Dirac Gamma Matrices}

As described for instance in \cite{Mannheim1984}, in terms of the standard Dirac $\gamma^{\mu}_{\rm D}$  basis for the Dirac gamma matrices 
\begin{eqnarray}
\gamma_{\rm D}^{0}=\left(\matrix{I&0\cr 0&-I\cr}\right),\qquad
\gamma_{\rm D}^{i}=\left(\matrix{0&\sigma_i\cr -\sigma_i&0\cr}\right),
\label{A1}
\end{eqnarray}
one constructs the Majorana  basis via 
\begin{eqnarray}
\gamma_{\rm M}^{\mu}=\frac{1}{\surd{2}}(1-\gamma^2_{\rm D})\gamma_{\rm D}^{\mu}\frac{1}{\surd{2}}(1+\gamma^2_{\rm D}),
\label{A2}
\end{eqnarray}
to yield 
\begin{eqnarray}
&&\gamma_{\rm M}^{0}=\left(\matrix{0&\sigma_2\cr \sigma_2&0\cr}\right),\qquad
\gamma_{\rm M}^{1}=-i\left(\matrix{\sigma_3&0\cr 0&\sigma_3\cr}\right),\qquad
\gamma_{\rm M}^{2}=\left(\matrix{0&\sigma_2\cr -\sigma_2&0\cr}\right),
\nonumber \\
&&
\gamma_{\rm M}^{3}=i\left(\matrix{\sigma_1&0\cr 0&\sigma_1\cr}\right),\qquad 
\gamma_{\rm M}^{5}=\left(\matrix{-\sigma_2&0\cr 0&\sigma_2\cr}\right),\qquad 
C_{\rm M}=\left(\matrix{0&\sigma_2\cr \sigma_2&0\cr}\right),
\label{A3}
\end{eqnarray}
where $\gamma_{\rm M}^{5}=i\gamma_{\rm M}^{0}\gamma_{\rm M}^{1}\gamma_{\rm M}^{2}\gamma_{\rm M}^{3}$ and $C_{\rm M}$ effects $C_{\rm M}\gamma_{\rm M}^{\mu}C_{\rm M}^{-1}=-\widetilde{\gamma_{\rm M}^{\mu}}$. These matrices obey the standard $\gamma^{\mu}_{\rm M}\gamma^{\nu}_{\rm M}+\gamma^{\nu}_{\rm M}\gamma^{\mu}_{\rm M}=2\eta^{\mu\nu}$, and as constructed, every non-zero element of every  $\gamma_{\rm M}^{\mu}$, of $\gamma_{\rm M}^5$, and of $C_{\rm M}$ is pure imaginary. In the Majorana basis $C_{\rm M}=\gamma_{\rm M}^{0}$.

With the gamma matrices one then constructs the six antisymmetric $M^{\mu\nu}=i[\gamma^{\mu},\gamma^{\nu}]/4$, to obtain 
\begin{eqnarray}
&&M_{\rm M}^{01}=\frac{i}{2}\left(\matrix{0&\sigma_1\cr \sigma_1&0\cr}\right),\qquad
M_{\rm M}^{02}=\frac{i}{2}\left(\matrix{-I&0\cr 0&I\cr}\right),\qquad
M_{\rm M}^{03}=\frac{i}{2}\left(\matrix{0&\sigma_3\cr \sigma_3&0\cr}\right),
\nonumber \\
&&M_{\rm M}^{12}=\frac{i}{2}\left(\matrix{0&-\sigma_1\cr \sigma_1&0\cr}\right),\qquad
M_{\rm M}^{23}=\frac{i}{2}\left(\matrix{0&\sigma_3\cr -\sigma_3&0\cr}\right),\qquad
M_{\rm M}^{31}=\frac{1}{2}\left(\matrix{\sigma_2&0\cr 0&\sigma_2\cr}\right).
\label{A4}
\end{eqnarray}
The six $M_{\rm M}^{\mu\nu}$ satisfy the infinitesimal Lorentz generator algebra given in (\ref{H29ab}), and as constructed every non-zero element of every  $M_{\rm M}^{\mu\nu}$ is pure imaginary. Consequently, for real $w_{\mu\nu}$ the transformation  $\exp(iw_{\mu\nu}M_{\rm M}^{\mu\nu})$ is purely real, and thus maintains the reality of a real Majorana spinor under a real Lorentz transformation.

In the vector representation of the Lorentz group the $M_{\rm V}^{\mu\nu}$ are given by
\begin{eqnarray}
&&M_{\rm V}^{01}=\left(\matrix{0&i&0&0\cr i&0&0&0\cr 0&0&0&0\cr 0&0&0&0\cr}\right),~~
M_{\rm V}^{02}=\left(\matrix{0&0&i&0\cr 0&0&0&0\cr i&0&0&0\cr 0&0&0&0\cr}\right),~~
M_{\rm V}^{03}=\left(\matrix{0&0&0&i\cr 0&0&0&0\cr 0&0&0&0\cr i&0&0&0\cr}\right),
\nonumber \\
&&M_{\rm V}^{12}=\left(\matrix{0&0&0&0\cr 0&0&-i&0\cr 0&i&0&0\cr 0&0&0&0\cr}\right),~~
M_{\rm V}^{23}=\left(\matrix{0&0&0&0\cr 0&0&0&0\cr 0&0&0&-i\cr 0&0&i&0\cr}\right),~~
M_{\rm V}^{31}=\left(\matrix{0&0&0&0\cr 0&0&0&i\cr 0&0&0&0\cr 0&-i&0&0\cr}\right).~
\label{A5}
\end{eqnarray}
These six $M_{\rm V}^{\mu\nu}$ also satisfy the Lorentz algebra given in (\ref{H29ab}), and as constructed every non-zero element of every  $M_{\rm V}^{\mu\nu}$ is pure imaginary. Consequently, for real $w^{\mu\nu}$ the transformation  $\exp(iw_{\mu\nu}M_{\rm V}^{\mu\nu})$ is also purely real, and thus maintains the reality of a real vector under a real Lorentz transformation.

\subsection{Quantization of Majorana Spinors}

To quantize fermionic fields one needs to specify the value of the equal time anticommutator $\{\psi_{\alpha}(\vec{x},t),\psi^{\dagger}_{\beta}(\vec{y},t)\}$. Since the combination $\{\psi_{\alpha}(\vec{x},t),\psi^{\dagger}_{\beta}(\vec{y},t)\}$ is Hermitian in quantum field space, quantization must set it equal to a real c-number times a delta function, which we write as $R_{\alpha\beta}\delta^3(\vec{x}-\vec{y})$. Moreover, since this anticommutator transforms as a $4 \otimes 4$ tensor product in the $(\alpha, \beta)$ Dirac gamma matrix space, we can write $R_{\alpha\beta}$ as $R_{\alpha\beta}=\sum_i a_i\Gamma^i_{\alpha\beta}$ as summed over the 16 $\Gamma^i$ of the form $I,~\gamma^5,~\gamma^{\mu},~\gamma^{\mu}\gamma^5,~[\gamma^{\mu},\gamma^{\nu}]$, with each different choice defining  its own quantization scheme. Since we can write any Dirac spinor as $\psi=(\psi+\psi^{\dagger})/2+(\psi-\psi^{\dagger})/2$, we can set $\psi=\psi^1+i\psi^2$ where $\psi^1$ and $\psi^2$ are Hermitian. Now while we can make such a decomposition in any basis for the Dirac gamma matrices, it is only in the Majorana basis that $\psi^1$ and $\psi^2$ are respectively self-conjugate and anti-self-conjugate. In the 16 $\Gamma^i$ expansion there are 10 symmetric $S_{\alpha\beta}$ matrices and 6 antisymmetric $A_{\alpha\beta}$ matrices. Thus in general, and on restricting to $\vec{x}=\vec{y}$,  the anticommutation relations take the form
\begin{eqnarray}
\psi^1_{\alpha}\psi^1_{\beta}+\psi^1_{\beta}\psi^1_{\alpha}+\psi^2_{\alpha}\psi^2_{\beta}+\psi^2_{\beta}\psi^2_{\alpha}
&=& S_{\alpha\beta}\delta^3(\vec{0}),
\nonumber\\
i[\psi^2_{\alpha}\psi^1_{\beta}+\psi^1_{\beta}\psi^2_{\alpha}-\psi^1_{\alpha}\psi^2_{\beta}-\psi^2_{\beta}\psi^1_{\alpha}]
&=&A_{\alpha\beta}\delta^3(\vec{0}).
\label{A6}
\end{eqnarray}

Given these relations we can evaluate the components of the scalar ($S=\bar{\psi}\psi$), pseudoscalar ($P=\bar{\psi}i\gamma^5\psi$), vector ($V^{\mu}=\bar{\psi}\gamma^{\mu}\psi$),  axial vector ($A^{\mu}=\bar{\psi}\gamma^{\mu}\gamma^5\psi$), and tensor ($T^{\mu\nu}=\bar{\psi}i[\gamma^{\mu},\gamma^{\nu}]\psi$) fermion bilinears that are of interest to us in this paper,  where $\bar{\psi}=(\psi^1-i\psi^2)\gamma^0$, $\psi=\psi^1+i\psi^2$. For any given choice of $S_{\alpha\beta}$ and $A_{\alpha\beta}$ some set of the fermion bilinears will acquire c-number $\delta^3(\vec{0})$ contributions arising from the fact that the bilinears are composed of singular products of fields at the same spacetime point. These infinities are removed by normal-ordering. And given (\ref{A6}), following some algebra we obtain normal-ordered bilinears of the form
\begin{eqnarray}
S&=&-2i[\psi^1_{1}\psi^1_{4}+\psi^1_{3}\psi^1_{2}+\psi^2_{1}\psi^2_{4}+\psi^2_{3}\psi^2_{2}],
\nonumber\\
P&=&2i[\psi^1_{1}\psi^1_{3}+\psi^1_{2}\psi^1_{4}+\psi^2_{1}\psi^2_{3}+\psi^2_{2}\psi^2_{4}].
\label{A7}
\end{eqnarray}
\begin{eqnarray}
V^0&=&i[\psi^1_{1}\psi^2_{1}-\psi^2_{1}\psi^1_{1}+\psi^1_{2}\psi^2_{2}-\psi^2_{2}\psi^1_{2}
+\psi^1_{3}\psi^2_{3}-\psi^2_{3}\psi^1_{3}+\psi^1_{4}\psi^2_{4}-\psi^2_{4}\psi^1_{4}],
\nonumber\\
V^1&=&2i[\psi^1_{1}\psi^2_{4}-\psi^2_{1}\psi^1_{4}+\psi^1_{2}\psi^2_{3}-\psi^2_{2}\psi^1_{3}],
\nonumber\\
V^2&=&-i[\psi^1_{1}\psi^2_{1}-\psi^2_{1}\psi^1_{1}+\psi^1_{2}\psi^2_{2}-\psi^2_{2}\psi^1_{2}
-\psi^1_{3}\psi^2_{3}+\psi^2_{3}\psi^1_{3}-\psi^1_{4}\psi^2_{4}+\psi^2_{4}\psi^1_{4}],
\nonumber\\
V^3&=&2i[\psi^1_{1}\psi^2_{3}-\psi^2_{1}\psi^1_{3}-\psi^1_{2}\psi^2_{4}+\psi^2_{2}\psi^1_{4}].
\label{A8}
\end{eqnarray}
\begin{eqnarray}
A^0&=&2i[\psi^1_{1}\psi^1_{2}+\psi^1_{4}\psi^1_{3}+\psi^2_{1}\psi^2_{2}+\psi^2_{4}\psi^2_{3}],
\nonumber\\
A^1&=&2i[\psi^1_{1}\psi^1_{3}+\psi^1_{4}\psi^1_{2}+\psi^2_{1}\psi^2_{3}+\psi^2_{4}\psi^2_{2}],
\nonumber\\
A^2&=&-2i[\psi^1_{1}\psi^1_{2}+\psi^1_{3}\psi^1_{4}+\psi^2_{1}\psi^2_{2}+\psi^2_{3}\psi^2_{4}],
\nonumber\\
A^3&=&-2i\psi^1_{1}\psi^1_{4}+\psi^1_{2}\psi^1_{3}+\psi^2_{1}\psi^2_{4}+\psi^2_{2}\psi^2_{3}].
\label{A9}
\end{eqnarray}
\begin{eqnarray}
T^{01}&=&2i [\psi^1_{1}\psi^2_{1}-\psi^2_{1}\psi^1_{1}-\psi^1_{2}\psi^2_{2}+\psi^2_{2}\psi^1_{2}
+\psi^1_{3}\psi^2_{3}-\psi^2_{3}\psi^1_{3}-\psi^1_{4}\psi^2_{4}+\psi^2_{4}\psi^1_{4}],
\nonumber\\
T^{02}&=&4i[\psi^1_{1}\psi^2_{4}-\psi^2_{1}\psi^1_{4}-\psi^1_{3}\psi^2_{2}+\psi^2_{3}\psi^1_{2}],
\nonumber\\
T^{03}&=&-4i[\psi^1_{1}\psi^2_{2}-\psi^2_{1}\psi^1_{2}+\psi^1_{3}\psi^2_{4}-\psi^2_{3}\psi^1_{4}],
\nonumber\\
T^{12}&=&2i[\psi^1_{1}\psi^2_{1}-\psi^2_{1}\psi^1_{1}-\psi^1_{2}\psi^2_{2}+\psi^2_{2}\psi^1_{2}
-\psi^1_{3}\psi^2_{3}+\psi^2_{3}\psi^1_{3}+\psi^1_{4}\psi^2_{4}-\psi^2_{4}\psi^1_{4}],
\nonumber\\
T^{23}&=&-4i[\psi^1_{1}\psi^2_{2}-\psi^2_{1}\psi^1_{2}-\psi^1_{4}\psi^2_{3}+\psi^2_{4}\psi^1_{3}],
\nonumber\\
T^{31}&=&4i[\psi^1_{1}\psi^2_{3}-\psi^2_{1}\psi^1_{3}+\psi^1_{2}\psi^2_{4}-\psi^2_{2}\psi^1_{4}],
\label{A10}
\end{eqnarray}
in the Majorana basis for the Dirac gamma matrices, regardless of the specific quantization scheme chosen. As a check on our calculations,  we apply the $\hat{\pi}\hat{\tau}=\hat{\Lambda}^{0}_{\phantom{0}3}(i\pi)\hat{\Lambda}^{0}_{\phantom{0}2}(i\pi)\hat{\Lambda}^{0}_{\phantom{0}1}(i\pi)$ Lorentz transformation of interest to us in this paper. With it implementing $\psi(x)\rightarrow\gamma^5 \psi(-x)$, $\tilde{\psi}(x)\rightarrow\tilde{\psi}(-x)\tilde{\gamma^5}$, we find that the bilinear products transform as $S(x)\rightarrow S(-x)$, $P(x)\rightarrow P(-x)$, $V^{\mu}(x)\rightarrow -V^{\mu}(-x)$, $A^{\mu}(x)\rightarrow -A^{\mu}(-x)$, $T^{\mu\nu}(x)\rightarrow T^{\mu\nu}(-x)$, just as required.

\subsection{Implications of Complex Conjugation}

In applying complex conjugation one ordinarily takes $K$ to act on c-numbers but not on q-numbers, so that for the typical $\psi^1+i\psi^2$, $K$ is taken to effect $K(\psi^1+i\psi^2)K=\psi^1-i\psi^2$. However, this is not a general rule, since if we apply $K$ to the $[\hat{x},\hat{p}]=i$ commutator we find that $K[\hat{x},\hat{p}]K=-i$. Hence one of $\hat{x}$ and $\hat{p}$ must conjugate into minus itself. Now both $\hat{x}$ and $\hat{p}$  are Hermitian, and given the $[\hat{x},\hat{p}]=i$ commutator, both $\hat{x}$ and $\hat{p}$ can be represented as infinite-dimensional matrices. If one sets $\hat{x}=(a+a^{\dagger})/\sqrt{2}$, $\hat{p}=i(a^{\dagger}-a)/\sqrt{2}$, so that $[a,a^{\dagger}]=1$, then in the Fock space with a vacuum that obeys $a|\Omega\rangle=0$, we find that $\hat{x}$ is represented by an infinite-dimensional matrix that is real and symmetric (analog of $\sigma_1$), while $\hat{p}$ is represented by an infinite-dimensional matrix that is pure imaginary and antisymmetric (analog of $\sigma_2$). Complex conjugation thus does see the $i$ factor in $\hat{p}$ and effects $K\hat{p}K=-\hat{p}$ while leaving $\hat{x}=K\hat{x}K$ untouched. 

For field theory exactly the same situation prevails for the canonical commutator $[\phi(\vec{x},t),\pi(\vec{y},t)]=i\delta^3(\vec{x}-\vec{y})$, and with one ordinarily taking the Hermitian $\phi(x)$ to be a real and symmetric infinite-dimensional matrix that obeys $K\phi(x)K=\phi(x)$, one must take the Hermitian $\pi(x)$ to be a pure imaginary and antisymmetric infinite-dimensional matrix that obeys $K\pi(x)K=-\pi(x)$. However, since one ordinarily only discusses how operations such as time reversal affect the fields that  appear in the Lagrangian, one does not  need to discuss how complex conjugation might affect their canonical conjugates. 

However, for fermions the situation can be different. Ordinarily one chooses to set $R_{\alpha\beta}=I_{\alpha\beta}$ (in any basis for the gamma matrices), to give 
\begin{eqnarray}
\psi^1_{\alpha}\psi^1_{\beta}+\psi^1_{\beta}\psi^1_{\alpha}+\psi^2_{\alpha}\psi^2_{\beta}+\psi^2_{\beta}\psi^2_{\alpha}
&=& I_{\alpha\beta}\delta^3(\vec{0}),
\nonumber\\
i[\psi^2_{\alpha}\psi^1_{\beta}+\psi^1_{\beta}\psi^2_{\alpha}-\psi^1_{\alpha}\psi^2_{\beta}-\psi^2_{\beta}\psi^1_{\alpha}]
&=&0.
\label{A11}
\end{eqnarray}
And even though the anticommutation relations are then consistent with each component of the Hermitian $\psi^1_{\alpha}$ and $\psi^2_{\beta}$ being represented by matrices that are real and symmetric, one could equally represent these relations by appropriately choosing some or even all of the components of $\psi^1_{\alpha}$ and $\psi^2_{\beta}$ to be pure imaginary and antisymmetric (cf. $\sigma_2^2=I$).

The above remarks also hold in the Dirac basis of the gamma matrices if one sets $R_{\alpha\beta}=(\gamma^0_{\rm D})_{\alpha\beta}$ since $\gamma^0_{\rm D}$ is real and diagonal, differing only from $I$ in the signs but not in the reality of its two lower components. However, if one sets $R_{\alpha\beta}=(\gamma^0_{\rm M})_{\alpha\beta}$ in the Majorana basis, one encounters two differences. First, one would have multiply by $i$ since $(\gamma^0_{\rm M})_{\alpha\beta}$ is pure imaginary, so as to give $R_{\alpha\beta}=i(\gamma^0_{\rm M})_{\alpha\beta}$. And second, $(\gamma^0_{\rm M})_{\alpha\beta}$ is antisymmetric in its $(\alpha,\beta)$ indices. Thus with this quantization scheme we obtain 
\begin{eqnarray}
i[\psi^2_{\alpha}\psi^1_{\beta}+\psi^1_{\beta}\psi^2_{\alpha}-\psi^1_{\alpha}\psi^2_{\beta}-\psi^2_{\beta}\psi^1_{\alpha}]
&=&i(\gamma^0_{\rm M})_{\alpha\beta}\delta^3(\vec{0}),
\nonumber\\
\psi^1_{\alpha}\psi^1_{\beta}+\psi^1_{\beta}\psi^1_{\alpha}+\psi^2_{\alpha}\psi^2_{\beta}+\psi^2_{\beta}\psi^2_{\alpha}
&=&0.
\label{A12}
\end{eqnarray}
Now, since the Hermitian $\gamma^0_{\rm M}$ is pure imaginary and antisymmetric, every term in $\psi^2_{\alpha}\psi^1_{\beta}+\psi^1_{\beta}\psi^2_{\alpha}-\psi^1_{\alpha}\psi^2_{\beta}-\psi^2_{\beta}\psi^1_{\alpha}$ must be pure imaginary, and thus must be affected by complex conjugation. Thus with the choices $R_{\alpha\beta}=I_{\alpha\beta}$, $R_{\alpha\beta}=(\gamma^0_{\rm D})_{\alpha\beta}$ some of the representations of the  fermion fields could be pure imaginary. However, with the choice $R_{\alpha\beta}=i(\gamma^0_{\rm M})_{\alpha\beta}$ some of the representations must be pure imaginary. Thus whether or not Hermitian fields are affected by complex conjugation is not an intrinsic property of the fields themselves, but is instead a property of the structure of the quantization conditions. Thus in general we see that complex conjugation can act non-trivially on q-number fields depending on how they are represented, with the general rule being that $K$ complex conjugates all factors of $i$ no matter where they might appear. Thus in imposing complex conjugation one does not need to differentiate between c-numbers and q-numbers at all.

For our purposes here we shall quantize using $R_{\alpha\beta}=i(\gamma^0_{\rm M})_{\alpha\beta}$. With $(\gamma^0_{\rm M})_{14}=-(\gamma^0_{\rm M})_{23}=(\gamma^0_{\rm M})_{32}=-(\gamma^0_{\rm M})_{41}=-i$, we can realize (\ref{A12}) with the Hermitian $\psi^1_{1}$, $\psi^1_{2}$, $\psi^2_{1}$, $\psi^2_{2}$ all being real matrices, and the Hermitian $\psi^1_{3}$, $\psi^1_{4}$, $\psi^2_{3}$, $\psi^2_{4}$ all being pure imaginary ones. With this realization we find that $S$, $P$, $V^1$, $V^3$, $A^1$, $A^3$, $T^{02}$, $T^{31}$ are all real, while  $V^0$, $V^2$, $A^0$, $A^2$, $T^{01}$, $T^{03}$, $T^{12}$, $T^{23}$ are all pure imaginary. Thus under $K$ they transform as
\begin{eqnarray}
&&KSK=S,~~KPK=P, 
\nonumber\\
&&K(V^0,V^1,V^2,V^3)K=(-V^0,V^1,-V^2,V^3),
\nonumber\\
&&K(A^0,A^1,A^2,A^3)K=(-A^0,A^1,-A^2,A^3), 
\nonumber\\
&&K(T^{01},T^{02},T^{03},T^{12}, T^{23}, T^{31})K
=(-T^{01},T^{02},-T^{03},-T^{12}, -T^{23}, T^{31}).
\label{A13}
\end{eqnarray}
While we see alternations in sign under $K$ within given Lorentz multiplets, such a pattern is familiar from the rotation group where $\sigma_1$ and $\sigma_3$ are real and $\sigma_2$ is imaginary.

Our interest in this paper is only in spin zero combinations as they are the only combinations that can appear in a Lorentz invariant Lagrangian. The needed combinations are thus $S$, $P$, $V^{\mu}V_{\mu}$, $V^{\mu}A_{\mu}$, $A^{\mu}A_{\mu}$, $T^{\mu\nu}T_{\mu\nu}$. Quite remarkably, the pattern of plus and minus signs in (\ref{A13}) is such that every single one of these spin zero combinations is completely real.\footnote{\label{F21} It was in order to achieve this reality condition that we took charge conjugation to be an antilinear operator in  \cite{Mannheim2016}. In this paper we take charge conjugation to be a linear operator, and derive this same reality condition via a judicious fermion anticommutator quantization condition.}. Thus all of these combinations are invariant under both Lorentz transformations and complex conjugation, just as needed for the derivation of the $CPT$ theorem presented in this paper.

While we have quantized the fermion fields so that $K$ changes the signs of the two lower components of the $\psi_{\alpha}$ spinor, this does not mean that time reversal does too. Rather time reversal must effect $\hat{T}\psi(\vec{x},t)\hat{T}^{-1}=\gamma^1\gamma^2\gamma^3\psi(\vec{x},-t)$ as this is the transformation that leaves the action for a free Dirac field invariant. Now the time reversal operator can be written as $\hat{U}K$ where $\hat{U}$ is unitary. Ordinarily one introduces the standard $\hat{U}_1$ that with $K$ effects $\hat{T}\psi(\vec{x},t)\hat{T}^{-1}=\gamma^1\gamma^2\gamma^3\psi(\vec{x},-t)$ when $K$ is taken not to affect q-numbers at all. Thus in our case we  set $\hat{U}=\hat{U}_1\hat{U}_2$ where $\hat{U}_2$ effects $\hat{U}_2\psi(\vec{x},t)\hat{U}_2^{-1}=\gamma^2\gamma^0\psi(\vec{x},t)$ as this also reverses the signs of the two lower components of the spinor. Thus with $\hat{T}=\hat{U}_1\hat{U}_2K$, the effect of time reversal on $\psi(\vec{x},t)$ is the standard one that effects $\hat{T}\psi(\vec{x},t)\hat{T}^{-1}=\gamma^1\gamma^2\gamma^3\psi(\vec{x},-t)$. And indeed, it was using this standard form for the time reversal transformation that the entries in Tables I and II given in Sec. \ref{cpt} were obtained.

\subsection{Comparing the Charge Conjugation Operator with the $PT$ Theory ${\cal C}$ Operator}

In quantum field theory  the charge conjugation operator obeys $[\hat{C},\hat{H}]=0$, $\hat{C}^2=I$, and in $PT$ theory there exists a ${\cal C}$ operator that obeys $[{{\cal C}},\hat{H}]=0$, ${{\cal C}}^2=I$. It was noted in \cite{Mannheim2016} that with every Hamiltonian being $CPT$ invariant, in the event that the Hamiltonian is also charge conjugation invariant, one would then have  a $PT$ invariant Hamiltonian that possesses an additional charge conjugation invariance, to thus suggest \cite{Mannheim2016} that the $\hat{C}$ and ${\cal C}$ operators could be one and the same. Attractive as this possibility is, we show here that this is not in fact the case. However, if it is not to be the case, then one has to ask where the ${\cal C}$ operator invariance comes from if it is not to be charge conjugation invariance, and need to ask why a Hamiltonian should then possess two separate $C$-type invariances. We address these issues here.

To see why there is a difference between the two $C$-type operators, it suffices to consider the simple matrix $M(s)$ given in (\ref{H1ab}). As noted in Sec. \ref{intro}, in its $s^2>1$ and $s^2<1$ realizations (energies real and energies in a complex pair) the $PT$ theory ${\cal C}$ operator is given by ${{\cal C}}(s^2>1)=(\sigma_1+i\sigma_3\cos\alpha)/\sin \alpha $ where $\sin \alpha=(s^2-1)^{1/2}/s$, and ${{\cal C}}(s^2<1)=(\sigma_1+i\sigma_3\cosh\beta)/i\sinh \beta $ where $\sinh\beta=(1-s^2)^{1/2}/s$. First, we note that these two expressions differ from each other, and second we note that both become singular when $s^2=1$, the point at which the Hamiltonian becomes Jordan block. Such a behavior cannot occur for charge conjugation, since a Hamiltonian is either charge conjugation invariant or it is not, and its status under charge conjugation or the structure of the charge conjugation operator cannot change as one varies c-number coefficients since charge conjugation only acts on q-number fields. Also, charge conjugation  is not sensitive to any possible Jordan-block structures, with a Jordan-block Hamiltonian being able to be charge conjugation invariant. 

However, before concluding definitively that the ${\cal C}$ operator does not exist in the Jordan-block case even though the charge conjugation operator does exist, we have to show that there is no other choice for ${\cal C}$ that might exist in this case. To this end we consider the $s^2=1$ structure of our simple model as given in the Jordan canonical form exhibited on the right hand side of (\ref{H2ab}), where $M=\sigma_0+(\sigma_1+i\sigma_2)/2$. If there is to be a ${\cal C}$ operator for it, the ${\cal C}$ operator must take the form ${{\cal C}}=c_0\sigma_0+c_i\sigma_i$, and if it is to square to one and not simply be the identity matrix the coefficients must obey $c_0=0$, $c_1^2+c_2^2+c_3^2=1$. On setting $[{{\cal C}},M]=0$ we obtain $-ic_2\sigma_3+ic_3\sigma_2-c_1\sigma_3+c_3\sigma_1=0$. Thus we need $c_1+ic_2=0$, $c_3=0$. Since these conditions are not compatible with $c_1^2+c_2^2+c_3^2=1$, we conclude that there is no solution  $[{{\cal C}},M]=0$, ${{\cal C}}^2=I$ in the Jordan-block case except the identity matrix, and only it would be continuous in the continuing through the three $s^2>1$, $s^2=1$ and $s^2<1$ regions. 

Even though we have only derived this result in the two-dimensional case, this result is in fact quite general for any antilinear operator for which we can continue parameters to go from the Jordan-block domain to the domain where energy eigenvalues appear in  complex conjugate pairs. In that domain we only need to look at each pair separately, and  since each such pair forms  a two-dimensional system, we can continue back to the Jordan-block case pair by pair, to thus establish that the only allowed ${\cal C}$ operator that is continuous in the Jordan-block limit is the identity matrix. That of course does not mean that we cannot use a non-trivial ${\cal C}$ operator away from the Jordan-block  limit, it is just that any such non-trivial ${\cal C}$ operator would have to be singular in the limit.  Moreover, since the charge conjugation operator would obey the same two conditions (commute with the Hamiltonian and square to one) as the ${\cal C}$ operator in the event the Hamiltonian is charge conjugation invariant, we can also conclude that for any charge conjugation invariant field-theoretic Hamiltonian that can be Jordan block, the charge conjugation operator must be the identity operator. In fact we have even met an example of this  -- the neutral scalar field theory with the action given in (\ref{H41ab}), as both the neutral scalar field and the associated Hamiltonian are  charge conjugation even, with the Hamiltonian becoming Jordan block when $M_1^2=M_2^2$. Since the gravitational field is charge conjugation even, similar remarks thus apply to the conformal gravity theory, since its Hamiltonian is non-diagonalizable.

We thus have to conclude that the charge conjugation operator $\hat{C}$ and the $PT$ theory ${\cal C}$ operator are different independent operators. Moreover, $\hat{C}$ is a spacetime based operator whose action on fields is intrinsic to the fields themselves no matter in what particular Hamiltonian they might appear, whereas the structure found for the ${\cal C}$ operator in our example shows it to depend intrinsically on the structure of the Hamiltonian, to thus change as one goes from one Hamiltonian to another. 

Since we did find that the ${\cal C}$ operator becomes singular in the Jordan-block limit, this suggests that when a ${\cal C}$ operator does exist it should be related to the Hamiltonian-dependent similarity transformation that brings a given diagonalizable Hamiltonian to a diagonal form, since this similarity transform must also become singular in the Jordan-block limit if the Hamiltonian is not to be diagonalizable in the limit. We now show that this is indeed the case. 

Thus consider a general diagonalizable Hamiltonian $H$ that is brought to diagonal form by the similarity transform $BHB^{-1}=H_{D}$. In the diagonal form one can always find a non-trivial operator ${{\cal C}}_{D}$ that will commute with $H_{D}$ and square to one. Specifically, one only needs every diagonal element of ${{\cal C}}_{D}$ to be $+1$ or $-1$, and this can always be achieved. If for instance $H_D$ is $N$-dimensional, we can use the $N$ diagonal $\lambda_i$ operators of $U(N)$ as a complete basis for any diagonal operator in that space. Since we can form $N$ independent linear combinations of the diagonal $\lambda_i$, we have just the right number of degrees of freedom to be able to specify the $N$ diagonal elements  of ${{\cal C}}_D$ in that space. In order to be definitive, we shall always define the  ${{\cal C}}_D$ operator of interest to be the one that has equal numbers of $+1$ and $-1$ diagonal elements when $N$ is even, and to have one  additional $+1$ element when $N$ is odd.\footnote{\label{F22} For $N=2$ for instance we can take ${{\cal C}}_D=\sigma_3$ (i.e. ${\rm diag}[C_D]= (1,-1)$). And for $N=3$ where ${\rm diag}[\lambda_0]=(\surd{2}/\surd{3}, \surd{2}/\surd{3}, \surd{2}/\surd{3})$, ${\rm diag}[\lambda_3]=(1,-1,0)$, ${\rm diag}[\lambda_8]=(1/\surd{3},1/\surd{3},-2/\surd{3})$, we can take ${{\cal C}}_D=\lambda_3+\lambda_0/\surd{6}-\lambda_8/\surd{3}$ (i.e. ${\rm diag}[C_D]= (1,-1,1)$).} Finally, now having defined the diagonal elements of ${{\cal C}}_D$, we can transform back to the original basis to identify ${\cal C}=B^{-1}{{\cal C}}_DB$. This then gives us the desired ${\cal C}$ operator for any diagonalizable Hamiltonian (with either real or complex pair eigenvalues), while showing that a non-trivial ${\cal C}$ operator must always exist in such cases, i.e. it must exist simply because of diagonalizability, even though it has no relation to the charge conjugation operator. Finally, since a Jordan-block Hamiltonian cannot be diagonalized, the $B$ operator must become singular in the Jordan-block limit, with ${\cal C}=B^{-1}{{\cal C}}_DB$ becoming undefined.

Some further constraints on ${\cal C}$ can be obtained in the event that all eigenvalues are real. Specifically, in this case all the eigenvalues of the diagonal $H_D$ are real and $H_D$ is Hermitian. Thus now we obtain $BHB^{-1}=H_D=H_D^{\dagger}=(B^{-1})^{\dagger}H^{\dagger}B^{\dagger}$, to yield $B^{\dagger}BHB^{-1}(B^{\dagger})^{-1}=H^{\dagger}$. Thus on defining $V=B^{\dagger}B$ we obtain $VHV^{-1}=H^{\dagger}$. We thus recognize the $V$ operator that transforms $H$ into $H^{\dagger}$ to be related to the $B$ operator that transforms $H$ into $H_D$. Now with $V$ being of the form $B^{\dagger}B$, $V$ is not only Hermitian, it is a positive operator of the type introduced by Mostafazadeh \cite{Mostafazadeh2002}, with all of its eigenvalues being positive. Since that is the case, we can write $V=G^2$ where $G$ is also a Hermitian operator. We thus obtain $(GHG^{-1})^{\dagger}=G^{-1}H^{\dagger}G=G^{-1}G^2HG^{-2}G=GHG^{-1}$, with $GHG^{-1}$ thus being Hermitian. Since one can bring a Hermitian operator to a diagonal form by a unitary transformation $U$, we can set $B=UG$, and can thus identify ${\cal C}=G^{-1}{{\cal C}}_UG$, where ${{\cal C}}_U=U^{-1}{{\cal C}}_DU$. We can thus express ${\cal C}$ in terms of the operator $G$ that effects $G^2HG^{-2}=H^{\dagger}$. With ${\cal C}=G^{-2}{{\cal C}}_U+G^{-2}[G{{\cal C}}_UG-{{\cal C}}_U]$, it is often the case  in $PT$ studies that $G{{\cal C}}_UG-{{\cal C}}_U=0$, in which case we can set ${{\cal C}}=G^{-2}{{\cal C}}_U=V^{-1}{{\cal C}}_U={{\cal C}}^{-1}={{\cal C}}_UV$.\footnote{\label{F23} For the example given in (\ref{H14ab}), $U=(\sigma_0+i\sigma_2)/\surd{2}$, $U(\sigma_0+\sigma_1\tan\alpha)U^{-1}=\sigma_0+\sigma_3\tan\alpha$, ${{\cal C}}_D=\sigma_3={{\cal C}}_D^{-1}$,  ${{\cal C}}_U=\sigma_1$, $G^{\pm 1}=A\sigma_0\pm B\sigma_2$, and $G\sigma_1G=\sigma_1$, to give ${{\cal C}}=G^{-2}\sigma_1=V^{-1}P={{\cal C}}^{-1}=PV$.} And since we have seen that in general we should use the $V$ norm, in those cases where $G{{\cal C}}_UG={{\cal C}}_U$ we can justify the use of the ${{\cal C}}$ operator norm that is used in $PT$ studies.\footnote{\label{F24} In those $PT$ symmetric cases in which $T=K$, and $H$ is symmetric, we have $H=PTHT^{-1}P^{-1}=PK\tilde{H}KP^{-1}=PH^{\dagger}P^{-1}$, and since $VHV^{-1}=H^{\dagger}$,  we see that ${{\cal C}}=PV$ commutes with $H$. Then in those cases in which in addition $V^{-1}P=PV$, we also have ${{\cal C}}^2=I$, with ${\cal C}$ again being related to $V$. In addition, when all energies are real one can also use $[PT,H]=0$ to find an operator that commutes with $H$. Specifically, if we set $PT=LK$ where $L$ is linear, we can write $H=PTHT^{-1}P^{-1}=LKB^{-1}H_DBKL^{-1}=L[B^{-1}]^*KH_DKB^*L^{-1}$. Then, since $KH_DK=H_D$ when all energies are real, we obtain $H=L[B^{-1}]^*BHB^{-1}B^*L^{-1}$. In terms of the operator $E=L[B^{-1}]^*B$ we thus obtain $H=EHE^{-1}$, and $E$ commutes with $H$.}

Having seen the utility of the ${{\cal C}}$ operator norm, we note that if we were to quantize the fermion field described earlier using $R_{\alpha\beta}=(\gamma^0_{\rm D})_{\alpha\beta}$, while the two lower components of $\psi_{\alpha}$ would be quantized with a negative sign, this could be compensated for in inner products by using a ${{\cal C}}$ operator norm, where ${\rm diag}[{{\cal C}}]=(1,1,-1,-1)$. Such an $R_{\alpha\beta}=(\gamma^0_{\rm D})_{\alpha\beta}$ quantization procedure thus has the structure of a $PT$ theory.

\subsection{Causality in a Non-Hermitian but $CPT$-Symmetric Fourth-Order Derivative Quantum Field Theory}

 Consider a fourth-order plus second-order derivative scalar field theory based on the action 
\begin{eqnarray}
I_S&=&\frac{1}{2}\int d^4x\bigg{[}\partial_{\mu}\partial_{\nu}\phi\partial^{\mu}
\partial^{\nu}\phi-(M_1^2+M_2^2)\partial_{\mu}\phi\partial^{\mu}\phi
+M_1^2M_2^2\phi^2\bigg{]},
\label{A14}
\end{eqnarray}
a theory which, as noted in Sec. \ref{implications}, is $CPT$ symmetric but not Hermitian. For this theory the propagator obeys
\begin{eqnarray}
&&(\partial_t^2-\nabla^2+M_1^2)(\partial_t^2-\nabla^2+M_2^2)
 D^{(4)}(x^2,M_1^2,M_2^2)=\delta^4(x).
\label{A15}
\end{eqnarray}
If we introduce a standard second-order theory propagator that obeys
\begin{eqnarray}
(\partial_t^2-\nabla^2+M^2)D^{(2)}(x^2,M^2)=\delta^4(x),
\label{A16}
\end{eqnarray}
we can reexpress the fourth-order propagator as 
\begin{eqnarray}
&&D^{(4)}(x^2,M_1^2,M_2^2)={1 \over (M_2^2-M_1^2)}
[D^{(2)}(x^2,M_1^2)-D^{(2)}(x^2,M_2^2)],
\label{A17}
\end{eqnarray}
and it can readily be checked that this form for $D^{(4)}(x^2,M_1^2,M_2^2)$ obeys (\ref{A15}). 

For the second-order case the standard retarded propagator is given 
\begin{eqnarray}
D^{(2)}(x^2,M^2)={1 \over 4\pi r}\theta(t)\delta(t-r)-
{M \over 4\pi(t^2-r^2)^{1/2}}\theta(t-r)J_1(M(t^2-r^2)^{1/2}).
\label{A18}
\end{eqnarray}
Since $D^{(2)}(x^2,M^2)$ does not take support outside the light cone but only on or inside it, there can be no response to a signal emitted at time $t=0$ that would register any point that obeys $r>t$, and thus there can be no incoming waves at spatial infinity. Given (\ref{A17}) and (\ref{A18}), one can construct a retarded fourth-order theory propagator of the form
\begin{eqnarray}
D^{(4)}(x^2,M_1^2,M_2^2)&=&-{\theta(t-r) \over 4\pi(t^2-r^2)^{1/2}(M_2^2-M_1^2)}
\nonumber\\
&\times&\bigg{[}M_1J_1(M_1(t^2-r^2)^{1/2})-M_2J_1(M_2(t^2-r^2)^{1/2})\bigg{]},
\label{A19}
\end{eqnarray}
and it also does not take support outside the light cone. Thus, as noted in \cite{Mannheim2007}, 
the relative minus sign between the two second-order terms in (\ref{A17}) has no effect on the causality of the fourth-order theory. And with $M_1^2$ and $M_2^2$ both being real, causality is completely standard and there is no response outside the light cone. (As noted in \cite{Bender2008a} and as discussed above, this relative minus sign does not lead to ghost states either.)

As constructed, $I_S$ involves both fourth-order and second-order derivative terms. Consider now the pure fourth-order theory case where $M_1^2$ and $M_2^2$ are both set to zero. In this limit the action is given by
\begin{eqnarray}
I_S=\frac{1}{2}\int d^4x\partial_{\mu}\partial_{\nu}\phi\partial^{\mu}\partial^{\nu}\phi,
\label{A20}
\end{eqnarray}
and remains $CPT$ symmetric. In this case  the propagator given in (\ref{A19}) readily limits to \cite{Mannheim2007}
\begin{eqnarray}
D^{(4)}(x^2,M_1^2=0,M_2^2=0)&=&{1 \over 8\pi}\theta(t-r),
\label{A21}
\end{eqnarray}
and it can readily be checked \cite{Mannheim2007} that it obeys
\begin{eqnarray}
(\partial_t^2-\nabla^2)^2 D^{(4)}(x^2,M_1^2=0,M_2^2=0)=\delta^4(x).
\label{A22}
\end{eqnarray}
As we see, causality is not lost, and the pure fourth-order propagator does not take support outside the light cone. Now just like the Pais-Uhlenbeck oscillator theory,  the pure fourth-order scalar field theory Hamiltonian is also of Jordan-block form, with the pure fourth-order theory time-dependent Schr\"odinger equation having runaway solutions that grow linearly in time. As we see, the presence of runaways does not lead to any violations of causality.\footnote{\label{F25} In the standard discussion of the electromagnetic radiation reaction problem one encounters a third-order differential equation of motion that also has linear runaways. However since radiation reaction is caused by the effect of Lienard-Wiechert potentials, there can be no violations of causality since these potentials are fully retarded and do not take support outside the light cone. (In some treatments of the radiation reaction problem one introduces an asymptotic future boundary condition that there be no runaways at $t=+\infty$. However, such an assumption is acausal as it requires that one know in advance what the solution is to look like at late times, and leads to preaccelerations. This acausality is an artifact of using a non-causal future boundary condition and does not entail that temporal runaways entail any loss of causality.)}

The mathematical reason why there is no loss of causality when $M_1^2$ and $M_2^2$ are both set to zero is that in (\ref{A18}) the factors that control the causal structure of the propagator are the mass-independent $\delta(t-r)$ and $\theta(t-r)$ factors, with all of the mass dependence being in the arguments of the Bessel functions. Since there is no causal sensitivity to mass terms, we can even consider the complex conjugate case where $M_1^2=M^2+iN^2$, $M_2^2=M^2-iN^2$, with $M^2$ and $N^2$ both real. Since the factors $M_1^2+M_2^2=2M^2$ and $M_1^2M_2^2=M^4+N^4$ given in (\ref{A14}) remain real, the action $I_S$ remains $CPT$ symmetric. In this case the retarded propagator is given by  
\begin{eqnarray}
D^{(4)}(x^2,M^2,N^2)&=&{\theta(t-r) \over 4\pi(t^2-r^2)^{1/2}2iN^2}
\bigg{[}(M^2+iN^2)^{1/2}J_1[(M^2+iN^2)^{1/2}(t^2-r^2)^{1/2}]
\nonumber\\
&-&(M^2-iN^2)^{1/2}J_1[(M^2-iN^2)^{1/2}(t^2-r^2)^{1/2}]\bigg{]},
\label{A23}
\end{eqnarray}
and does not take support outside the light cone, with the propagator being causal even though solutions to a wave equation with complex masses grow exponentially in time. Thus in all of three of its realizations (real, zero, and complex conjugate $M_1^2$ and $M_2^2$) the action $I_S$ is non-Hermitian but $CPT$ symmetric, and in each case propagation is causal.


\bigskip
\begin{thebibliography}{99}

\bibitem{Bender1998}Bender  C M and Boettcher S Phys. Rev. Lett. 1998 \textbf{80}, 5243

\bibitem{Bender1999} Bender C M, Boettcher S and Meisinger P N J. Math. Phys. 1999 \textbf{40}, 2201

\bibitem{Bender2007} Bender C M Rep. Prog. Phys. 2007 \textbf{70}, 947 

\bibitem{Special2012} \textit{Special issue on quantum physics with non-Hermitian operators}, Bender C, Fring A, G\"unther U and Jones H (Guest Editors) J. Phys. A: Math. Theor. 2012 \textbf{45}, 444001 - 444036 

\bibitem{Theme2013} \textit{Theme issue on PT quantum mechanics}, Bender C M, DeKieviet M and Klevansky S P (Guest Editors) Phil. Trans. R. Soc. A 2013 \textbf{371}, issue 1989  

\bibitem{Guo2009} Guo A et. al. Phys. Rev. Lett. 2009 \textbf{103}, 093902

\bibitem{Bender2008a} Bender C M and Mannheim P  D Phys. Rev. Lett. 2008 \textbf{100}, 110422

\bibitem{Bender2008b} Bender C M and Mannheim P D Phys. Rev. D 2008 \textbf{78}, 025022

\bibitem{Mannheim2011} Mannheim P D Gen. Rel. Gravit. 2011 \textbf{43}, 703

\bibitem{Mannheim2012} Mannheim P D Found. Phys. 2012 \textbf{42}, 388 


\bibitem{Mannheim2013} Mannheim P D Phil. Trans. R. Soc. A 2013 \textbf{371}, 20120060

\bibitem{Lee1969} Lee T D and Wick G C Nucl. Phys. B 1969 \textbf{9}, 209; Nucl. Phys. B 1969 \textbf{10}, 1; Phys. Rev. D 1970 \textbf  {2}, 1033

\bibitem{Scholtz1992} Scholtz F G, Geyer H B and Hahne F J W, Ann. Phys.
(N.Y.) 1992 \textbf{213}, 74 

\bibitem{Mostafazadeh2002} Mostafazadeh A J. Math. Phys. 2002 \textbf{43}, 205, 2814, 3944 

\bibitem{Bender2002} Bender C M, Berry M V and Mandilara A J. Phys. A: Math. Gen. 2002 \textbf{35}, L467 

\bibitem{Bender2010} Bender C M and Mannheim P D Phys. Lett. A 2010 \textbf{374}, 1616

\bibitem{Siegl2012}  Siegl  P and Krejcirik D Phys. Rev. D 2012 \textbf{86}, 121702 

\bibitem{Mannheim2016} Mannheim P D Phys. Lett. B 2016 \textbf{753}, 288 

\bibitem{Streater1964} Streater R F and Wightman A S \textit{PCT, Spin and Statistics, and all that} 1964, W. A. Benjamin, New York 

\bibitem{Weinberg1995} Weinberg S \textit{The Quantum Theory of Fields: Volume I} 1995,  Cambridge University Press Cambridge, U. K.

\bibitem{Selover2013} Selover M and Sudarshan E C G \textit{Derivation of the TCP Theorem using Action Principles}, 
arXiv:1308.5110 [hep-th], August 2013.

\bibitem{Mannheim1984} Mannheim P D Int. J. Theor. Phys. 1984 \textbf{23}, 643

\bibitem{Mannheim1985} Mannheim P D Phys. Rev. D 1985 \textbf{32}, 898 



\bibitem{Mannheim2007} Mannheim P D Found. Phys. 2007 \textbf{37}, 415 

\bibitem{Mannheim2014} Mannheim P D~\textit{$PT$  Symmetry, Conformal Symmetry, and the Metrication of Electromagnetism}, Found. Phys. 2016 (in press).

\bibitem{Mannheim2015b} Mannheim P D J. Phys: Conf. Ser. 2015 \textbf{615}, 012004 

\bibitem{Mannheim2016b} Mannheim P D  Int. J. Mod. Phys. D 2016 \textbf{25}, 1644003

\bibitem{Mannheim2015} Mannheim P D \textit{Living Without Supersymmetry -- the Conformal Alternative and a Dynamical Higgs Boson}, arXiv:1506.01399 [hep-ph], June 2015.

\bibitem{tHooft2014} 't Hooft G Int. J. Mod. Phys. D 2015 \textbf{24}, 1543001

\bibitem{tHooft2015} 't Hooft G \textit{Singularities, Horizons, Firewalls, and Local Conformal Symmetry}, arXiv:1511.04427 [gr-qc], November 2015.

\bibitem{tHooft2010a} 't Hooft G \textit{Probing the Small Distance Structure of Canonical Quantum Gravity using the Conformal Group}, arXiv:1009.0669 [gr-qc], September, 2010.

\bibitem{tHooft2011} 't Hooft G Found. Phys. 2011 \textbf{41}, 1829 

\end{thebibliography}
\end{document}